\documentstyle[tighten,preprint,prd,floats,aps]{revtex}

\font\tenscr=rsfs10 scaled \magstep1
\font\sevenscr=rsfs7 scaled \magstep1
\font\fivescr=rsfs5 scaled \magstep1
\skewchar\tenscr='177 \skewchar\sevenscr='177 \skewchar\fivescr='177
\newfam\scrfam \textfont\scrfam=\tenscr \scriptfont\scrfam=\sevenscr
\scriptscriptfont\scrfam=\fivescr
\def\scr{\fam\scrfam}

\begingroup\makeatletter\ifx\SetFigFont\undefined
\def\x#1#2#3#4#5#6#7\relax{\def\x{#1#2#3#4#5#6}}%
\expandafter\x\fmtname xxxxxx\relax \def\y{splain}%
\ifx\x\y   % LaTeX or SliTeX?
\gdef\SetFigFont#1#2#3{%
  \ifnum #1<17\tiny\else \ifnum #1<20\small\else
  \ifnum #1<24\normalsize\else \ifnum #1<29\large\else
  \ifnum #1<34\Large\else \ifnum #1<41\LARGE\else
     \huge\fi\fi\fi\fi\fi\fi
  \csname #3\endcsname}%
\else
\gdef\SetFigFont#1#2#3{\begingroup
  \count@#1\relax \ifnum 25<\count@\count@25\fi
  \def\x{\endgroup\@setsize\SetFigFont{#2pt}}%
  \expandafter\x
    \csname \romannumeral\the\count@ pt\expandafter\endcsname
    \csname @\romannumeral\the\count@ pt\endcsname
  \csname #3\endcsname}%
\fi
\fi\endgroup

\input eepic.sty
\def\Bu{\lower.6ex\hbox{\kern.1em$\bullet$}}

\def\Tr{\mathop{\rm Tr}}

\begin{document}
\preprint{IFUP-TH 63/94}
\title{Strong-coupling expansion of chiral models}
\author{Massimo Campostrini, Paolo Rossi, and Ettore Vicari}
\address{Dipartimento di Fisica dell'Universit\`a and I.N.F.N.,
I-56126 Pisa, Italy}
\maketitle

\begin{abstract}
The strong-coupling character expansion of lattice models is
reanalyzed in the perspective of its complete algorithmization.  The
geometric problem of identifying, counting, and grouping together all
possible contributions is disentangled from the group-theoretical
problem of weighting them properly.  The first problem is completely
solved for all spin models admitting a character-like expansion and
for arbitrary lattice connectivity.  The second problem is reduced to
the evaluation of a class of invariant group integrals defined on
simple graphs.  Since these integrals only depend on the global
symmetry of the model, results obtained for principal chiral models
can be used without modifications in lattice gauge theories.

By applying the techniques and results obtained we study the
two-dimensional principal chiral models on the square and honeycomb
lattice.  These models are a prototype field theory sharing with QCD
many properties.  Strong-coupling expansions for Green's functions are
derived up to 15th and 20th order respectively.  Large-$N$ and
$N=\infty$ results are presented explicitly.  Related papers are
devoted to a discussion of the results.
\end{abstract}

\section{Introduction}
\label{introduction}

It is certainly appropriate to consider two-dimensional principal
chiral models as a theoretical physics laboratory.  These models
display a rich physical structure, and share with four-dimensional
gauge theories a number of fundamental properties: nonabelian symmetry
with fields in the matrix representation, asymptotic freedom,
dynamical mass generation.  Moreover, principal chiral models admit a
$1/N$ expansion and a large-$N$ limit which is a sum over planar
diagrams, in total analogy with nonabelian gauge theories.

However, the absence of local gauge invariance and the reduced number
of dimensions make chiral models much simpler to handle both by
analytical and by numerical methods.  Moreover, the on-shell solution
of the models is known by Bethe-Ansatz methods: a factorized
$S$-matrix exists and the particle spectrum is explicitly known.  We
can therefore try to make progress, both in analytical and in
numerical techniques, by testing these methods on chiral models and,
in case of success, applying them to four-dimensional gauge theories.

The spirit of this approach is well expressed in the papers by Green
and Samuel \cite{Green-Samuel-1,Green-Samuel-2,Green-Samuel-3}, who
advocated a systematic study of lattice chiral models as a preliminary
step towards an understanding of lattice gauge theories, especially in
the large-$N$ limit.  One of the techniques favored by the
above-mentioned authors was the strong-coupling character expansion.
However, the existence of a large-$N$ phase transition from strong to
weak coupling phase seemed to indicate at that time an obstruction to
further pushing this method of investigation.

In much more recent times, a few facts came to suggest that this
``no-go'' result might be over-pessimistic.  It was indeed
observed by the present authors \cite{Rossi-Vicari-chiral1,%
Rossi-Vicari-chiral2,Campostrini-Rossi-Vicari-strong} that scaling of
physical observables is present in finite-$N$ chiral models already in
a coupling region within the convergence radius of the strong-coupling
expansion.  Moreover, a change of variables corresponding to adopting
the so-called ``energy scheme'' for the definition of the temperature
smoothens the lattice $\beta$-function to the point that asymptotic
scaling is observed within the strong-coupling region.  These patterns
are unaffected by growing $N$, and therefore survive the large-$N$
phase transition.  These ``experimental'' observations led us to
reconsidering the possibility that a strong-coupling approach could be
turned into a predictive method for the evaluation of physical
quantities in the neighborhood of the continuum fixed point of the
models.

A second theoretical motivation for a renewed effort towards extending
strong-coupling series of chiral models, especially for large $N$,
comes in connection with the possibility that the above-mentioned
transition, while uninteresting for the standard continuum physics,
may be related to a description of quantum gravity by
the so-called ``double scaling limit'' \cite{Douglas-Shenker,%
Gross-Migdal,Brezin-Kazakov,Periwal-Shevitz}.  In simple models,
this limit is studied by analytical techniques, but more complex
situations might need perturbative methods, and strong-coupling seems
well suited for such an analysis, which corresponds to exploring the
region in the vicinity of the first singularity in the complex
coupling constant plane.

Another significant change, of a completely different nature, has
occurred since the original studies on the strong-coupling character
expansion (cf.\ Ref.~\cite{Drouffe-Zuber} for a review) were performed.
The increased availability of symbolic-manipulation computer programs
and the enormous increase in performance of computers have now made
the strong-coupling expansion a plausible candidate for an algorithmic
implementation, that might extend series well beyond the level that
can be reached by purely human resources, while granting a definitely
higher reliability of results.

The purpose of the present work is to set the stage and make a
significant effort towards a complete algorithmization of the
strong-coupling character expansion.

Two major classes of problems must be handled and solved.  The first
has to do with counting the multiplicities of terms appearing in the
expansion.  It is basically a geometrical problem and it leads to the
definition of a ``geometrical factor''.  We must stress the fact that
this geometrical factor depends only on the lattice connectivity, and
therefore applies without any modification to the strong-coupling
expansion of all spin models admitting a character-like expansion,
including ${\rm O}(N)$ and ${\rm CP}^{N-1}$ models with
nearest-neighbour interactions \cite{Rabinovici-Samuel}.
We have completely solved this problem, % for lattice chiral models
with no conceptual restrictions on the dimensionality and connectivity
of the lattice.  We have not addressed the corresponding problem for
lattice gauge theories, but we are confident that no major conceptual
obstruction should arise in pursuing that program.

The second class of problems is related to the evaluation of group
integrals that appear as coefficients of the expansion.  Evaluating
group integrals is an algebraic problem, and in principle a solved
one.  However, algorithmic implementation is not in practice a trivial
task, and therefore we limited ourselves to a general classification
and to an explicit evaluation of the cases of direct interest to our
calculations, with a few useful generalizations.  We stress that the
evaluation of ``group-theoretical factors'' is universal, and results
may be applied as they stand to lattice gauge theories.

The representation of the strong-coupling expansion in terms of
explicitly evaluated geometrical factors and symbolically denoted
group-theoretical factors can be achieved by a fully computerized
approach, and applies as it stands to all nonlinear sigma models
defined on group manifolds.  This is probably the main result of the
present paper.  However we shall not exhibit here the explicit general
formulae resulting from our approach, because they are so long that
their pratical use does necessarily involve computer manipulation;
therefore we shall make available our results in form of computer
files, publicly available by anonymous ftp on the host
{\tt ftp.difi.unipi.it}, in the directory
{\tt pub/campo/StrongCoupling}.

The application of our results to ${\rm O}(N)$ and ${\rm CP}^{N-1}$
models is definitely simpler then the case discussed here, since the
evaluation of group-theoretical factors lends itself to a completely
algorithmic implementation.  The corresponding results will be
presented in a forthcoming paper.

The present paper is organized as follows.

In Sect.~\ref{generalities} we review the character expansion, fix our
notation and present some useful formulae.

In Sect.~\ref{outline} we outline the procedure of the expansion by
identifying the logical steps and defining the relevant geometrical
and algebraic objects entering the computation.  Among these we
introduce the basic notion of a {\it skeleton diagram}, whose
multiplicity is the geometrical factor and whose connected value, or
{\it potential}, is the group-theoretical factor.

In Sect.~\ref{geometrical-factor} we explain how one may
algorithmically evaluate the geometrical factor.

In Sect.~\ref{group-theoretical-factor} we introduce the problem of
computing the group-theoretical factor.

Sect.~\ref{technical-remarks} is devoted to some technical remarks on
group integration.

Sect.~\ref{computing-potentials} offers some details on the
computation of potentials for principal chiral models.

In Sect.~\ref{Greens-functions} we analyze the main features of the
strong-coupling expansion of the two-point fundamental Green's
functions, introducing a parametrization for the propagator in the
case of a two-dimensional square lattice.

In Sect.~\ref{honeycomb-lattice} we discuss the relevant features of
the honeycomb lattice, and we present a few results for physical
quantities.

Appendix \ref{values-potentials} is devoted to a presentation of some
of our results concerning the explicit evaluation of potentials.

Appendix \ref{directory-potentials} is a list of potentials ordered
according to their appearance in the strong-coupling expansion.

Appendix \ref{N-square-results} is a presentation our results for large
but finite $N$ in the square-lattice formulation of the models.

Appendix \ref{honeycomb-gaussian} clarifies some non-standard features
of honeycomb-lattice models using the Gaussian model as a guide.

Appendix \ref{N-honeycomb-results} is the same as
App.~\ref{N-square-results} for the honeycomb-lattice formulation.

The present paper is the first of a series of papers devoted to the
strong-coupling analysis of two-dimensional lattice chiral models.
In a second paper we will present our analysis of the large-$N$
strong-coupling series by series-resummation techniques, while a
third paper will be devoted to a comparison with Monte Carlo studies
of the large-$N$ critical behavior.

\section{The character expansion: generalities}
\label{generalities}

The strong-coupling expansion of field theories involving
matrix-valued fields and enjoying $G \times G$ group symmetry is best
performed applying the character expansion, which reduces the number
of contributions to a given order in the expansion and decouples the
geometrical counting of configurations from the group-theoretical
factor.

The whole subject is reviewed in detail in Ref.~\cite{Drouffe-Zuber},
and we recall here only those properties that are essential in order
to make our presentation as far as possible self-contained.  We shall
only discuss the symmetry groups $G={\rm U}(N)$: extensions to
${\rm SU}(N)$ can be achieved following Ref.~\cite{Green-Samuel-1} and
applying the results presented in Ref.~\cite{Rossi-Vicari-chiral2}.

In the theory described by the lattice action
\begin{equation}
S_L = - N \beta \sum_{x,\mu} \bigl[
\Tr\{U(x)\,U^\dagger(x{+}\mu)\} +
\Tr\{U(x{+}\mu)\,U^\dagger(x)\}\bigr],
\end{equation}
the character expansion is achieved by replacing the Boltzmann factors
with their Fourier decomposition
\begin{eqnarray}
&&\exp\bigl\{N\beta\bigl[
\Tr\{U(x)\,U^\dagger(x{+}\mu)\} +
\Tr\{U(x{+}\mu)\,U^\dagger(x)\}\bigr]\bigr\}
\nonumber \\ && \quad=\;
\exp\bigl\{N^2 F(\beta)\bigr\} \sum_{(r)}
d_{(r)} {\tilde z}_{(r)}(\beta)\,\chi_{(r)}
\bigl(U(x)\,U^\dagger(x{+}\mu)\bigr),
\label{Fourier}
\end{eqnarray}
where
\begin{equation}
F(\beta) = {1\over N^2}\,\ln\int dU \exp\bigl\{N\beta \bigl(
\Tr U + \Tr U^\dagger\bigr)\bigl\}
= {1\over N^2}\,\ln\det I_{j-i}(2N\beta)
\end{equation}
is the free energy of the single-matrix model, $\sum_{(r)}$ denotes
the sum over all irreducible representations of $G$, $\chi_{(r)}(U)$
and $d_{(r)}$ are the corresponding characters and dimensions
respectively, and $I_{j-i}$ ($i,j=1,...,N$) is a $N{\times}N$ matrix
of modified Bessel functions.  We recall here the orthogonality
relations for representations:
\begin{equation}
\int dU {\scr D}^{(r)}_{ab}(U)\,{\scr D}^{(s)\,*}_{cd}(U) =
{1\over d_{(r)}}\,\delta_{(r),(s)}\,\delta_{a,c}\,\delta_{b,d},
\qquad  \chi_{(r)}(U) = {\scr D}^{(r)}_{aa}(U).
\label{ortho}
\end{equation}

In ${\rm U}(N)$ groups $(r)$ is characterized by two sets of
decreasing positive integers $\{l\} = l_1,...,l_s$,
$\{m\} = m_1,...,m_t$ and we define the order $n$ of $(r)$ by
\begin{equation}
n = n_+ + n_-, \quad n_+ = \sum_{i=1}^s l_i,
\quad n_- = \sum_{j=1}^t m_j \,.
\label{n}
\end{equation}
We may define the ordered set of integers
$\{\lambda\} = \lambda_1,...,\lambda_N$ by the relationships
\begin{eqnarray}
\lambda_k =& l_k, \quad& k \le s; \nonumber \\
\lambda_k =& 0, \quad& s<k<N-t+1; \nonumber \\
\lambda_k =& -m_{N-k+1}, \quad& k \ge N-t+1.
\end{eqnarray}
It is then possible to write down explicit representations of all
characters and dimensions:
\begin{eqnarray}
\chi_{\{\lambda\}}(U) &=&
{\det\bigl|\exp\{i\phi_i(\lambda_j+N-j)\}\bigr| \over
 \det\bigl|\exp\{i\phi_i(N-j)\}\bigr|} \,, \\
d_{\{\lambda\}} &=& {\prod_{i<j} (\lambda_i-\lambda_j+j-i) \over
 \prod_{i<j} (j-i)} = \chi_{\{\lambda\}}(1),
\label{characters-raw}
\end{eqnarray}
where $\phi_i$ are the eigenvalues of the matrix $U$.  As a
consequence, it is possible to evaluate explicitly the character
coefficients ${\tilde z}_{(r)}$ by the orthogonality relations
\begin{eqnarray}
d_{\{\lambda\}} {\tilde z}_{\{\lambda\}} &=& \int dU \exp\bigl\{
N\beta\bigl(\Tr U + \Tr U^\dagger\big)\bigr\} \chi_{\{\lambda\}}(U)
\exp\bigl\{-N^2 F(\beta)\bigr\} \nonumber \\
&=& {\det I_{\lambda_i + j - i}(2N\beta) \over
     \det I_{j - i}(2N\beta)}\,.
\label{dz}
\end{eqnarray}

Eq.~(\ref{dz}) becomes rapidly useless with growing $N$, due to the
difficulty of evaluating determinants of large matrices.  It is however
possible to obtain considerable simplifications, in the strong
coupling regime and for sufficiently large $N$, when we consider
representations such that $n<N$.  In this case, character coefficients
are simply expressed by \cite{Green-Samuel-1}
\begin{equation}
d_{\{l; m\}} {\tilde z}_{\{l; m\}} =
{1\over n_+!}\, {1\over n_-!}\,
\sigma_{\{l\}}\sigma_{\{m\}} (N\beta)^n
\bigr[1 +  O\bigr(\beta^{2N}\bigl)\bigl],
\label{dze}
\end{equation}
where we have introduced the quantity $\sigma_{\{l\}}$, the dimension
of the representation $l_1,...,l_s$ of the permutation group, enjoying
the property
\begin{equation}
\int \chi_{\{l\}}(U) (\Tr U^\dagger)^p dU =
\sigma_{\{l\}} \delta_{p,n_+} \,.
\end{equation}

It is important to notice that the strong-coupling expansion and the
large-$N$ limit do not commute: large-$N$ character coefficients have
jumps and singularities at $\beta=\case{1}{2}$ \cite{Green-Samuel-1},
and therefore the relevant region for a strong-coupling character
expansion is $\beta<\case{1}{2}$.

A consequence of Eq.~(\ref{dze}) is the relationship
\begin{equation}
z \equiv {\tilde z}_{1;0}(\beta) = \beta + O\bigl(\beta^{2N+1}\bigr),
\label{z1-beta}
\end{equation}
and in turn, because of the property
\begin{equation}
z(\beta) = {1\over2}\,{\partial\over\partial\beta} F(\beta),
\end{equation}
one may obtain the large-$N$ relationship
\begin{equation}
F(\beta) = \beta^2 + O\bigl(\beta^{2N+2}\bigr).
\end{equation}
According to the above observations, at $N=\infty$ the relationship
$F=\beta^2$ may only hold when $\beta<\case{1}{2}$, even if the
function $\beta^2$ is perfectly regular for all $\beta$.

For the purpose of actual computations it is convenient to have
expressions in closed form for the quantities $\sigma_{\{l\}}$
and $d_{\{l; m\}}$ not involving infinite sums or
products even in the $N\to\infty$ limit.  We found such expressions in
the form
\begin{equation}
{1\over n_+!} \sigma_{l_1,...,l_s} =
{\prod_{1\le j<k\le s} (l_j-l_k+k-j)! \over \prod_{i=1}^s (l_i+s-i)!}
\end{equation}
and a similar relationship for $\sigma_{\{m\}}$.  Notice that
these quantities are independent of $N$.  Now by manipulating appropriately
Eq.~(\ref{characters-raw}) we can show that
\begin{equation}
d_{\{l; m\}} = {\sigma_{\{l\}}\over n_+!}\,
{\sigma_{\{m\}}\over n_-!}\, C_{\{l; m\}}\,,
\end{equation}
where
\begin{equation}
C_{\{l; m\}} = \prod_{i=1}^s {(N-t-i+l_i)!\over(N-t-i)!}\,
\prod_{j=1}^t {(N-s-j+m_j)!\over(N-s-j)!}\,
\prod_{i=1}^s \prod_{j=1}^t {N+1-i-j+l_i+m_j\over N+1-i-j}\,.
\label{C-cooked}
\end{equation}
The essential feature of Eq.~(\ref{C-cooked}) is the possibility of
extracting results with a finite number of operations even in the
large-$N$ limit.  As a byproduct we obtain the large-$N$ character
coefficients in the useful form
\begin{equation}
{{\tilde z}_{\{l; m\}}\over z^n} \to
{N^n\over C_{\{l; m\}}}\,.
\label{large-N-character}
\end{equation}

\section{Outline of the procedure}
\label{outline}

The general purpose of the strong-coupling expansion is an evaluation
of the Green's functions of the model as power series in $\beta$.
If we are interested in the mass spectrum of the model, we may focus
on the class of two-point Green's functions defined by
\begin{equation}
G_{(r)}(x) = {1\over d_{(r)}}
\left<\chi_{(r)}\bigl(U^\dagger(x)\,U(0)\bigl)\right>,
\label{Gr}
\end{equation}
and even more specifically we may decide to evaluate the fundamental
two-point Green's function
\begin{equation}
G(x) = {1\over N} \left<\Tr\bigl(U^\dagger(x)\,U(0)\bigl)\right>.
\end{equation}

Evaluating such expectation values by the character expansion involves
performing the group integrals that are generated from choosing an
arbitrary representation for each link of the lattice.  As a
consequence of Eq.~(\ref{dze}), only a finite number of nontrivial
representations contribute to any definite order in the series
expansion of $G(x)$ in powers of $\beta$; we must however find a
systematic way of identifying the relevant contributions.

As a preliminary condition for the definition of an algorithmic
approach to the strong-coupling expansion of $G(x)$, it is convenient
to identify explicitly all the logical steps of such a computation and
define a number of objects that play a special r\^ole in it.

\subsection{Assignments}

Each lattice integration variable $U(y)$ can only appear in the
integrand either through the representation characters defined on
links terminating on the lattice site $y$ or through the observable
whose expectation value is to be evaluated.  According to the rules of
group integration, nontrivial contributions are obtained only if the
product of all representations involving $U(y)$ contains the identity
(the trivial representation).

We define an {\it assignment\/} $\{r\}$ to be a choice of a
representation for each link of the lattice that is consistent with
the above requirement.  Necessary conditions for an assignment can be
obtained by a close examination of the rules for the composition of
two irreducible representations of ${\rm U}(N)$.  When we consider
Green's functions in the class defined by Eq.~(\ref{Gr}), we recognize
that the operator whose expectation value we are evaluating, when
considered from the point of view of group integration, plays the
r\^ole of a unit length link connecting the sites $x$ and $0$,
weighted with a factor $d_{(r)}^{-2}$.  Therefore all the relevant
group integrals can be put into correspondence with integrals
appearing in the character expansion of the partition function
(possibly in higher dimensions).

Changing the convention for the orientation of links changes
each representation $r$ into its conjugate $\bar r$
($l\leftrightarrow m$), but, since
${\tilde z}_{(r)} = {\tilde z}_{(\bar r)}$, it does not affect the
expansion.  Hence we can consider all links terminating in a given
site as ``ingoing''.  It is now possible to prove that an assignment
must satisfy the following conditions at each lattice site:
\begin{eqnarray}
&&\sum_i(n_+^i - n_-^i) = 0,  \label{assi1}\\
&&n_\pm^j \le \sum_{i\ne j} n_\mp^i
\ \ \hbox{(non-backtracking condition)},  \label{assi2}
\end{eqnarray}
where the summation is extended to all ingoing links.

Order by order in the strong-coupling expansion, the relevant
assignments involve nontrivial representations only on a finite number
of links, which allows the possibility of drawing on the lattice the
{\it diagram\/} of each assignment.  Such a diagram is characterized
by vertices, where more then two nontrivial representations meet, and
paths, i.e.\ chains of links connecting vertices.  Orthogonality of
representations implies that the choice of representation along a
given path cannot change.  We will denote the length of each path $p$
by $L_p$, and the corresponding (nontrivial) representation by $r_p$.
The topology $\scr S$ of a diagram may be represented by the
connectivity matrix between its vertices.  As we shall show later, the
value of the group integral associated with each assignment can only
be a function of $r_p$, $L_p$, and $\scr S$; we shall denote it by
$R^{(\scr S)}_{\{r, L\}}$.

\subsection{Configurations}
\label{configurations}

The set $(n_+,n_-)$ does not in general identify uniquely a
representation.  It is convenient to define {\it oriented
configurations\/}: they are the sets of all assignments having the
same $(n_+^l,n_-^l)$ for each link of the lattice.  The relevance of
oriented configurations in the context of the strong-coupling
expansion stays in the fact that they are in a one-to-one
correspondence with the monomials one would obtain in the integrand
after the series expansion of the Boltzmann factor in powers of
$\beta$.  They are therefore the simplest objects admitting a
meaningful definition for their connected contributions.

Eq.~(\ref{dze}) tells us that the lowest-order contribution of any
character coefficient to the strong-coupling series depends only on
$n = n_+ + n_-$.  Hence it is useful to define (unoriented)
{\it configurations\/} by summing up all the oriented configurations
characterized by the same value of $n_l$ for each link of the lattice.
The set $\{n\} = (n_1,...)$ uniquely identifies a configuration.  We
might have defined configurations directly as the sets of all
assignments sharing the same $\{n\}$; our procedure insures us about
the possibility of defining the connected contribution of a
configuration.

\begin{figure}[tb]
\centerline{\setlength{\unitlength}{0.009in}
\begin{picture}(674,168)(0,-10)
\put(31.500,121.000){\arc{61.000}{1.3895}{4.8937}}
\put(59.500,121.000){\arc{75.000}{2.2143}{4.0689}}
\put(56.500,121.000){\arc{75.000}{2.2143}{4.0689}}
\put(14.500,121.000){\arc{75.000}{5.3559}{7.2105}}
\put(17.500,121.000){\arc{75.000}{5.3559}{7.2105}}
\put(42.500,121.000){\arc{61.000}{4.5311}{8.0353}}
\put(151.500,121.000){\arc{61.000}{1.3895}{4.8937}}
\put(177.500,121.000){\arc{61.000}{4.5311}{8.0353}}
\put(152.500,121.000){\arc{75.000}{5.3559}{7.2105}}
\put(149.500,121.000){\arc{75.000}{5.3559}{7.2105}}
\put(194.500,121.000){\arc{75.000}{2.2143}{4.0689}}
\put(179.500,121.000){\arc{75.000}{2.2143}{4.0689}}
\put(31.500,31.000){\arc{61.000}{1.3895}{4.8937}}
\put(42.500,31.000){\arc{61.000}{4.5311}{8.0353}}
\put(92.500,31.000){\arc{75.000}{5.3559}{7.2105}}
\put(89.500,31.000){\arc{75.000}{5.3559}{7.2105}}
\put(134.500,31.000){\arc{75.000}{2.2143}{4.0689}}
\put(131.500,31.000){\arc{75.000}{2.2143}{4.0689}}
\put(196.500,31.000){\arc{61.000}{1.3895}{4.8937}}
\put(239.500,31.000){\arc{75.000}{2.2143}{4.0689}}
\put(194.500,31.000){\arc{75.000}{5.3559}{7.2105}}
\put(237.500,31.000){\arc{61.000}{4.5311}{8.0353}}
\put(331.500,31.000){\arc{61.000}{1.3895}{4.8937}}
\put(314.500,31.000){\arc{75.000}{5.3559}{7.2105}}
\put(432.500,31.000){\arc{61.000}{4.5311}{8.0353}}
\put(359.500,31.000){\arc{75.000}{5.3559}{7.2105}}
\put(404.500,31.000){\arc{75.000}{2.2143}{4.0689}}
\put(449.500,31.000){\arc{75.000}{2.2143}{4.0689}}
\put(224.500,31.000){\arc{75.000}{2.2143}{4.0689}}
\put(209.500,31.000){\arc{75.000}{5.3559}{7.2105}}
\put(271.500,121.000){\arc{61.000}{1.3895}{4.8937}}
\put(299.500,121.000){\arc{75.000}{2.2143}{4.0689}}
\put(296.500,121.000){\arc{75.000}{2.2143}{4.0689}}
\put(254.500,121.000){\arc{75.000}{5.3559}{7.2105}}
\put(297.500,121.000){\arc{61.000}{4.5311}{8.0353}}
\put(269.500,121.000){\arc{75.000}{5.3559}{7.2105}}
\put(391.500,121.000){\arc{61.000}{1.3895}{4.8937}}
\put(374.500,121.000){\arc{75.000}{5.3559}{7.2105}}
\put(447.500,121.000){\arc{61.000}{4.5311}{8.0353}}
\put(422.500,121.000){\arc{75.000}{5.3559}{7.2105}}
\put(467.500,121.000){\arc{75.000}{2.2143}{4.0689}}
\put(464.500,121.000){\arc{75.000}{2.2143}{4.0689}}
\put(597.500,121.000){\arc{61.000}{4.5311}{8.0353}}
\put(614.500,121.000){\arc{75.000}{2.2143}{4.0689}}
\put(541.500,121.000){\arc{61.000}{1.3895}{4.8937}}
\put(566.500,121.000){\arc{75.000}{2.2143}{4.0689}}
\put(521.500,121.000){\arc{75.000}{5.3559}{7.2105}}
\put(524.500,121.000){\arc{75.000}{5.3559}{7.2105}}
\put(538.500,31.000){\arc{61.000}{1.3895}{4.8937}}
\put(549.500,31.000){\arc{61.000}{4.5311}{8.0353}}
\put(635.500,31.000){\arc{75.000}{2.2143}{4.0689}}
\put(590.500,31.000){\arc{75.000}{5.3559}{7.2105}}
\put(680.500,31.000){\arc{75.000}{2.2143}{4.0689}}
\put(635.500,31.000){\arc{75.000}{5.3559}{7.2105}}
\path(109,1)(115,1)
\path(109,61)(115,61)
\put(97,118){\makebox(0,0)[b]{\smash{{{\SetFigFont{12}{14.4}{rm}$\to$}}}}}
\put(223,118){\makebox(0,0)[b]{\smash{{{\SetFigFont{12}{14.4}{rm}+}}}}}
\put(343,118){\makebox(0,0)[b]{\smash{{{\SetFigFont{12}{14.4}{rm}+}}}}}
\put(493,118){\makebox(0,0)[b]{\smash{{{\SetFigFont{12}{14.4}{rm}+}}}}}
\put(658,115){\makebox(0,0)[b]{\smash{{{\SetFigFont{12}{14.4}{rm}+}}}}}
\put(148,28){\makebox(0,0)[b]{\smash{{{\SetFigFont{12}{14.4}{rm}+}}}}}
\put(283,28){\makebox(0,0)[b]{\smash{{{\SetFigFont{12}{14.4}{rm}+}}}}}
\put(487,28){\makebox(0,0)[b]{\smash{{{\SetFigFont{12}{14.4}{rm}+
$1\over2$}}}}}
\end{picture}
}
\caption{All the non-trivial disconnections of a diagram.}
\label{disconnections}
\end{figure}
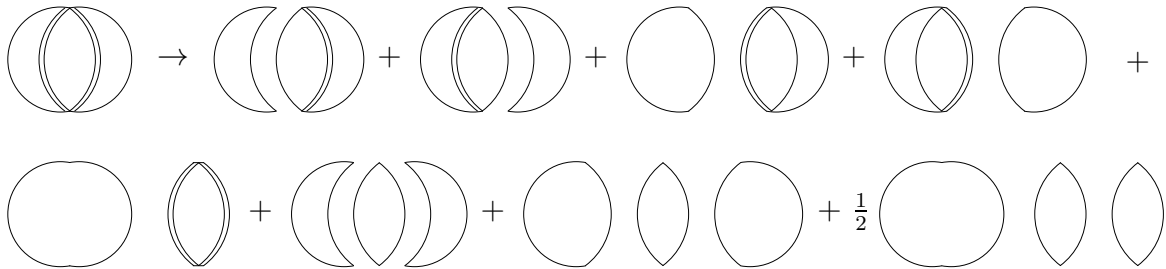

We may introduce the diagrammatic representation of oriented
configurations, by drawing each oriented link $(n_+,n_-)$ as a bundle
of $n$ links, of which $n_+$ bear a positively-oriented arrow and
$n_-$ bear a negatively-oriented arrow.  Removing the arrows leaves us
with a diagrammatic representation of (unoriented) configurations.
One may easily get convinced that the algebraic notion of
disconnection turns out to coincide with the geometrical one.  In this
representation, a disconnection is a set of subdiagrams such that
their superposition reproduces the original diagram.
An example of disconnection is drawn in Fig.~\ref{disconnections}.

Without belaboring on this topic, which is widely discussed in the
literature \cite{Parisi-book,Itzikson-Drouffe}, we only remind that
the connected part of a collection of $n$ (abstract) objects is
recursively defined by the condition that the set of $n$ objects
coincides with the sum of the connected parts of all its partitions,
including the collection itself.  In presence of multiple copies of
the same object, in standard perturbation theory a combinatorial
factor appears, which is hidden in the character expansion; as a
consequence, when subtracting disconnections one must take care of
dividing by the corresponding symmetry factors in order to restore the
correct normalization.

The definitions imply that the geometric features $\scr S$ and $\{L\}$
of all assignments belonging to a given configuration are the same;
therefore the path $p$ of a configuration is characterized by $L_p$
and by the value $n_p$ of the order of $r_p$.

\subsection{Skeleton diagrams}

It is convenient to reduce each configuration to its {\it skeleton
diagram\/}, whose links are the paths of the configuration.  The
topology $\scr S$ is obviously unchanged, and each link is
characterized by the pair of numbers $(n_p,L_p)$.

In order to clarify the relevance of such a definition, let us
consider the problem of evaluating the group integrals
$R^{(\scr S)}_{\{r, L\}}$ for the assignments belonging to a given
configuration.

An elementary consequence of Eq.~(\ref{ortho}) is the evaluation of
the simplest nontrivial group integral entering our calculations:
\begin{equation}
\int dU \chi_{(r)}(A^\dagger U)\,\chi_{(r)}(U^\dagger B) =
{1\over d_{(r)}}\,\chi_{(r)}(A^\dagger B).
\label{simplest}
\end{equation}
By applying repeatedly Eq.~(\ref{simplest}) along the paths we easily
obtain
\begin{equation}
R^{({\scr S})}_{\{r,L\}} = \Biggl\{\prod_{p=1}^\nu
 \bigl[{\tilde z}_{(r_p)}\bigl]^{L_p} \Biggr\}
S^{({\scr S})}_{\{r\}}\,,
\label{val-skeleton}
\end{equation}
where $\nu$ the number of paths of the configuration (with $n\ne0$),
and $S^{({\scr S})}_{\{r\}}$ is the value of the group integral
associated with the skeleton diagram, in which all links are assigned
unit length and weight, and representations are chosen according to
the assignment.  Further simplification is obtained by noticing that
the effective strong-coupling variable in the character expansion is
$z(\beta)$ (for large $N$ actually $z(\beta)\approx\beta$ because of
Eq.~(\ref{z1-beta})).  Therefore by replacing the character
coefficients ${\tilde z}_{(r)}$ with the ratios
\begin{equation}
z_{(r)} = {{\tilde z}_{(r)} \over z^n}
\end{equation}
we may express the strong coupling series as a series in powers of
$z$, with coefficients that are functions of $z_{(r)}$;
by the way, these quantities for large enough $N$ are pure numbers,
dependent on $N$ but independent of $\beta$, because of
Eq.~(\ref{large-N-character}).  We can rewrite
Eq.~(\ref{val-skeleton}) as
\begin{equation}
R^{({\scr S})}_{\{r,L\}} = z^{\sum_p n_p L_p}
\Biggl\{\prod_{p=1}^\nu z_{(r_p)}^{L_p} \Biggr\}
S^{({\scr S})}_{\{r\}}\,;
\label{RS}
\end{equation}
since $z_{1;0}\equiv1$, there is no dependence on the lengths of the
links with $n=1$, apart from the overall factor of $z$, depending only
on the total length of the configuration $L = \sum_p n_p L_p$.
Therefore the corresponding $L_p$ indices can be dropped, thus
defining a {\it reduced skeleton}.  The contribution of a
configuration to the functional integral is simply the sum of the
contributions of all its assignments.  It then follows from
Eq.~(\ref{RS}) that whenever two configurations can be reduced to the
same skeleton, they will give the same contribution.

An exchange in the ordering of the vertices will not change the
topology of a skeleton diagram; therefore configurations that are
related by this symmetry will give the same contribution.  Moreover,
configurations sharing the same reduced skeleton will give
contributions differing only by an overall proportionality factor,
depending on the total length $L$.  We can group together all
configurations with the same reduced skeleton (taking into account the
abovementioned symmetry) and the same value of $L$: their number is
what we call the {\it geometrical factor}.  The common value of each
of these configurations is proportional to the {\it group-theoretical
  factor\/} of the reduced skeleton:
\begin{equation}
T^{({\scr S})}_{\{n,L\}} = \sum_{\stackrel{\scriptstyle\{r\}}%
{\scriptstyle\{n\}{\rm fixed}}} \prod_{p=1}^\nu
\left[z_{(r_p)}\right]^{L_p} S^{({\scr S})}_{\{r\}}
\end{equation}
with a proportionality factor $z^L$.

The strong-coupling character expansion of a group integral can
therefore be organized as a series in the powers of
$z \equiv {\tilde z}_{1;0}$, with coefficients obtained by taking sums
of products of geometrical and group-theoretical factors.  In order to
understand the computational simplifications achieved at this stage,
let us only notice that, at different orders in the expansion, the
same reduced skeletons may appear again and again in association with
different values of $L$; their group-theoretical factors
however are computed once and for all, while extracting the
geometrical factors is a task that, as we shall show later, can be
completely automatized.

\subsection{Superskeletons}
\label{superskeletons}

Both for the purpose of bookkeeping and in view of the problem of
actually computing the group-theoretical factors, at this stage we
need a classification and labeling of (reduced) skeleton diagrams,
which must keep track of their topological properties and try to put
into evidence whatever further simplification we may conceive.  We
found it convenient to isolate for each topology $\scr S$ a ``core''
topology $\scr T$ which we call {\it superskeleton}, defined by the
condition that each vertex in it is connected by at most a single link
to any other vertex (i.e.\ the entries in the connectivity matrix are
either 0 or 1).

The essential ingredient for the reduction of a skeleton to a
superskeleton is the extraction of {\it bubbles}, defined as sets of
two links in a skeleton connecting the same pair of vertices.  Let us
now recall the decomposition rule for a product of characters:
\begin{equation}
\chi_{(r)}(U)\,\chi_{(s)}(U) = \sum_t C_{(rst)}\,\chi_{(t)}(U)\,,
\label{decomp}
\end{equation}
where $C_{(rst)}$ is a set of integer numbers counting the
multiplicity of $(t)$ in the product of representations
$(r)\otimes(s)$.  For all assignments of $(r)$, $(s)$ consistent with
a given skeleton, $(t)$ must be such that the triplet $(r)$, $(s)$,
$(\bar t)$ satisfies Eqs.~(\ref{assi1}) and (\ref{assi2}).  Therefore
replacing a bubble with a single link and allowing for all
$\chi_{(t)}$ obtained from Eq.~(\ref{decomp}) to be inserted in it
defines new consistent assignments.  Notice however that in general we
may not expect all these assignments to belong to the same skeleton,
since $n$ may vary within the class of admissible $(t)$.

We can repeat the procedure, replacing paths with links when needed,
consistently with orthogonality of representations and
Eq.~(\ref{simplest}), until all the bubbles in the skeleton have
disappeared.  The resulting diagram is the superskeleton of our
original diagram.  We must stress that a superskeleton is {\it not\/}
a skeleton diagram, because it does not make sense to assign a value
of $(n,L)$ to its links.  It is however important to observe that the
value $S^{(\scr S)}_{\{r\}}$ of the group integral corresponding to any
assignment $\{r\}$ on the skeleton $\scr S$ can be expressed as a
weighted sum of factors $S^{(\scr T)}_{\{t\}}$ corresponding to the
consistent assignments of the superskeleton $\scr T$, with weights
that are related to the factors
\begin{equation}
C_{(rst)}\,{d_{(r)}d_{(s)}\over d_{(t)}}
\label{Cddd}
\end{equation}
obtained by replacing bubbles with single links.

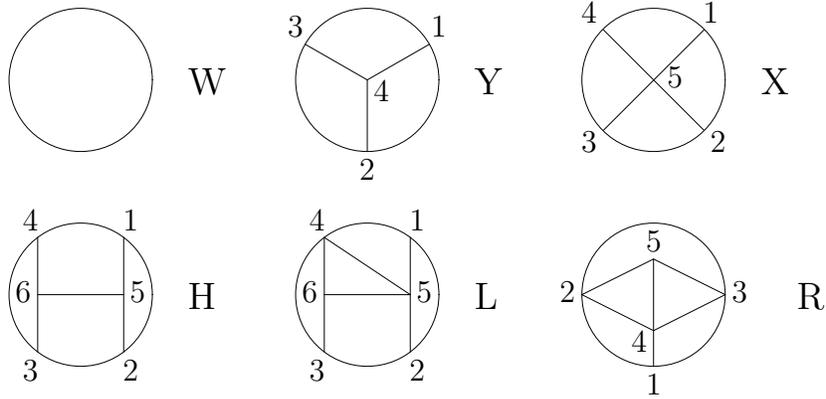
\begin{figure}[tb]
\centerline{\setlength{\unitlength}{0.0125in}
\begin{picture}(339,186)(0,-10)
\put(30,132){\ellipse{60}{60}}
\put(150,132){\ellipse{60}{60}}
\put(270,132){\ellipse{60}{60}}
\put(30,42){\ellipse{60}{60}}
\put(150,42){\ellipse{60}{60}}
\put(270,42){\ellipse{60}{60}}
\path(150,132)(150,102)
\path(291,111)(249,153)
\path(249,111)(291,153)
\path(150,132)(124,147)
\path(150,132)(176,147)
\path(12,66)(12,18)
\path(12,42)(48,42)
\path(48,66)(48,18)
\path(132,66)(132,18)
\path(132,42)(168,42)
\path(168,66)(168,18)
\path(132,66)(168,42)
\path(270,57)(270,12)
\path(240,42)(270,57)(300,42)
\path(240,42)(270,27)(300,42)
\put(75,126){\makebox(0,0)[lb]{\smash{{{\SetFigFont{14}{16.8}{rm}W}}}}}
\put(120,150){\makebox(0,0)[b]{\smash{{{\SetFigFont{12}{14.4}{rm}3}}}}}
\put(180,150){\makebox(0,0)[b]{\smash{{{\SetFigFont{12}{14.4}{rm}1}}}}}
\put(156,123){\makebox(0,0)[b]{\smash{{{\SetFigFont{12}{14.4}{rm}4}}}}}
\put(150,90){\makebox(0,0)[b]{\smash{{{\SetFigFont{12}{14.4}{rm}2}}}}}
\put(195,126){\makebox(0,0)[lb]{\smash{{{\SetFigFont{14}{16.8}{rm}Y}}}}}
\put(243,156){\makebox(0,0)[b]{\smash{{{\SetFigFont{12}{14.4}{rm}4}}}}}
\put(294,156){\makebox(0,0)[b]{\smash{{{\SetFigFont{12}{14.4}{rm}1}}}}}
\put(279,129){\makebox(0,0)[b]{\smash{{{\SetFigFont{12}{14.4}{rm}5}}}}}
\put(315,126){\makebox(0,0)[lb]{\smash{{{\SetFigFont{14}{16.8}{rm}X}}}}}
\put(243,102){\makebox(0,0)[b]{\smash{{{\SetFigFont{12}{14.4}{rm}3}}}}}
\put(297,102){\makebox(0,0)[b]{\smash{{{\SetFigFont{12}{14.4}{rm}2}}}}}
\put(129,69){\makebox(0,0)[b]{\smash{{{\SetFigFont{12}{14.4}{rm}4}}}}}
\put(171,69){\makebox(0,0)[b]{\smash{{{\SetFigFont{12}{14.4}{rm}1}}}}}
\put(9,69){\makebox(0,0)[b]{\smash{{{\SetFigFont{12}{14.4}{rm}4}}}}}
\put(51,69){\makebox(0,0)[b]{\smash{{{\SetFigFont{12}{14.4}{rm}1}}}}}
\put(6,39){\makebox(0,0)[b]{\smash{{{\SetFigFont{12}{14.4}{rm}6}}}}}
\put(54,39){\makebox(0,0)[b]{\smash{{{\SetFigFont{12}{14.4}{rm}5}}}}}
\put(75,36){\makebox(0,0)[lb]{\smash{{{\SetFigFont{14}{16.8}{rm}H}}}}}
\put(9,6){\makebox(0,0)[b]{\smash{{{\SetFigFont{12}{14.4}{rm}3}}}}}
\put(51,6){\makebox(0,0)[b]{\smash{{{\SetFigFont{12}{14.4}{rm}2}}}}}
\put(126,39){\makebox(0,0)[b]{\smash{{{\SetFigFont{12}{14.4}{rm}6}}}}}
\put(174,39){\makebox(0,0)[b]{\smash{{{\SetFigFont{12}{14.4}{rm}5}}}}}
\put(129,6){\makebox(0,0)[b]{\smash{{{\SetFigFont{12}{14.4}{rm}3}}}}}
\put(171,6){\makebox(0,0)[b]{\smash{{{\SetFigFont{12}{14.4}{rm}2}}}}}
\put(195,36){\makebox(0,0)[lb]{\smash{{{\SetFigFont{14}{16.8}{rm}L}}}}}
\put(330,36){\makebox(0,0)[lb]{\smash{{{\SetFigFont{14}{16.8}{rm}R}}}}}
\put(264,18){\makebox(0,0)[b]{\smash{{{\SetFigFont{12}{14.4}{rm}4}}}}}
\put(270,60){\makebox(0,0)[b]{\smash{{{\SetFigFont{12}{14.4}{rm}5}}}}}
\put(234,39){\makebox(0,0)[b]{\smash{{{\SetFigFont{12}{14.4}{rm}2}}}}}
\put(306,39){\makebox(0,0)[b]{\smash{{{\SetFigFont{12}{14.4}{rm}3}}}}}
\put(270,0){\makebox(0,0)[b]{\smash{{{\SetFigFont{12}{14.4}{rm}1}}}}}
\end{picture}
}
\caption{Superskeleton topologies.}
\label{topologies}
\end{figure}

A superskeleton is completely identified by its topology, and it is
worth mentioning that, as in the case of skeletons, supersksletons
differing only by a permutation of vertices are equivalent, and
therefore they can be reduced to a standard form.  The number of
different superskeletons that are relevant to a given order of the
strong-coupling expansion is bound to grow with the order; however
for sufficiently low orders their number is so small that we found it
convenient to label superskeletons by capital letters, in many cases
related to their actual shapes.  A provisional list of labelings is
provided by Fig.~\ref{topologies}.

This is the starting point of our classification scheme for skeletons.
Reduced skeletons are named by the symbol denoting the topology of
their superskeleton; the full information concerning superskeleton
links, denoted by $\sigma$, will appear as arguments; using the pair
of integers $ij$ to denote the link connecting node $i$ to node $j$,
with node numbering fixed by Fig.~\ref{topologies}, the skeletons will
be named
\[
\begin{array}{ll}
{\rm W}(\sigma), \ &
{\rm Y}(\sigma_{12};\sigma_{23};\sigma_{31};
\sigma_{14};\sigma_{24};\sigma_{34}), \\
{\rm X}(\sigma_{12};\sigma_{23};\sigma_{34};\sigma_{41};\sigma_{15};
\sigma_{25};\sigma_{35};\sigma_{45}), \ &
{\rm H}(\sigma_{12};\sigma_{23};\sigma_{34};\sigma_{41};\sigma_{15};
\sigma_{25};\sigma_{36};\sigma_{46};\sigma_{56}), \\
{\rm L}(\sigma_{12};\sigma_{23};\sigma_{34};\sigma_{41};\sigma_{36};
\sigma_{46};\sigma_{15};\sigma_{25};\sigma_{56};\sigma_{45}), \ &
{\rm R}(\sigma_{12};\sigma_{23};\sigma_{31};\sigma_{14};\sigma_{24};
\sigma_{25};\sigma_{35};\sigma_{34};\sigma_{45}).
\end{array}
\]
$\sigma$ contains
information about $n$; for $n\ne1$, also about the length $L$ in the
original skeleton and the bubble content.  For reasons to be clarified
later, we need not consider bubbles along $n=1$ lines.

\begin{figure}[tb]
\centerline{\setlength{\unitlength}{0.0125in}
\begin{picture}(544,96)(0,-10)
\put(53,36){\ellipse{32}{32}}
\put(315,36){\ellipse{46}{46}}
\path(23,36)(37,36)
\path(69,36)(83,36)
\path(278,36)(292,36)
\path(290,36)(338,36)
\put(53,57){\makebox(0,0)[b]{\smash{{{\SetFigFont{12}{14.4}{rm}$n_1,L_1$}}}}}
\put(80,42){\makebox(0,0)[b]{\smash{{{\SetFigFont{12}{14.4}{rm}$n,q$}}}}}
\put(53,9){\makebox(0,0)[b]{\smash{{{\SetFigFont{12}{14.4}{rm}$n_2,L_2$}}}}}
\put(113,33){\makebox(0,0)[lb]{\smash{{{\SetFigFont{12}{14.4}{rm}$n,p+q,[n_1,L_1;n_2,L_2]$}}}}}
\put(26,42){\makebox(0,0)[b]{\smash{{{\SetFigFont{12}{14.4}{rm}$n,p$}}}}}
\put(23,32){\makebox(0,0)[b]{\smash{{{\SetFigFont{14}{16.8}{rm}$\bullet$}}}}}
\put(83,32){\makebox(0,0)[b]{\smash{{{\SetFigFont{14}{16.8}{rm}$\bullet$}}}}}
\put(281,42){\makebox(0,0)[b]{\smash{{{\SetFigFont{12}{14.4}{rm}$n,p$}}}}}
\put(278,32){\makebox(0,0)[b]{\smash{{{\SetFigFont{14}{16.8}{rm}$\bullet$}}}}}
\put(368,33){\makebox(0,0)[lb]{\smash{{{\SetFigFont{12}{14.4}{rm}$n,p,[n_1,L_1;[n_2,L_2;n_3,L_3]]$}}}}}
\put(338,32){\makebox(0,0)[b]{\smash{{{\SetFigFont{14}{16.8}{rm}$\bullet$}}}}}
\put(314,66){\makebox(0,0)[b]{\smash{{{\SetFigFont{12}{14.4}{rm}$n_1,L_1$}}}}}
\put(314,24){\makebox(0,0)[b]{\smash{{{\SetFigFont{12}{14.4}{rm}$n_2,L_2$}}}}}
\put(314,0){\makebox(0,0)[b]{\smash{{{\SetFigFont{12}{14.4}{rm}$n_3,L_3$}}}}}
\end{picture}
}
\caption{Examples of bubbles.}
\label{bubbles}
\end{figure}
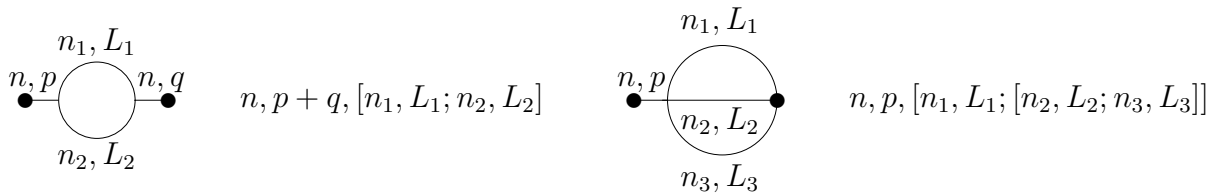

In general, a bubble will be denoted by $[\sigma_1;\sigma_2]$, where
$\sigma_1$ and $\sigma_2$ contains the information about the bubble
links.  In summary, a link information will take one of the forms
\begin{eqnarray}
\sigma &=& 1 \qquad \hbox{($n=1$),} \\
\sigma &=& n,L \qquad \hbox{($n\ne1$, no bubble insertions),} \\
\sigma &=& n,L,[\sigma_{1,1};\sigma_{1,2}][\sigma_{2,1};\sigma_{2,2}]...
 \qquad \hbox{(one or more bubble insertions on a line),} \\
\sigma &=& [\sigma_{1,1};\sigma_{1,2}][\sigma_{2,1};\sigma_{2,2}]...
 \qquad \hbox{(one or more bubble between two nodes),}
\end{eqnarray}
the $\sigma_{i,j}$ themselves taking one of the above forms; insertion
of $b$ identical bubbles will be denoted by exponential notation,
i.e.\ $[\sigma_1;\sigma_2]^b$.  Examples of this notation are
illustrated in Fig.~\ref{bubbles}.

\subsection{Potentials}

As we mentioned before, the possibility of defining the skeletons as
sets of oriented configurations insures us about the fact that we may
consistently define the connected contribution of each skeleton
diagram to the vacuum expectation value of an observable.

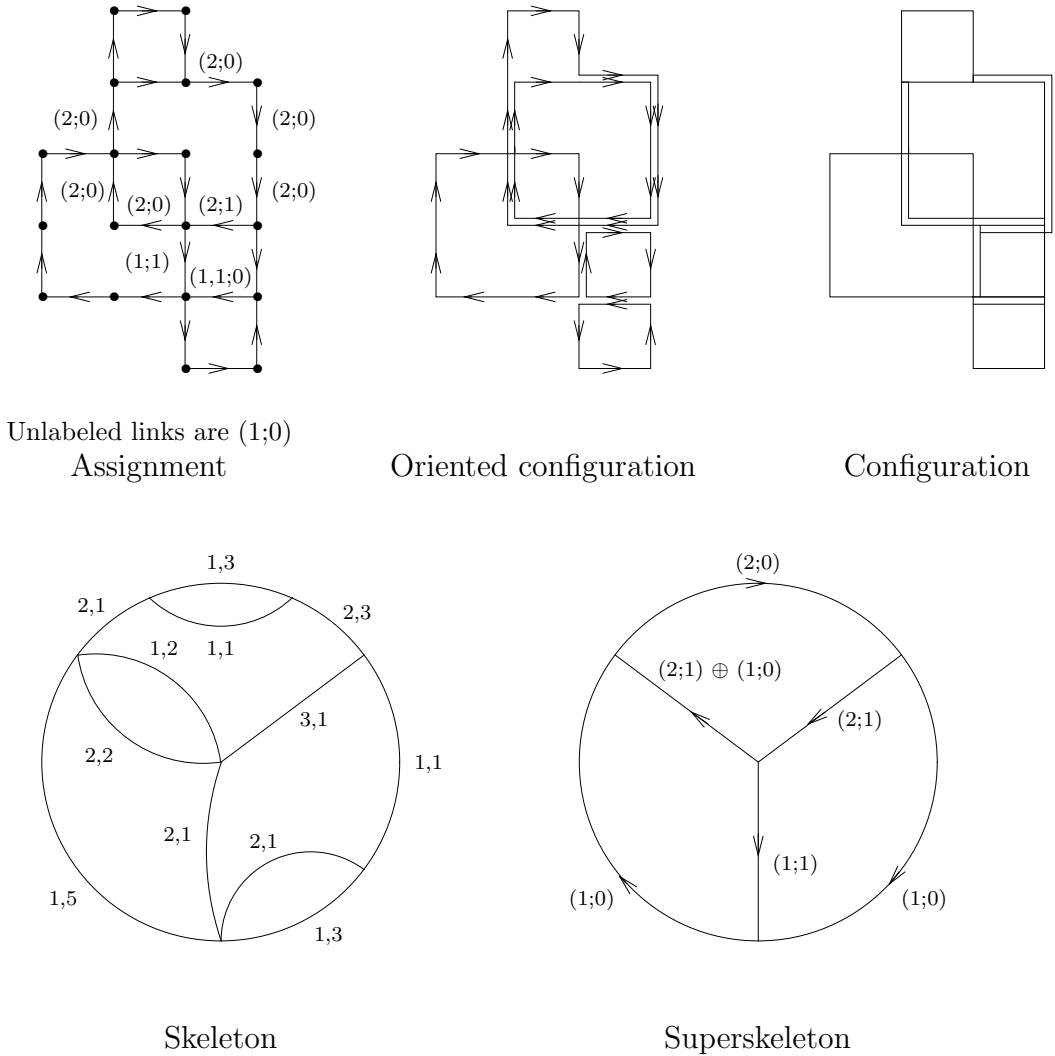
\begin{figure}[tb]
\centerline{\setlength{\unitlength}{0.0125in}
\begin{picture}(434,460)(0,-10)
\put(33.500,112.500){\arc{106.066}{4.5705}{6.1413}}
\put(123.500,45.000){\arc{75.000}{3.1416}{5.3559}}
\put(78.500,172.500){\arc{106.066}{1.4289}{2.9997}}
\put(195.500,82.500){\arc{231.487}{2.8116}{3.4715}}
\put(86.000,220.500){\arc{87.000}{0.8098}{2.3318}}
\put(86,120){\ellipse{150}{150}}
\path(86,120)(146,165)
\put(86,0){\makebox(0,0)[b]{\smash{{{\SetFigFont{12}{14.4}{rm}Skeleton}}}}}
\put(86,201){\makebox(0,0)[b]{\smash{{{\SetFigFont{8}{9.6}{rm}1,3}}}}}
\put(86,165){\makebox(0,0)[b]{\smash{{{\SetFigFont{8}{9.6}{rm}1,1}}}}}
\put(137,180){\makebox(0,0)[lb]{\smash{{{\SetFigFont{8}{9.6}{rm}2,3}}}}}
\put(167,117){\makebox(0,0)[lb]{\smash{{{\SetFigFont{8}{9.6}{rm}1,1}}}}}
\put(119,135){\makebox(0,0)[lb]{\smash{{{\SetFigFont{8}{9.6}{rm}3,1}}}}}
\put(125,45){\makebox(0,0)[lb]{\smash{{{\SetFigFont{8}{9.6}{rm}1,3}}}}}
\put(26,60){\makebox(0,0)[rb]{\smash{{{\SetFigFont{8}{9.6}{rm}1,5}}}}}
\put(41,120){\makebox(0,0)[rb]{\smash{{{\SetFigFont{8}{9.6}{rm}2,2}}}}}
\put(38,183){\makebox(0,0)[rb]{\smash{{{\SetFigFont{8}{9.6}{rm}2,1}}}}}
\put(56,165){\makebox(0,0)[lb]{\smash{{{\SetFigFont{8}{9.6}{rm}1,2}}}}}
\put(74,87){\makebox(0,0)[rb]{\smash{{{\SetFigFont{8}{9.6}{rm}2,1}}}}}
\put(311,120){\ellipse{150}{150}}
\path(371,435)(371,345)(431,345)
\path(374,375)(374,405)(401,405)
\path(431,285)(431,312)(401,312)
\path(401,315)(341,315)(341,375)
        (401,375)(401,285)(431,285)
\path(401,405)(431,405)(431,348)
        (374,348)(374,378)(374,375)
\path(311,120)(371,165)
\path(251,165)(311,120)(311,45)
\path(431,345)(431,312)
\path(431,345)(431,312)
\path(401,315)(431,315)
\path(401,315)(431,315)
\path(431,348)(431,345)
\path(431,348)(431,345)
\path(401,408)(401,405)
\path(401,408)(401,405)
\path(371,405)(374,405)
\path(371,405)(374,405)
\path(86,405)(89,405)
\path(81.000,403.000)(89.000,405.000)(81.000,407.000)
\path(41,387)(41,393)
\path(43.000,385.000)(41.000,393.000)(39.000,385.000)
\path(23,375)(29,375)
\path(21.000,373.000)(29.000,375.000)(21.000,377.000)
\path(53,375)(59,375)
\path(51.000,373.000)(59.000,375.000)(51.000,377.000)
\path(56,315)(53,315)
\path(61.000,317.000)(53.000,315.000)(61.000,313.000)
\path(26,315)(23,315)
\path(31.000,317.000)(23.000,315.000)(31.000,313.000)
\path(11,327)(11,333)
\path(13.000,325.000)(11.000,333.000)(9.000,325.000)
\path(11,357)(11,363)
\path(13.000,355.000)(11.000,363.000)(9.000,355.000)
\path(56,345)(53,345)
\path(61.000,347.000)(53.000,345.000)(61.000,343.000)
\path(41,357)(41,363)
\path(43.000,355.000)(41.000,363.000)(39.000,355.000)
\path(101,393)(101,390)(101,387)
\path(99.000,395.000)(101.000,387.000)(103.000,395.000)
\path(101,363)(101,357)
\path(99.000,365.000)(101.000,357.000)(103.000,365.000)
\path(101,330)(101,327)
\path(99.000,335.000)(101.000,327.000)(103.000,335.000)
\path(71,303)(71,297)
\path(69.000,305.000)(71.000,297.000)(73.000,305.000)
\path(83,285)(89,285)
\path(81.000,283.000)(89.000,285.000)(81.000,287.000)
\path(101,297)(101,303)
\path(103.000,295.000)(101.000,303.000)(99.000,295.000)
\path(53,405)(59,405)
\path(51.000,403.000)(59.000,405.000)(51.000,407.000)
\path(53,435)(59,435)
\path(51.000,433.000)(59.000,435.000)(51.000,437.000)
\path(71,423)(71,417)
\path(69.000,425.000)(71.000,417.000)(73.000,425.000)
\path(86,345)(83,345)
\path(91.000,347.000)(83.000,345.000)(91.000,343.000)
\path(86,315)(83,315)
\path(91.000,317.000)(83.000,315.000)(91.000,313.000)
\path(41,417)(41,423)
\path(43.000,415.000)(41.000,423.000)(39.000,415.000)
\path(71,363)(71,357)
\path(69.000,365.000)(71.000,357.000)(73.000,365.000)
\path(71,333)(71,330)
\path(69.000,338.000)(71.000,330.000)(73.000,338.000)
\path(206,435)(206,345)(266,345)
\path(251,405)(254,405)
\path(246.000,403.000)(254.000,405.000)(246.000,407.000)
\path(206,387)(206,393)
\path(208.000,385.000)(206.000,393.000)(204.000,385.000)
\path(188,375)(194,375)
\path(186.000,373.000)(194.000,375.000)(186.000,377.000)
\path(218,375)(224,375)
\path(216.000,373.000)(224.000,375.000)(216.000,377.000)
\path(221,315)(218,315)
\path(226.000,317.000)(218.000,315.000)(226.000,313.000)
\path(191,315)(188,315)
\path(196.000,317.000)(188.000,315.000)(196.000,313.000)
\path(176,327)(176,333)
\path(178.000,325.000)(176.000,333.000)(174.000,325.000)
\path(176,357)(176,363)
\path(178.000,355.000)(176.000,363.000)(174.000,355.000)
\path(221,345)(218,345)
\path(226.000,347.000)(218.000,345.000)(226.000,343.000)
\path(206,357)(206,363)
\path(208.000,355.000)(206.000,363.000)(204.000,355.000)
\path(266,393)(266,390)(266,387)
\path(264.000,395.000)(266.000,387.000)(268.000,395.000)
\path(266,363)(266,357)
\path(264.000,365.000)(266.000,357.000)(268.000,365.000)
\path(266,330)(266,327)
\path(264.000,335.000)(266.000,327.000)(268.000,335.000)
\path(236,303)(236,297)
\path(234.000,305.000)(236.000,297.000)(238.000,305.000)
\path(248,285)(254,285)
\path(246.000,283.000)(254.000,285.000)(246.000,287.000)
\path(266,297)(266,303)
\path(268.000,295.000)(266.000,303.000)(264.000,295.000)
\path(218,405)(224,405)
\path(216.000,403.000)(224.000,405.000)(216.000,407.000)
\path(218,435)(224,435)
\path(216.000,433.000)(224.000,435.000)(216.000,437.000)
\path(236,423)(236,417)
\path(234.000,425.000)(236.000,417.000)(238.000,425.000)
\path(251,345)(248,345)
\path(256.000,347.000)(248.000,345.000)(256.000,343.000)
\path(206,417)(206,423)
\path(208.000,415.000)(206.000,423.000)(204.000,415.000)
\path(236,363)(236,357)
\path(234.000,365.000)(236.000,357.000)(238.000,365.000)
\path(236,333)(236,330)
\path(234.000,338.000)(236.000,330.000)(238.000,338.000)
\path(209,387)(209,393)
\path(211.000,385.000)(209.000,393.000)(207.000,385.000)
\path(209,375)(209,405)(236,405)
\path(206,435)(236,435)(236,408)
        (269,408)(269,345)(266,345)
\path(248,408)(254,408)
\path(246.000,406.000)(254.000,408.000)(246.000,410.000)
\path(269,396)(269,387)
\path(267.000,395.000)(269.000,387.000)(271.000,395.000)
\path(269,366)(269,357)
\path(267.000,365.000)(269.000,357.000)(271.000,365.000)
\path(254,348)(248,348)
\path(256.000,350.000)(248.000,348.000)(256.000,346.000)
\path(221,348)(218,348)
\path(226.000,350.000)(218.000,348.000)(226.000,346.000)
\path(236,405)(266,405)(266,348)
        (209,348)(209,378)(209,375)
\path(209,357)(209,363)
\path(211.000,355.000)(209.000,363.000)(207.000,355.000)
\path(266,285)(266,312)(236,312)
\path(251,312)(248,312)
\path(256.000,314.000)(248.000,312.000)(256.000,310.000)
\path(239,330)(239,333)
\path(241.000,325.000)(239.000,333.000)(237.000,325.000)
\path(251,315)(248,315)
\path(256.000,317.000)(248.000,315.000)(256.000,313.000)
\path(248,342)(254,342)
\path(246.000,340.000)(254.000,342.000)(246.000,344.000)
\path(266,342)(266,315)(239,315)
        (239,342)(266,342)
\path(236,375)(176,375)(176,315)
        (236,315)(236,375)
\path(236,312)(236,285)(266,285)
\path(101,315)(11,315)(11,375)
        (71,375)(71,285)(101,285)
\path(41,405)(71,405)
\path(41,435)(41,345)(101,345)
\path(41,435)(71,435)(71,405)
        (101,405)(101,285)
\path(404,345)(404,315)
\path(404,342)(434,342)(434,408)
        (401,408)(401,435)(371,435)
\path(290,136)(283,141)
\path(290.672,137.978)(283.000,141.000)(288.347,134.723)
\path(311,195)(314,195)
\path(306.000,193.000)(314.000,195.000)(306.000,197.000)
\path(339,141)(332,136)
\path(337.347,142.277)(332.000,136.000)(339.672,139.022)
\path(256,69)(253,72)
\path(260.071,67.757)(253.000,72.000)(257.243,64.929)
\path(370,74)(366,69)
\path(369.436,76.496)(366.000,69.000)(372.559,73.998)
\path(311,90)(311,81)
\path(309.000,89.000)(311.000,81.000)(313.000,89.000)
\put(386,240){\makebox(0,0)[b]{\smash{{{\SetFigFont{12}{14.4}{rm}Configuration}}}}}
\put(371,60){\makebox(0,0)[lb]{\smash{{{\SetFigFont{8}{9.6}{rm}(1;0)}}}}}
\put(317,75){\makebox(0,0)[lb]{\smash{{{\SetFigFont{8}{9.6}{rm}(1;1)}}}}}
\put(110,84){\makebox(0,0)[rb]{\smash{{{\SetFigFont{8}{9.6}{rm}2,1}}}}}
\put(86,411){\makebox(0,0)[b]{\smash{{{\SetFigFont{8}{9.6}{rm}(2;0)}}}}}
\put(35,387){\makebox(0,0)[rb]{\smash{{{\SetFigFont{8}{9.6}{rm}(2;0)}}}}}
\put(65,327){\makebox(0,0)[rb]{\smash{{{\SetFigFont{8}{9.6}{rm}(1;1)}}}}}
\put(86,321){\makebox(0,0)[b]{\smash{{{\SetFigFont{8}{9.6}{rm}(1,1;0)}}}}}
\put(56,240){\makebox(0,0)[b]{\smash{{{\SetFigFont{12}{14.4}{rm}Assignment}}}}}
\put(107,387){\makebox(0,0)[lb]{\smash{{{\SetFigFont{8}{9.6}{rm}(2;0)}}}}}
\put(107,357){\makebox(0,0)[lb]{\smash{{{\SetFigFont{8}{9.6}{rm}(2;0)}}}}}
\put(56,351){\makebox(0,0)[b]{\smash{{{\SetFigFont{8}{9.6}{rm}(2;0)}}}}}
\put(38,357){\makebox(0,0)[rb]{\smash{{{\SetFigFont{8}{9.6}{rm}(2;0)}}}}}
\put(86,351){\makebox(0,0)[b]{\smash{{{\SetFigFont{8}{9.6}{rm}(2;1)}}}}}
\put(221,240){\makebox(0,0)[b]{\smash{{{\SetFigFont{12}{14.4}{rm}Oriented
configuration}}}}}
\put(269,156){\makebox(0,0)[lb]{\smash{{{\SetFigFont{8}{9.6}{rm}(2;1) $\oplus$
(1;0)}}}}}
\put(344,135){\makebox(0,0)[lb]{\smash{{{\SetFigFont{8}{9.6}{rm}(2;1)}}}}}
\put(311,0){\makebox(0,0)[b]{\smash{{{\SetFigFont{12}{14.4}{rm}Superskeleton}}}}}
\put(56,255){\makebox(0,0)[b]{\smash{{{\SetFigFont{10}{12.0}{rm}Unlabeled links
are (1;0)}}}}}
\put(41,435){\makebox(0,0)[b]{\smash{{{\SetFigFont{8}{9.6}{rm}\Bu}}}}}
\put(71,435){\makebox(0,0)[b]{\smash{{{\SetFigFont{8}{9.6}{rm}\Bu}}}}}
\put(11,375){\makebox(0,0)[b]{\smash{{{\SetFigFont{8}{9.6}{rm}\Bu}}}}}
\put(11,345){\makebox(0,0)[b]{\smash{{{\SetFigFont{8}{9.6}{rm}\Bu}}}}}
\put(11,315){\makebox(0,0)[b]{\smash{{{\SetFigFont{8}{9.6}{rm}\Bu}}}}}
\put(41,315){\makebox(0,0)[b]{\smash{{{\SetFigFont{8}{9.6}{rm}\Bu}}}}}
\put(71,285){\makebox(0,0)[b]{\smash{{{\SetFigFont{8}{9.6}{rm}\Bu}}}}}
\put(71,315){\makebox(0,0)[b]{\smash{{{\SetFigFont{8}{9.6}{rm}\Bu}}}}}
\put(101,285){\makebox(0,0)[b]{\smash{{{\SetFigFont{8}{9.6}{rm}\Bu}}}}}
\put(71,345){\makebox(0,0)[b]{\smash{{{\SetFigFont{8}{9.6}{rm}\Bu}}}}}
\put(101,405){\makebox(0,0)[b]{\smash{{{\SetFigFont{8}{9.6}{rm}\Bu}}}}}
\put(71,405){\makebox(0,0)[b]{\smash{{{\SetFigFont{8}{9.6}{rm}\Bu}}}}}
\put(41,405){\makebox(0,0)[b]{\smash{{{\SetFigFont{8}{9.6}{rm}\Bu}}}}}
\put(41,375){\makebox(0,0)[b]{\smash{{{\SetFigFont{8}{9.6}{rm}\Bu}}}}}
\put(71,375){\makebox(0,0)[b]{\smash{{{\SetFigFont{8}{9.6}{rm}\Bu}}}}}
\put(41,345){\makebox(0,0)[b]{\smash{{{\SetFigFont{8}{9.6}{rm}\Bu}}}}}
\put(101,315){\makebox(0,0)[b]{\smash{{{\SetFigFont{8}{9.6}{rm}\Bu}}}}}
\put(101,375){\makebox(0,0)[b]{\smash{{{\SetFigFont{8}{9.6}{rm}\Bu}}}}}
\put(101,345){\makebox(0,0)[b]{\smash{{{\SetFigFont{8}{9.6}{rm}\Bu}}}}}
\put(251,60){\makebox(0,0)[rb]{\smash{{{\SetFigFont{8}{9.6}{rm}(1;0)}}}}}
\put(311,201){\makebox(0,0)[b]{\smash{{{\SetFigFont{8}{9.6}{rm}(2;0)}}}}}
\end{picture}
}
\caption{Steps showing that a sample assignment contributes to the
potential ${\rm Y}(1,[1;2,1];1;2,4,1;1;1;[1;2,2]) = $
${\rm W}(2,1)\,{\rm Y}(1;1;2,4,1;1;1;[1;2,2])$.}
\label{assignment-to-potential}
\end{figure}

Since the geometrical notion of a disconnection only depends on the
topology of a diagram, as a consequence of definitions we can define
the (algebraic) connected contribution of a skeleton starting from its
geometrical formulation.  As a matter of fact, it is most convenient
to exploit the fact that $n=1$ lines cannot be split, and define the
connected contribution of a reduced skeleton, i.e.\ the connected
group-theoretical factor, which we shall call {\it potential\/}:
\begin{equation}
P^{(\scr S)}_{\{n,L\}} =
\left[\sum_{\stackrel{\scriptstyle\{r\}}%
{\scriptstyle\{n\}{\rm fixed}}}
\prod_p\left[z_{(r_p)}\right]^{L_p}
S^{(\scr S)}_{\{r\}}\right]_{\rm connected}.
\end{equation}
An example of the chain leading from an assignment to the
superskeleton and to the potential is illustrated in
Fig.~\ref{assignment-to-potential}.

When we are evaluating the skeletons contributing to the partition
function, the sum of their potentials with the same geometrical
factors is just the free energy.
Unfortunately, there will be in general no correspondence between the
connected contributions to an arbitrary Green's function and the
corresponding contributions to the free energy.  A notable exception
is that of the fundamental two-point function $G(x)$.  In this case no
disconnection of the vacuum diagram can split the $n=1$ line
associated with the fundamental character $\Tr U^\dagger(x)\,U(0)$,
and there is therefore a one-to-one correspondence between the
connected contribution of a given skeleton diagram and the
contribution of the associated vacuum diagram to the free energy.
Moreover the weight $d_{1;0}^{-2} = 1/N^2$ is the correct
normalization, insuring that in the large-$N$ limit finite
contributions to the Green's functions correspond to finite
contributions to the free energy.  From now on we may therefore focus
on the evaluation of potentials related to vacuum skeleton diagrams.

It is worth mentioning that we might have defined oriented potentials,
but this notion, while conceptually useful, does not find any use in
our actual computations.

A final observation concern notations: we shall label potentials with
the same symbols adopted in the labeling of the corresponding reduced
skeletons.

We must draw some attention to the fact that our definition of
potentials, although referred to unoriented diagrams, is originated by
the problem of evaluating Green's functions.  Therefore we are
assuming that the orientation of one of the links has been fixed.  By
a trivial symmetry of conjugate representations, our potentials will
be one half of the corresponding vacuum contributions to the free
energy.

Including this factor of 2, the disconnections drawn in
Fig.~\ref{disconnections} can be written as
\begin{eqnarray}
&&{\rm disc}\bigl(2 {\rm W}(2,L_1,[2,L_2;2,0,1])\bigr) =
2\times 2^2 {\rm W}(1)\,{\rm W}(2,L_1,1)
+ 2\times 2^2 {\rm W}(1)\,{\rm W}(2,L_2,1) \nonumber \\
&&\quad+\; 2^2 {\rm W}(1)\,{\rm W}(2,L_1{+}L_2)
+ \case{5}{2}\times 2^3 {\rm W}(1)^3.
\end{eqnarray}

\section{Computing the geometrical factor}
\label{geometrical-factor}

The enumeration of all configurations possessing the same reduced
skeleton can be completely automatized by the following considerations
and procedures.

Eqs.~(\ref{assi1}) and (\ref{assi2}) insure us about the existence of
a (non necessarily unique) non-backtracking random walk of length
$\sum_p n_p L_p$ reproducing the diagrammatic representation of each
configuration.  We therefore generate all non-backtracking random
walks with fixed length, fixed origin $0$, and fixed end $x$, and we
compute the corresponding configuration $\{n\}$, i.e.\ we compute
$n_l$ (the number of times each link is visited) for each link of the
lattice.  We now compare the generated configurations, and discard
multiple copies, choosing one (and only one) walk for each different
configuration.

The total (bulk) free energy can be computed by summing over all the
configurations.  Therefore the free energy {\it per site\/} can be
computed by summing over all the configuration that are not related by
a translation.  These are easily obtained by generating all
non-backtracking closed random walks touching a given site,
identifying the corresponding configurations, and chosing one
configuration for each equivalence class under translation symmetry.
{}From this point on, the computation is identical both for the Green's
function and for the free energy.

We must notice that at this point we have generated all the sets
$\{n\}$ obeying Eqs.~(\ref{assi1}) and (\ref{assi2}), but not all of
them lead to nonvanishing group integrals; we get rid of these
``null'' configurations by defining their group-theoretical factor to
be zero.  Our computer program recognizes and automatically discards
two classes of null diagrams:

Diagrams that can be disconnected by removing a single node.
A very simple property of invariant group integration allows for the
possibility of setting a single integration variable to 1.  As a
consequence, one may prove that, whenever the removal of a vertex in a
skeleton leaves us with disconnected subdiagrams, the value of the
group integral factorizes into a product of terms that are just the
values of its disconnected parts.  Therefore, the corresponding
potentials vanish identically.

Diagrams that can be disconnected by removing two links, unless the
links share the same value of $n$.  Such diagrams vanish as a trivial
consequence of the orthogonality of representations.

\begin{figure}[tb]
\centerline{\setlength{\unitlength}{0.0125in}
\begin{picture}(210,75)(0,-10)
\path(0,60)(0,30)(60,30)
        (60,0)(30,0)(30,60)(0,60)
\path(120,60)(120,30)(210,30)
        (210,60)(120,60)
\path(150,60)(150,27)(180,27)(180,60)
\path(150,57)(123,57)(123,33)
        (207,33)(207,57)(180,57)
\end{picture}
}
\caption{Two null configurations: the first can be disconnected by
  removing a single node; the second can be disconnected by removing a
  $n=1$ and a $n=3$ link.}
\label{null-configurations}
\end{figure}
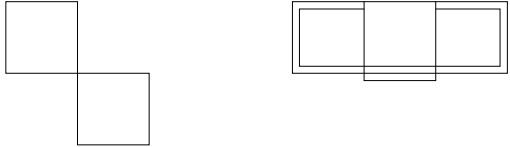

Examples of this phenomena are drawn in
Fig.~\ref{null-configurations}.

We compute the reduced skeleton of each of these configurations.  We
now group together all the configurations originating equivalent
reduced skeletons (i.e.\ which are equal apart from a permutation of
vertices); the geometrical factor is the number of such
configurations, and we choose one representative configuration for
each group.

We factorize each skeleton ``cutting'' along $n=1$ paths,
identify bubbles according to the scheme of
Subs.~\ref{superskeletons}, and compute the connectivity of the
corresponding superskeleton.  The superskeleton is then either
identified as in Fig.~\ref{topologies}, or shown to originate from a
null configuration.  Finally we put together all this information
to obtain the potential, and use the superskeleton symmetry to bring
it in a standard form.

While the data needed in the intermediate stages of this computation
can be extremely large, the results of the last step (potentials and
geometrical factors) are rather compact and can be stored for further
processing.

The limiting factor in this procedure is the available RAM.  On a
workstation with 140 Mbytes of RAM we were able to generate Green's
functions up to 18th order and the free energy up to 20th order on the
square lattice (they of course involve several new superskeletons
beside those drawn in Fig.~\ref{topologies}).  Computer time is not a
limiting factor, since the longest computations take about one CPU
hour on a HP-730/125.

At this stage we must clarify what we mean by standard form of a
superskeleton.  In sufficiently complex cases, an ambiguity may arise
as a consequence of different sequences of elimination of the bubbles.
While the resulting superskeleton is always the same, equivalent
skeletons may receive superficially inequivalent labelings.  We have
not made an effort to reduce completely all these different namings to
a standard form, but we were satisfied with the reduction to a common
form in most cases.  We checked explicitly that the computed values of
differently labeled equivalent potentials are equal.

\section{Computing the group-theoretical factor}
\label{group-theoretical-factor}

In contrast with the previous Section, we must say that the evaluation
of potentials is not yet fully automatized.

We can routinely generate all the sets $\{l; m\}$ needed to identify
the representations of ${\rm U}(N)$ to a definite order.  The closed
formulae presented in Sect.~\ref{generalities} enable us to evaluate
automatically of their dimensions and their large-$N$ character
coefficients.

We can perform the decomposition of the products of these
representations, thus identifying the coefficients $C_{(rst)}$ and the
factors defined in Eq.~(\ref{Cddd}).  We can therefore reduce the
evaluation of the group-theoretical factors, by computer
manipulations, to a linear combination with known coefficients of
factors $S^{({\scr T})}_{\{r\}}$, that are nothing but group integrals
corresponding to consistent assignments of representations on the
(unit length, unit weight) links of a superskeleton with topology
$\scr T$.

Computing the factors $S^{({\scr T})}_{\{r\}}$ is basically a
sophisticated exercise in group integration, and it is therefore
completely solved from a conceptual point of view.  The group
integration over a multiple product of representations can always be
performed by decomposing the product into sums of representations, via
the introduction of appropriate Clebsch-Gordan coefficients, and
applying orthogonality of representations (Eq.~(\ref{ortho})) in the
last step.  This may however become a very inconvenient procedure,
essentially because of the fantastic proliferation of indices (all to
be finally contracted, but appearing at intermediate stages already in
the simplest examples) resulting from writing higher-order
representations in the basis of polynomials of the fundamental
representation.

We have not seriously tried to overcome this problem in general, i.e.\
we have no algorithm capable of generating the Clebsch-Gordan
coefficients for the decomposition of the product of two arbitrary
representations of ${\rm U}(N)$, which would allow to implement the
relevant group integrations in a computer program.  Instead we
followed a slightly different approach, more limited in purpose and
simpler to implement, within our self-imposed limits, without fully
computerizing the computation.

The essentials of our approach are the following.

We observed that, for not very high orders of strong coupling, only a
small number of superskeletons and low-order representations enter the
calculation.  Therefore, by making use of a few well known results of
group integration (that can basically be reported to the knowledge of
the six-matrix deWit-'t~Hooft integral \cite{DeWit-THooft}), we
managed to compute explicitly all the factors $S^{({\scr T})}_{\{r\}}$
entering in our calculations.

However, the possibility of inserting bubbles and varying the lengths
$L_s$ allows the generation of a huge number of different skeletons
even starting from a very small set of assignments on a superskeleton.
The group-theoretical factors of these skeletons can thus be evaluated
symbolically on wide classes, as functions of the above parameters
(which are the same entering the labeling of skeletons), and the
explicit evaluation of the potentials entering an actual calculation
can be implemented in a computer algebra program.

The procedure consisting in the generalization of each new object
occurring at a definite order in the expansion to a whole family of
more complicated objects and the symbolic evaluation of all the
members of the family insures a considerable reduction in the number
of new objects appearing at each further step in the extension of the
series.

A final comment concerns the opportunity of applying the above
strategy directly to the computation of connected group-theoretical
factors, i.e.\ of the potentials.  The generation of disconnections
can be performed algorithmically; however we did not develop a
specific computer program, resorting to geometric arguments in the
cases we analyzed explicitly.  All these cases were simple enough for
us to be able to write down compact symbolic expressions referring
directly to the potentials.  Some of our results will be presented in
detail in the following Sections.

\section{Technical remarks on group integration}
\label{technical-remarks}

In evaluating quantities like $S^{({\scr T})}_{\{r\}}$, one may take
advantage of the invariance properties of the Haar measure for group
integration
\begin{equation}
d\mu(U) = d\mu(UA) = d\mu(AU)
\end{equation}
in order to eliminate (``gauge'') one of the variables (defined on the
nodes of the diagram).  A judicious use of gauging can induce notable
simplifications in the actual computations, by replacing ``open
indices'' (representations) with ``closed indices'' (characters) in
the integrands, and decoupling many variables from each other.

As an illustrative example, let us consider the simplest nontrivial
superskeleton.  In principle we must evaluate
\begin{eqnarray}
S^{(\rm Y)}_{r_1,r_2,r_3,r_4,r_5,r_6} &\propto&
\int \chi_{(r_1)}(A B^\dagger)\,\chi_{(r_2)}(B C^\dagger)\,
     \chi_{(r_3)}(C A^\dagger)\,\chi_{(r_4)}(A^\dagger D)\,
     \chi_{(r_5)}(B^\dagger D)\,\chi_{(r_6)}(C^\dagger D)\,
\nonumber \\ &&\qquad\qquad
     dA\,dB\,dC\,dD\,.
\end{eqnarray}
However, by gauging the variable $D$ we can reduce the previous
expression to the factorized integral
\begin{eqnarray}
S^{(\rm Y)}_{r_1,r_2,r_3,r_4,r_5,r_6} &\propto&
\int \chi_{(r_4)}(A^\dagger)\,{\scr D}^{\alpha\beta}_{(r_1)}(A)\,
{\scr D}^{\mu\nu}_{(r_3)}(A)\,dA
\nonumber \\ &\times&
\int \chi_{(r_5)}(B^\dagger)\,{\scr D}^{\gamma\delta}_{(r_2)}(B)\,
{\scr D}^{\beta\alpha}_{(r_1)}(B)\,dB
\int \chi_{(r_6)}(C^\dagger)\,{\scr D}^{\nu\mu}_{(r_3)}(C)\,
{\scr D}^{\delta\gamma}_{(r_2)}(C)\,dC \,,
\end{eqnarray}
whose factors in turn will be expressible in terms of the
representations of the identity via the relationship
\begin{eqnarray}
\int \chi_{(t)}(A^\dagger)\,{\scr D}^{\alpha\beta}_{(r)}(A)\,
{\scr D}^{\gamma\delta}_{(s)}(A)\,dA &=& \sum_{(u)}
\int \chi_{(t)}(A^\dagger)\,{\scr D}^{\mu\nu}_{(u)}(A)\,
{C_{(rsu)}}^{\mu\nu}_{\alpha\beta\gamma\delta}\,dA
\nonumber \\ &=&
{1\over d_{(t)}}\,\delta^{\mu\nu}_{(t)}
{C_{(rst)}}^{\mu\nu}_{\alpha\beta\gamma\delta}
= {1\over d_{(t)}}\,\delta^{(t)}_{\alpha\gamma,\beta\delta} \,,
\end{eqnarray}
where ${C_{(rst)}}^{\mu\nu}_{\alpha\beta\gamma\delta}$ are the
Clebsch-Gordan coefficients and
$\delta^{(t)}_{\alpha\gamma,\beta\delta}$ are the (not necessarily
irreducible) representations of the identity.

We shall call these factors ``gauged vertices'', and present a few
explicit examples, because they are essential ingredients of most of
the actual computations we have performed; it is immediately apparent
that proper gauging can reduce the evaluation of all $\rm X$ (as well
as $\rm Y$) superskeletons to contractions of gauged vertices.

The simplest nontrivial vertex involves two $n=1$ representations and
one $n=2$ representations.  There are three $n=2$ representation,
which we write down adopting the notation
\begin{eqnarray}
{\scr D}^{ik,jl}_\pm(A) &=& \case{1}{2}
\bigl[A_{ij}A_{kl} \pm A_{il}A_{kj}\bigr],
\label{D+-} \\
{\scr D}^{ik,jl}_{1;1}(A) &=& A_{ij}A^\dagger_{lk}
- {1\over N}\,\delta_{ik}\delta_{jl}\,,
\label{D1;1}
\end{eqnarray}
where ${\scr D}_+ = {\scr D}_{2;0}$ and
${\scr D}_- = {\scr D}_{1,1;0}$.  One can easily show that the
(ungauged) vertices are
\begin{eqnarray}
\int {\scr D}^{ik,jl}_\pm(A) A^\dagger_{ab} A^\dagger_{cd} &=&
{1\over d_\pm}\,\delta^{(\pm)}_{ik,bd}\,\delta^{(\pm)}_{ac,jl}
\nonumber \\ &=& {1\over4d_\pm}
\bigl(\delta_{ib}\delta_{kd} \pm \delta_{id}\delta_{kb}\bigr)
\bigl(\delta_{aj}\delta_{cl} \pm \delta_{al}\delta_{cj}\bigr),
\label{D+-v} \\
\int {\scr D}^{ik,jl}_{1;1}(A) A_{ba} A^\dagger_{cd} &=&
{1\over d_{1;1}}\,\delta^{(1;1)}_{ik,db}\delta^{(1;1)}_{ca,jl}
\nonumber \\ &=& {1\over d_{1;1}}
\left(\delta_{id}\delta_{bk} - {1\over N} \delta_{ik}\delta_{db}\right)
\left(\delta_{cj}\delta_{al} - {1\over N} \delta_{ca}\delta_{jl}\right).
\label{D1;1v}
\end{eqnarray}
The gauged vertices are trivially obtained by contraction of indices,
and correspond to the representations of the identity matrix in the
form (\ref{D+-}), (\ref{D1;1}).
Eqs.~(\ref{D+-v}) and (\ref{D1;1v}) may also be used in the evaluation
of a few integrals belonging to superskeletons with topology $\rm H$.

The next vertices in order of difficulty involve one each of the
$n=1$, $n=2$, and $n=3$ representations.
Adopting for $n=3$ representations the notation
\begin{equation}
\chi^{(3)}_+ = \chi_{3;0}, \qquad \chi^{(3)}_- = \chi_{1,1,1;0},  \qquad
\chi^{(2,1)}_+ = \chi_{2;1}, \qquad \chi^{(2,1)}_- = \chi_{1,1;1},
\end{equation}
we may express the corresponding vertices in the form
\begin{eqnarray}
\int \chi^{(3)}_\pm(A)\,{\scr D}^{ik,jl}_\pm(A^\dagger)\,
A^\dagger_{mn}\,dA &=&
{1\over d_\pm^{(3)}}\,\delta_\pm^{(3)}(ikm,jln), \\
\int \chi_{2,1;0}(A)\,{\scr D}^{ik,jl}_\pm(A^\dagger)\,
A^\dagger_{mn}\,dA &=&
{1\over d^{(1,2;0)}}\,\delta_\pm^{(1,2;0)}(ikm,jln), \\
\int \chi^{(2;1)}_\pm(A)\,{\scr D}^{ik,jl}_{1;1}(A)\,
A^\dagger_{mn}\,dA &=&
{1\over d_\pm^{(2;1)}}\,\delta_\pm^{(2;1)}(ikn,jlm), \\
\int \chi^{(2;1)}_\pm(A)\,{\scr D}^{lm,kn}_\pm(A^\dagger)\,
A_{ij}\,dA &=&
{1\over d_\pm^{(2;1)}}\,\delta_\pm^{(2;1)}(ikn,jlm),
\end{eqnarray}
where
\begin{eqnarray}
\delta_\pm^{(3)}(ikm,jln) &=& \case{1}{6}\bigl[
\delta_{ij}\delta_{kl}\delta_{mn} + \delta_{il}\delta_{kn}\delta_{mj} +
\delta_{in}\delta_{kj}\delta_{ml}
\nonumber \\ &&\quad\pm\;
\delta_{il}\delta_{kj}\delta_{mn}\pm
\delta_{ij}\delta_{kn}\delta_{ml}\pm\delta_{in}\delta_{kl}\delta_{mj}
\bigr], \\
\delta_\pm^{(1,2;0)}(ikm,jln) &=& \case{1}{6}\bigl[
2\delta_{ij}\delta_{kl}\delta_{mn} -  \delta_{il}\delta_{kn}\delta_{mj} -
 \delta_{in}\delta_{kj}\delta_{ml}
\nonumber \\  &&\quad\pm\;
2\delta_{il}\delta_{kj}\delta_{mn}\mp
 \delta_{ij}\delta_{kn}\delta_{ml}\mp \delta_{in}\delta_{kl}\delta_{mj}
\bigr], \\
\delta_\pm^{(2;1)}(ikn,jlm) &=& \case{1}{2}\bigl[
\delta_{ij}\delta_{kl}\delta_{nm}\pm\delta_{ij}\delta_{km}\delta_{nl}
\bigr] \nonumber \\ &&\quad-\; {1\over2(N\pm1)}\bigl[
\delta_{ik}\delta_{jl}\delta_{nm}\pm\delta_{ik}\delta_{jm}\delta_{nl}\pm
\delta_{in}\delta_{jl}\delta_{km} + \delta_{in}\delta_{jm}\delta_{kl}
\bigr].
\end{eqnarray}

Aside from a few technicalities, the results from group integration
presented in this section are essentially all that is needed for an
evaluation of the full 15th-order strong-coupling contribution to the
fundamental two-point Green's functions of the two-dimensional chiral
model on the square lattice.

\section{Computing the potentials}
\label{computing-potentials}

The quantities that we have denoted with the general symbol
$P^{({\scr S})}_{\{n,L\}}$ and called {\it potentials\/} are the
connected parts of sums over the sets of representations consistent
withy the geometry of a given skeleton diagram.  Needless to say,
knowledge of compact analytic expressions for wide classes of
potentials can only dramatically simplify the task of evaluating
explicitly high orders of the character expansion.  In turn, since the
reduction of any diagram to its superskeleton can be performed
algorithmically, simplifying the problem of diagram recognition, it
would be obviously pleasant to possess expressions for potentials
general enough to be referred to superskeletons instead of individual
skeletons.

We made some progress in this direction, classifying all and
evaluating most of the skeleton diagrams whose superskeletons are
drawn in Fig.~\ref{topologies} and obey the constraint $n\le3$ for all
links.  In this section we shall present some general considerations
and all the results that are needed for an explicit evaluation of all
$G(x)$ up to 12th order.  We computed many more potentials, but often
results are too cumbersome to make their presentation useful in any
sense; they are available upon request from the authors.

We recall that the potentials are labeled by the same symbols
attributed to the corresponding skeleton diagrams.

We already mentioned that the length of the $n=1$ links does not enter
the definition of the potentials.  Moreover, the bubble content of the
$n=1$ links is factorized, i.e.\ the connected value of the full
diagram is simply the product of the connected values of the diagram
without bubble insertion and the diagram obtained by closing the
bubble on itself and dividing by $N^2$; both these quantities are just
lower-order potentials.  The proof of factorization is very simple,
and can be obtained immediately by gauging one of the vertices of the
bubble and integrating over the second vertex variable.

This explains why we decided not to have a notation for skeletons with
bubbles along $n=1$ paths: their name and value are expressed by the
product of their factors.

Let us now consider bubble insertions on nontrivial links $n\ne1$, in
order of difficulty. The simplest case involves insertion of a bubble
formed by two $n=1$ lines between two vertices.  Let us work out this
example in detail in order to explain the general procedure.
We take the product of representations
\begin{eqnarray}
&&\bigl((1;0) \oplus (0;1)\bigr) \otimes \bigl((1;0) \oplus (0;1)\bigr)
\nonumber \\
&=& (2;0) \oplus (1,1;0) \oplus (0;2) \oplus (0;1,1) \oplus
    (1;1) \oplus (1;1) \oplus (0;0) \oplus (0;0).
\end{eqnarray}
We must recognize that the existence of such a bubble implies the
possibility of two disconnections of the total diagram, corresponding
of the two orientations of the closed path running around the bubble.
Therefore the connected contribution of the bubble is obtained by
removing the two $(0;0)$ representations from the product, and amounts
simply to replacing the bubble with a single $n=2$ line (of length
$L=0$), with weight obtained from Eq.~(\ref{Cddd}) and expressible in
the form
\begin{equation}
B_\pm \equiv {N^2\over d_\pm} \quad\hbox{for}\;{\scr D}_\pm\,, \qquad
B_{1;1} \equiv {2 N^2\over d_{1;1}} \quad\hbox{for}\;{\scr D}_{1;1}\,.
\end{equation}
Given the ubiquitous presence of insertions of such bubbles along
$n=2$ lines, we will adopt the shorthand notation
\begin{equation}
\sigma = 2,L,b,... \equiv 2,L,[1;1]^b...
\end{equation}
for the insertion of $b$ $[1;1]$ bubbles.
Such insertions imply the replacements
\begin{eqnarray}
d_\pm &\to& d_\pm (B_\pm)^b , \\
d_{1;1} &\to& d_{1;1} (B_{1;1})^b
\end{eqnarray}
in the expression for the value of the corresponding superskeleton,
and the inclusion of a factor $2^b$ in front of all the disconnections
corresponding to a splitting of the $n=2$ line.

We may now consider the insertion of a bubble $[1;2]$ between two
vertices.  According to the rules for the product of representations,
the corresponding contribution is obtained by replacing the bubble
either with a $n=1$ line or with a $n=3$ line (with different weights
attached to different $n=3$ representations).  In the first case we
may apply the previously discussed factorization, while in the second
case it is convenient to define the {\it bubble factors\/} $B(a,b)$ as
\begin{eqnarray}
d^{(3)}_\pm B^{(3)}_\pm(a,b) &=& N d_\pm z_\pm^a B_\pm^b , \\
d_{2,1;0} B_{2,1;0}(a,b) &=&
N d_+ z_+^a B_+^b + N d_- z_-^a B_-^b , \\
d^{(2,1)}_\pm B^{(2,1)}_\pm(a,b) &=&
N d_\pm z_\pm^a B_\pm^b + N d_{1;1} z_{1;1}^a B_{1;1}^b .
\end{eqnarray}
The insertion of the set of $k$ bubbles
$[2,a_1,b_1;1]...[2,a_k,b_k;1]$ along a $n=3$ line can now be
accounted for by the following substitutions in the expression of the
superskeleton:
\begin{eqnarray}
d^{(3)}_\pm &\to& d^{(3)}_\pm \prod_{i=1}^k B^{(3)}_\pm(a_i,b_i), \\
d_{2,1;0} &\to& d_{2,1;0} \prod_{i=1}^k B_{2,1;0}(a_i,b_i), \\
d^{(2,1)}_\pm &\to& d^{(2,1)}_\pm \prod_{i=1}^k B^{(2,1)}(a_i,b_i).
\end{eqnarray}
Moreover one must introduce factors of $2^{b_i}$ in the disconnections
involving the splitting of the $i$th $n=2$ line, and a factor $3^k$ in
the disconnections involving the full splitting of the $n=3$ line into
$n=1$ lines.

Next in order of difficulty are the rules concerning the insertions of
$[1,3]$ and $[2;2]$ bubbles.  In each case the allowed replacements
involve either a $n=2$ or a $n=4$ line.

The bubble factors to be inserted along a $n=4$ line are essentially
trivial generalizations of our previous examples whose expressions we
shall not exhibit explicitly.

The $n=2$ case is more interesting, because it is the first instance
of a new phenomenon: the occurrence of disconnections of the skeleton
diagram not corresponding to disconnections of the superskeleton.  As
one may easily understand, these disconnections correspond to
lower-order bubbles that may be removed from the skeleton turning it
into another acceptable skeleton.  This possibility can be
systematically taken into account by defining {\it connected\/} bubble
insertions.

Let us therefore introduce the bubble factors
$B(p;a_1,b_1;...;a_r,b_r)$, corresponding to the insertion of
$[1;3,p,[1;2,a_1,b_1]...[1;2,a_r,b_r]]$, and $C(a_1,b_1;a_2,b_2)$,
corresponding to the insertion of $[2,a_1,b_1;2,a_2,b_2]$:
\begin{eqnarray}
d_\pm B_\pm(p;a_1,b_1;...;a_r,b_r) &=&
N d^{(3)}_\pm \bigl(z^{(3)}_\pm\bigr)^p \prod_{i=1}^r B^{(3)}_\pm(a_i,b_i) +
N d_{2,1;0} z_{2,1;0}^p \prod_{i=1}^r B_{2,1;0}(a_i,b_i)
\nonumber \\ &&\quad +\;
N d^{(2,1)}_\pm \bigl(z^{(2,1)}_\pm\bigr)^p
\prod_{i=1}^r B^{(2,1)}_\pm(a_i,b_i)
\nonumber \\ &&\quad -\;
2 N^2 d_\pm z_\pm^p \prod_{i=1}^r
\left(z_\pm^{a_i} B_\pm^{b_i} + 2^{b_i} B_\pm\right),\\
d_{1;1} B_{1;1}(p;a_1,b_1;...;a_r,b_r) &=&
N d^{(2,1)}_+ \bigl(z^{(2,1)}_+\bigr)^p \prod_{i=1}^r
B^{(2,1)}_+(a_i,b_i)
\nonumber \\ &&\quad +\;
N d^{(2,1)}_- \bigl(z^{(2,1)}_-\bigr)^p \prod_{i=1}^r B^{(2,1)}_-(a_i,b_i)
\nonumber \\ &&\quad -\;
N^2 d_{1;1} z_{1;1}^p \prod_{i=1}^r
\left(z_{1;1}^{a_i} B_{1;1}^{b_i} + 2^{b_i} B_{1;1}\right),\\
d_\pm C_\pm(a_1,b_1;a_2,b_2) &=&
d_{1;1} z_{1;1}^{a_1} B_{1;1}^{b_1}
\left[d_+ z_+^{a_2} B_+^{b_2} + d_- z_-^{a_2} B_-^{b_2}\right]
\nonumber \\ &&\quad +\;
d_{1;1} z_{1;1}^{a_2} B_{1;1}^{b_2}
\left[d_+ z_+^{a_1} B_+^{b_1} + d_- z_-^{a_1} B_-^{b_1}\right] -
2 N^4 2^{b_1+b_2}, \\
d_{1;1} C_{1;1}(a_1,b_1;a_2,b_2) &=&
d_{1;1}^2 z_{1;1}^{a_1+a_2} B_{1;1}^{b_1+b_2}
\nonumber \\ &&\quad +\;
\left[d_+ z_+^{a_1} B_+^{b_1} + d_- z_-^{a_1} B_-^{b_1}\right]
\left[d_+ z_+^{a_2} B_+^{b_2} + d_- z_-^{a_2} B_-^{b_2}\right]
\nonumber \\ &&\quad -\;
2 N^4 2^{b_1+b_2}.
\end{eqnarray}
When considering disconnections of these diagrams, one must be careful
to include only those that have a corresponding term among the
disconnections of the superskeleton.

\begin{figure}[tb]
\centerline{\setlength{\unitlength}{0.0125in}
\begin{picture}(330,164)(0,-10)
\put(300.000,30.438){\arc{53.124}{3.9153}{5.5094}}
\put(300.000,29.562){\arc{53.124}{0.7737}{2.3679}}
\put(30.000,30.154){\arc{53.692}{3.6835}{5.7413}}
\put(210.000,30.300){\arc{53.400}{3.5952}{5.8296}}
\put(30,119){\ellipse{60}{60}}
\put(120,119){\ellipse{60}{60}}
\put(120,119){\ellipse{54}{54}}
\put(210,119){\ellipse{60}{60}}
\put(210,119){\ellipse{54}{54}}
\put(210,119){\ellipse{48}{48}}
\put(120,30){\ellipse{60}{60}}
\put(30,30){\ellipse{60}{60}}
\put(300,30){\ellipse{60}{60}}
\put(210,30){\ellipse{60}{60}}
\path(210,30)(210,0)
\path(120,27)(93,42)
\path(321,9)(279,51)
\path(279,9)(321,51)
\path(30,30)(4,45)
\path(30,30)(56,45)
\path(32,31)(32,0)
\path(29,31)(29,0)
\path(122,31)(122,0)
\path(121,30)(94,45)
\path(120,27)(147,42)
\path(119,30)(146,45)
\path(119,31)(119,0)
\path(210,30)(237,43)
\path(210,26)(239,40)
\path(210,30)(183,43)
\path(210,22)(239,36)
\path(213,0)(213,31)
\end{picture}
}
\caption{Potentials.}
\label{all-superskeletons}
\end{figure}
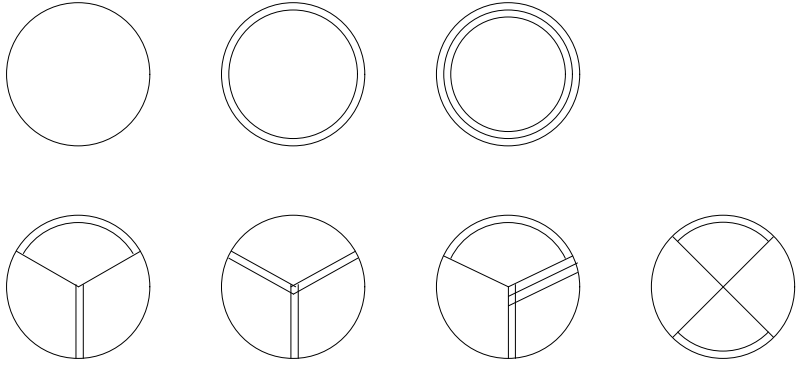

These rules are the essential ingredients for the construction of the
connected contributions of all the skeleton diagrams entering our
15th-order calculations.  In particular, all potentials entering
12th-order calculations can be obtained by the abovementioned
insertions into the superskeletons drawn in
Fig.~\ref{all-superskeletons}.

The values of these potentials are reported in
Appendix~\ref{values-potentials}.  Here we will only report the
results concerning ${\rm W}(2,...)$, for reference and illustration of
our formalism.

We first recall that ${\rm W}(1)$ is completely trivial:
${\rm W}(1)=1$ and the associate geometrical factor is related to the
number of self-avoiding random walks of length equal to the power of
$z$.

For the most general potential related to ${\rm W}(2,L)$ we are
interested in, the main $n=2$ line splits into $q$ $[1;1]$ bubbles,
$r$ $[2;2]$ bubbles (the bubble links themselves splitting into
$b_{i,1}$ and $b_{i,2}$ $[1;1]$ bubbles), and $s$ $[3;1]$ bubbles (the
$n=3$ link splitting into $u_j$ $[2;1]$ bubbles, each $n=2$ link
splitting into $b'_{jk}$ $[1;1]$ bubbles).  We obtained the value of
the potentials in the form
\begin{eqnarray}
&&N^2 {\rm W}(2,L,q, [2,a_{1,1},b_{1,1};2,a_{1,2},b_{1,2}]...
[2,a_{r,1},b_{r,1};2,a_{r,2},b_{r,2}]
\nonumber \\ &&\qquad\times\;
[1;3,p_1,[2,a'_{1,1},b'_{1,1};1]...[2,a'_{1,u_1},b'_{1,u_1};1]]...
[1;3,p_s,[2,a'_{s,1},b'_{s,1};1]...[2,a'_{s,u_s},b'_{s,u_s};1]])
\nonumber \\
&=&
z_+^L d_+^2 B_+^q \prod_{j=1}^s
B_+(p_j,a'_{j,1},b'_{j,1},...,a'_{j,u_j},b'_{j,u_j})
\prod_{i=1}^r C_+(a_{i,1},b_{i,1},a_{i,2},b_{i,2})
\nonumber \\
&+&
z_-^L d_-^2 B_-^q \prod_{j=1}^s
B_-(p_j,a'_{j,1},b'_{j,1},...,a'_{j,u_j},b'_{j,u_j})
\prod_{i=1}^r C_-(a_{i,1},b_{i,1},a_{i,2},b_{i,2})
\nonumber \\
&+&\case{1}{2} z_{1;1}^L d_{1;1}^2 B_{1;1}^q \prod_{j=1}^s
B_{1;1}(p_j,a'_{j,1},b'_{j,1},...,a'_{j,u_j},b'_{j,u_j})
\prod_{i=1}^r C_{1;1}(a_{i,1},b_{i,1},a_{i,2},b_{i,2})
\nonumber \\
&-&
2^{q+r+s} N^4 \prod_{j=1}^s \sum_{{\scr P}(u_j)}
2^{\sum_{m\not\in{\scr P}(u_j)} b'_{j,m}}
{\rm W}\bigl(2,p_j + {\textstyle\sum_{k\in{\scr P}(u_j)}}a'_{j,k},
u_j + {\textstyle\sum_{k\in{\scr P}(u_j)}} (b'_{j,k}-1)\bigr)
\nonumber \\
&&\quad\times\;
\prod_{i=1}^r
\bigl(2^{b_{i,2}} {\rm W}(2;a_{i,1},b_{i,1}) +
      2^{b_{i,1}} {\rm W}(2;a_{i,2},b_{i,2})\bigr),
\end{eqnarray}
where ${\scr P}(u_j)$ are all the subsets of $\{1,...,u_j\}$,
and $\sum_{m\not\in{\scr P}(u_j)}$ is a shorthand for
$\sum_{m\in\{1,...,u_j\}\setminus{\scr P}(u_j)}$.

\section{The two-point Green's functions and the inverse propagator}
\label{Greens-functions}

The techniques and results presented in the previous Sections set up
the stage for the evaluation of the strong-coupling series for the
two-point Green's functions $G(x)$ of ${\rm U}(N) \times {\rm U}(N)$
principal chiral models on a two-dimensional square lattice, as
functions of $x$, $z(\beta)$, and of the potentials.  At any finite
order $q$ of the strong-coupling expansion, only a finite number of
coordinate space Green's functions are nonzero, owing to the fact that
the leading contribution comes from the shortest walk connecting $x$
with the origin, which is proportional to $z^{|x_1|+|x_2|}$; therefore
all the Green's functions such that $|x_1|+|x_2|>q$ vanish.  The
number of nontrivial Green's functions, exploiting discrete
symmetries, is therefore
\begin{equation}
\case{1}{4}(q+2)^2\quad\hbox{($q$ even),}\qquad
\case{1}{4}(q+1)(q+3)\quad\hbox{($q$ odd).}
\end{equation}

Coordinate space Green's functions are the natural output of a
strong-coupling computation.  Listing their individual strong-coupling
series is however by no means the most compact and physically most
appealing way of presenting the results.  It is certainly convenient
to introduce the lattice momentum transform
\begin{equation}
\tilde G(p) = \sum_x G(x)\, \exp(ip \cdot x)
\label{G-order}
\end{equation}
which, because of the lattice symmetries, turns out to be a function of
the symmetric combinations of $\cos n_1 p_1$ and  $\cos n_2 p_2$, with
$n_1,n_2\le q$.

A really dramatic simplification however occurs only when we take into
consideration the inverse lattice propagator $\tilde G^{-1}(p)$.
Indeed, due to the recursive nature of the path-generating process,
any strong-coupling expansion admitting a reinterpretation as a
summation over paths can be seen, at any definite order in the
expansion, as originated by a generalized Gaussian model in which the
appearance of new structures violating lower-order recursion equations
can be seen as the effect of quasi-local interactions that appear in
the inverse propagator as Fourier transforms of non-nearest neighbor
couplings.  A new structure capable of violating the recursion must
correspond to a nontrivial path topology, with the property of
multiply connecting the endpoints.  Such a path must necessarily be at
least three times as long as the minimal path.  This arguments shows
that, in contrast with Eq.~(\ref{G-order}), in the inverse propagator
combinations of $\cos n_1 p_1$ and $\cos n_2 p_2$ may appear only for
$n_1,n_2\le q/3$.  A more refined analysis shows that the highest
values of $n_1$ and $n_2$ generated in $\tilde G^{-1}(p)$ to order $q$
in the expansion are
\begin{eqnarray}
n_1,n_2\le u-2 &\quad& \hbox{($q=3u-2$),} \qquad
n_1,n_2\le u-1\quad\hbox{($q=3u-1$),} \nonumber \\
n_1,n_2\le u   &\quad& \hbox{($q=3u$),} \qquad
\hbox{($u$ integer).}
\end{eqnarray}

A more immediate physical interpretation of the results is obtained by
introducing the traditional function
\begin{equation}
\hat p_\mu = 2\sin{p_\mu\over2}\,,
\end{equation}
and expressing $\tilde G^{-1}(p)$ as a function of $\hat p_\mu^2 =
2(1-\cos p_\mu)$.  One may easily get convinced that the number of
independent symmetric combinations of powers of $\hat p_\mu^2$
entering a given order in the expansion of $\tilde G^{-1}(p)$ is equal
to the number of independent effective couplings one might define at
the same order, consistently with the abovementioned considerations.
This is in turn related to the number of lattice sites, not related by
a lattice symmetry transformation, such that $|x_1|+|x_2|\le u$.  We
found that a natural basis for the parametrization of these
independent combinations is offered by
\begin{equation}
\hat p^{2s} \bigl(\bigl(\hat p^2\bigr)^2 - \hat p^4\bigr)^t,
\qquad s+2t \le u,
\end{equation}
where
\begin{equation}
\hat p^0 = 1,\qquad
\hat p^{2s} = \sum_\mu \hat p_\mu^{2s}\ \ (s\ge1).
\end{equation}
We also found that terms with $t\ne0$ appear in $\tilde G^{-1}(p)$ at
order $q=3(s+2t)$, while terms with $t=0$ appear only at order
$q=3s+2$; the implication of this phenomenon will be discussed later.

We will therefore make use of the parametrization
\begin{equation}
\tilde G^{-1}(p) = A_0 + A_1 \hat p^2 + \sum_{u=2}^\infty
\sum_{\stackrel{\scriptstyle s=0}{\scriptstyle u-s\ \rm even}}^u
A_{u,s}
\hat p^{2s} \bigl(\bigl(\hat p^2\bigr)^2 - \hat p^4\bigr)^{(u-s)/2},
\label{G-A}
\end{equation}
to present our strong-coupling results for the inverse propagator in
the form of expansions for the coefficients $A_{u,s}$.  As already
mentioned, the expansions of $A_{u,s}$ will be power series on $z$,
starting with $z^{3u}$ when $s\ne u$ and with $z^{3u+2}$ when $s=u$,
with coefficients that are polynomials in the potentials.

The large-$N$ limit of the $A_{u,s}$ is
\begin{mathletters}
\begin{eqnarray}
A_{0} &=&
1 - 4\,z + 4\,{z^2} - 4\,{z^3} + 12\,{z^4} - 28\,{z^5} + 52\,{z^6} -
132\,{z^7} + 324\,{z^8} - 908\,{z^9} + 2020\,{z^{10}}
\nonumber \\ &&\quad -\; 6284\,{z^{11}}
+ 15284\,{z^{12}} - 48940\,{z^{13}} + 116612\,{z^{14}} -
393132\,{z^{15}} + O\bigl(z^{16}\bigr),
\\
A_{1} &=&
z + {z^3} + 7\,{z^5} + 4\,{z^6} + 33\,{z^7} + 32\,{z^8} + 243\,{z^9} +
324\,{z^{10}} + 1819\,{z^{11}} + 2520\,{z^{12}}
\nonumber \\ &&\quad +\; 14859\,{z^{13}}
+ 23084\,{z^{14}} + 123883\,{z^{15}}
+ O\bigl(z^{16}\bigr),
\\
A_{2,0} &=&
-{z^6} - 6\,{z^8} - 8\,{z^9} - 57\,{z^{10}} - 116\,{z^{11}} -
500\,{z^{12}} - 1152\,{z^{13}} - 5155\,{z^{14}} - 11632\,{z^{15}}
\nonumber \\ &&\quad +\; O\bigl(z^{16}\bigr),
\\
A_{2,2} &=&
- 2\,{z^8} - 4\,{z^9} - 24\,{z^{10}} - 70\,{z^{11}} - 242\,{z^{12}} -
816\,{z^{13}} - 2824\,{z^{14}} - 8528\,{z^{15}}
\nonumber \\ &&\quad +\; O\bigl(z^{16}\bigr),
\\
A_{3,1} &=&
{z^9} + {\case{29}{2}\,{z^{11}}} + 26\,{z^{12}} + 144\,{z^{13}} +
482\,{z^{14}} + 1806\,{z^{15}} + O\bigl(z^{16}\bigr),
\\
A_{3,3} &=&
2\,{z^{11}} + 4\,{z^{12}} + 40\,{z^{13}} + 140\,{z^{14}} +
548\,{z^{15}} + O\bigl(z^{16}\bigr),
\\
A_{4,0} &=&
-{\case{5}{2}\,{z^{12}}} - 37\,{z^{14}} - 84\,{z^{15}} +
O\bigl(z^{16}\bigr),
\\
A_{4,2} &=&
-{z^{12}} - 31\,{z^{14}} - 64\,{z^{15}} + O\bigl(z^{16}\bigr),
\\
A_{4,4} &=&
-2\,{z^{14}} - 4\,{z^{15}} + O\bigl(z^{16}\bigr),
\\
A_{5,1} &=&
7\,{z^{15}} + O\bigl(z^{16}\bigr),
\\
A_{5,3} &=&
{z^{15}} + O\bigl(z^{16}\bigr).
\end{eqnarray}
\end{mathletters}

We have computed all $A_{u,s}$ to ${\rm O}(z^{15})$ as functions of
the potentials, but the results will not be presented here, for
reasons explained in the Introduction.  We shall limit ourselves to
the presentation in Appendix~\ref{N-square-results} of 15th-order
expressions as explicit functions of $z$ and $N$.  These functions
are obtained by substituting Eq.~(\ref{dze}) for the character
expansion coefficients, obtaining the values of the potentials as
$N$-dependent coefficients, and summing up all homogeneous
contributions.

The limitations of such a procedure can easily be identified:
$q$th-order expressions are correct for ${\rm U}(N)$ groups with $N\ge
q/2$ and for ${\rm SU}(N)$ groups with $N\ge q+2$.  Even symbolic
expressions for potentials suffer from some limitations, essentially
because for small $N$ not all the representations formally introduced
are really nontrivial or independent.  A manifestation of this fact is
the appearance of the so-called 't Hooft-DeWit poles, which plague
${\rm U}(N)$ strong-coupling expressions when $N\le(q-2)/4$.  In
${\rm SU}(N)$ another limitation comes from the occurrence of
self-dual representations, which spoil the applicability of
${\rm U}(N)$ results already for $N\le q/2$.

In practice the results we have presented hold as they stand for all
${\rm U}(N)$ groups with $N>7$, while by using 12th-order expressions
in terms of potentials one might obtain with a minor effort 15th-order
expressions correct for all $N>3$.  ${\rm SU}(N)$ groups are correctly
reproduced for $N>16$, and by use of 8th-order potentials one might
obtain all $N>7$.

We must however stress that in their most abstract formulation, i.e.\
when expressed as weighted combinations of connected group-theoretical
factors, our results are fully general and apply not only to principal
chiral models but also to all nonlinear sigma models on group
manifolds admitting a character expansion, including ${\rm O}(N)$ and
${\rm CP}^{N-1}$ models.

A number of physically interesting quantities can be extracted from
$\tilde G^{-1}(p)$ by appropriate manipulations.  In the present
Section we will only present the large-$N$ limit of some of them.  In
particular we obtain the magnetic susceptibility
\begin{eqnarray}
\chi &=& \sum_x G(x) = {1\over A_0} \nonumber \\
&=& 1 + 4\,z + 12\,{z^2} + 36\,{z^3} + 100\,{z^4} +
  284\,{z^5} + 796\,{z^6} + 2276\,{z^7} + 6444\,{z^8} +
  18572\,{z^9} \nonumber \\
&&\quad+\;  53292\,{z^{10}}
  + 155500\,{z^{11}} + 451516\,{z^{12}} + 1330796\,{z^{13}} +
  3904908\,{z^{14}} + 11617356\,{z^{15}}
\nonumber \\ &&\quad+\; O\bigl(z^{16}\bigr).
\end{eqnarray}
By defining the second moment of the correlation functions
\begin{equation}
\chi\left<x^2\right>_G = {1\over4} \sum_x x^2 G(x)
= {A_1\over A_0^2}\,,
\end{equation}
we can introduce the second-moment definition of the correlation
length
\begin{eqnarray}
M^2_G &=& {1\over\left<x^2\right>_G} = {A_0\over A_1} \nonumber \\
&=&
{1\over z} - 4 + 3\,z + 2\,{z^3} - 4\,{z^4} + 12\,{z^5} - 40\,{z^6} +
84\,{z^7} - 296\,{z^8} + 550\,{z^9} - 1904\,{z^{10}} \nonumber \\
&&\quad+\;
3316\,{z^{11}} - 15248\,{z^{12}} + 27756\,{z^{13}} +
O\bigl(z^{14}\bigr)
\end{eqnarray}
and the corresponding wavefunction renormalization
\begin{eqnarray}
Z_G &=& {1\over A_1} =
  z + {z^3} + 7\,{z^5} + 4\,{z^6} + 33\,{z^7} + 32\,{z^8} + 243\,{z^9} +
  324\,{z^{10}}\nonumber \\
&&\quad+\;
  1819\,{z^{11}} + 2520\,{z^{12}} + 14859\,{z^{13}} +
  23084\,{z^{14}} + 123883\,{z^{15}} + O\bigl(z^{16}\bigr).
\end{eqnarray}

The true mass gap should in principle be extracted from the
long-distance behavior of the two-point Green's function:
\begin{equation}
\mu = -\lim_{|x|\to\infty} \ln{G(x)\over|x|} \,.
\label{mu-def}
\end{equation}
This quantity is however related by standard analyticity properties to
the imaginary momentum pole singularity of $\tilde G(p)$, and we can
therefore extract the mass gap by solving the equation
\begin{equation}
\tilde G^{-1}(p_1{=}i\mu_s, p_2{=}0) = 0.
\label{mus-eq}
\end{equation}
In absence of strict rotation invariance, this quantity is to be
interpreted as the wall-wall correlation length.  Let us notice that
Eq.~(\ref{mus-eq}) involves only the coefficients $A_{u,u}$ in the
expansion (\ref{G-A}) of $\tilde G^{-1}(p)$.  the power series of
these coefficient in turn start with $z^{3u+2}$ and are
associated with factors $\hat p^{2u}$.  Eq.~(\ref{mus-eq}) is
therefore, order by order in the strong-coupling expansion, algebraic
and series-expandable in the variable
\begin{equation}
zM^2_s = 2z(\cosh\mu_s-1).
\end{equation}

By recalling the properties of $A_{u,u}$ one can easily get convinced
that knowledge of $\tilde G^{-1}(p)$ to ${\rm O}(z^{3u-1})$ and
${\rm O}(z^{3u})$ allows for the determination of $\mu_s$ to
${\rm O}(z^{2u})$ and ${\rm O}(z^{2u+1})$ respectively.
There is a deep connection between the orders of the strong-coupling
expansion ``lost'' in the evaluation of $\mu_s$ and the
above-mentioned considerations on the appearance of structures
violating the recursive relationships among paths.  Indeed these
structures break down the exponentiation of the wall-wall correlation
functions at short distances \cite{Brower-Rossi-Tan-1,%
Brower-Rossi-Tan-2,Green-Samuel-2}, and one is easily convinced
that loss of exponentiation at a distance $\sim u$ implies from
Eq.~(\ref{mu-def}) a residual precision $\sim 2u$ in the determination
of $\mu$.

The resulting value is
\begin{eqnarray}
\mu_s &=& -\log z -2\,z -\case{2}{3}\,{{z^3}} -2\,{z^4}
-\case{42}{5}\,{{z^5}} -8\,{z^6} -\case{310}{7}\,{{z^7}} -70\,{z^8}
-\case{3188}{9}\,{{z^9}} -520\,{z^{10}} -\case{28778}{11}\,{{z^{11}}}
\nonumber \\
&&\quad-\;\case{13154}{3}\,{{z^{12}}} + O\bigl(z^{13}\bigr).
\end{eqnarray}

In full analogy with the discussion above, we may consider the
equation for the diagonal mass gap (i.e.\ the diagonal wall-wall
correlation length)
\begin{equation}
\tilde G^{-1}(p_1{=}i\mu_d/\sqrt2, p_2{=}i\mu_d/\sqrt2) = 0.
\label{mud-eq}
\end{equation}
Eq.~(\ref{mud-eq}) is algebraic and series-expandable in
\begin{equation}
zM^2_d = 4z\left(\cosh{\mu_d\over\sqrt2}-1\right).
\end{equation}
Moreover one may show that $zM^2_d$ is an even function of
$z$, and knowledge of $\tilde G^{-1}(p)$ to
${\rm O}(z^{3u})$ allows for the determination of $\mu_d$ to
${\rm O}(z^{2u})$.
The result is
\begin{equation}
\mu_d=-\log 2z-{z^2}-3\,{z^4}-\case{119}{6}\,{{z^6}}
-136\,{z^8}-\case{41963}{40}\,{{z^{10}}} + O\bigl(z^{12}\bigr).
\end{equation}

Our results for the side and diagonal mass gap in terms of potentials,
up to ${\rm O}(z^{12})$ and ${\rm O}(z^{11})$ respectively, are
available as expained in the Introduction.  They are presented in form
of explicit functions of $N$ and $z$ in
Appendix~\ref{N-square-results}.  In order to compute the 12th-order
contribution to $\mu_s$, we evaluated a few long-distance Green's
functions to ${\rm O}(z^{16})$ and ${\rm O}(z^{17})$.

The analytic properties of the strong-coupling series (radius of
convergence, zeroes of partition function, critical behavior) are best
studied by considering such bulk quantities as the free energy
\begin{equation}
F(\beta) = {1\over N^2 V}\ln\int\prod_x dU(x) \exp(-S_L),
\end{equation}
the internal energy
\begin{equation}
E(\beta) = 1 - {1\over4}\,{\partial F\over\partial\beta} = 1 - G((1,0)),
\label{E-def}
\end{equation}
and the specific heat
\begin{equation}
C(\beta) =  - \beta^2\,{\partial E\over\partial\beta} =
{1\over4}\,\beta^2\,{\partial^2 F\over\partial\beta^2}\,.
\end{equation}
We were able to generate 20th-order series for the free energy per
site $F$ in terms of the potentials introduced in
Fig.~\ref{all-superskeletons}; the results are available as explained
in the Introduction.  We shall only present here the explicit
expression for large $N$ up to 18th order, and report the
result in terms of $N$ and $\beta$ in Appendix~\ref{N-square-results},
with the usual warning that they hold only for ${\rm U}(N)$, $N\ge9$,
and ${\rm SU}(N)$, $N>18$.
\begin{eqnarray}
F &=& 2\,{z^2} + 2\,{z^4} + 4\,{z^6} + 19\,{z^8} + 96\,{z^{10}} +
604\,{z^{12}} + 4036\,{z^{14}} + \case{58471}{2}\,{z^{16}} +
\case{663184}{3}\,{z^{18}}
\nonumber \\ &&\quad+\; O\bigl(z^{20}\bigr).
\end{eqnarray}

For large but finite $N$, we can use Eq.~(\ref{z1-beta}) to replace
the derivative with respect to $\beta$ in Eq.~(\ref{E-def}) with a
derivative with respect to $z$, thus obtaining a relationship between
$F$ and $G((1,0))$, which we verified explicitly.  It should be
noticed that no simple relationship between $F$ and $G((1,0))$ can be
obtained in terms of potentials.

\section{Strong coupling expansion on the honeycomb lattice.}
\label{honeycomb-lattice}

On the honeycomb lattice we consider the action with nearest-neighbor
interaction, which can be written as a sum over all links of the
lattice
\begin{equation}
S_{\rm h} = -2 N \beta \sum_{\rm links}
{\mathop{\rm Re}}\,{\mathop{\rm Tr}}\,[ U_l\,U^\dagger_r] ,
\label{exaction}
\end{equation}
where $l,r$ indicate the sites at the ends of each link.  As on the
square lattice, the link length $a$ is choosen as lattice spacing,
i.e.\ as length unit.  The continuum action of chiral models is
obtained from the $a\rightarrow 0$ limit of $S_{\rm h}$ by identifying
\begin{equation}
T =  {\sqrt{3}\over N\beta}\;.
\label{tempex}
\end{equation}

For what concerns the strong-coupling expansion, we only mention that
the determination of the geometrical factor is straightforward in the
light of our general discussion.  The only difference with the square
lattice concerns the generation of non-backtracking random walks.  It
is easy to see that the honeycomb lattice can be mapped in a subset of
the square lattice, having the same sites, the same links in $y$
direction, and only the links in $x$ directions starting from even
sites: therefore the walks on the honeycomb lattice are a subset of
the walks on the square lattice.  The value of the potentials is
obviously unchanged, but since the computation can be pushed some
orders further on the honeycomb lattice a few new calculations are
needed.  The only subtle point is that, since only half of the sites
of the lattice are related by translation invariance, the free energy
per site is one half of the quantity computed according to
Sect.~\ref{geometrical-factor}.

We generated strong coupling series of the free energy up to
$O\bigl(z^{26}\bigr)$, and of the fundamental Green's function
up to $O\bigl(z^{20}\bigr)$.  Our results as functions of the
potentials are available as explained in the Introduction.  We present
here only the large-$N$ results; we refer to
Appendix ~\ref{N-honeycomb-results} for results for large but finite
$N$.

In analogy wityh the square lattice, we evaluated the strong coupling
series of the fundamental correlation function $
G(x)={1\over N}\left<{\rm Tr}\,[U(x) U^\dagger(0)]\right>$ as function
of $x$, $z(\beta)$, and of the potentials.  The number of nontrivial
components of $G(x)$ which must evaluated at a given order $q$ is
\begin{eqnarray}
&&{q^2+3q+4\over 4} \quad {\rm if} \quad q=4k,4k+1\;,\nonumber \\
&&{q^2+3q+2\over 4} \quad {\rm if} \quad q=4k+2,4k+3\;,
\label{nocomp}
\end{eqnarray}
with $k$ integer.

The magnetic susceptibility $\chi$ and second-moment correlation
length $\xi_G$ are defined on the honeycomb lattice in perfect analogy
with square lattice definitions: $\chi=\sum_x G(x)$,
$\chi \xi^2_G=\case{1}{4}\sum_x x^2 G(x)$.

The analysis of models on honeycomb lattices presents some
complications, which will be illustrated in some detail in
Appendix~\ref{N-honeycomb-results} by considering a simple Gaussian
model of random walk.  The point is that, unlike square and triangular
lattices, not all sites are related by a translation; this fact does
not allow a straightforward definition of a Fourier transform.  Only
sites at an even distance (in the number of links) are related by a
translation.  We therefore define even and odd fields $U_e$, $U_o$;
$U_e(x)=U(x)$ for even $x$, zero for odd $x$, and $U_o(x)=U(x)$ for
odd $x$, zero for even $x$ (the parity is defined with respect to an
arbitrarily chosen origin).  We then define and even and odd
correlation functions
\begin{eqnarray}
G_e(x-y)&=& {1\over N}\left<{\rm Tr}\,[U_e(x) U^\dagger_e(y)]\right>
 =  {1\over N}\left<{\rm Tr} \,[U_o(x) U^\dagger_o(y)]\right> ,
\nonumber \\
G_o(x-y)&=& {1\over N}\left<{\rm Tr}\,[U_e(x) U^\dagger_o(y)]\right> =
{1\over N} \left<{\rm Tr}\,[U_o(x)U^\dagger_e(y)]\right> .
\label{hexcorrfunc}
\end{eqnarray}
Since even and odd sites lie on two distinct triangular sublattices,
it is possible to define consistent Fourier transforms on each
sublattice.

Guided by the analysis of the Gaussian model, we considered two
orthogonal wall-wall correlation functions:
\begin{eqnarray}
G^{{\rm(w)}}_1(x) &=& \sum_y G_e(x,y), \\
G^{{\rm(w)}}_2(x) &=& \sum_y \left[G_e(x,y) + G_o(x,y)\right].
\end{eqnarray}
In the strong-coupling domain both $G^{{\rm(w)}}_1(x)$ and
$G^{{\rm(w)}}_2(x)$ enjoy exponentiation for sufficiently large
lattice distance, allowing the definition of two corresponding masses
$\mu_1$ and $\mu_2$:
\begin{eqnarray}
G^{{\rm(w)}}_1(x) &\propto& \exp\bigl(-\case{3}{2}\mu_1 x\bigr), \\
G^{{\rm(w)}}_2(x) &\propto& \exp\bigl(-\case{1}{2}\sqrt{3}\mu_2 x\bigr).
\end{eqnarray}
In the continuum limit $\mu_1=\mu_2$ and they
both should reproduce the physical mass $M$ propagating in the
fundamental channel.  The ratio $\mu_1/\mu_2$ allows a test of
rotational invariance, in analogy with the side/diagonal mass ratio of
the square lattice.  It is also possible to define the quantities
\begin{eqnarray}
zM^2_1 = \case{4}{9}z\bigl(\cosh\case{3}{2}\mu_1-1\bigr), \\
zM^2_2 = \case{4}{3}z\bigl(\cosh\case{1}{2}\sqrt{3}\mu_2-2\bigr),
\end{eqnarray}
which play the r\^ole of $zM^2_s$ and $zM^2_d$ in the determination of
imaginary momentum pole of the inverse Fourier-transformed Green's
function.

In full analogy with the square lattice, we define the magnetic
susceptibility
\begin{eqnarray}
\chi &=& \sum_x G(x) =
1 +3\,z +6\,{z^2} +12\,{z^3} +24\,{z^4} +48\,{z^5} +90\,{z^6}
+174\,{z^7} + 348\,{z^8}
+702\,{z^9}
\nonumber \\&+&\; 1392\,{z^{10}}
+ 2814\,{z^{11}} +5658\,{z^{12}}
+ 11532\,{z^{13}} +23706\,{z^{14}} +49368\,{z^{15}} +101436\,{z^{16}}
\nonumber \\ &+&\; 211290\,{z^{17}}
+ 440598\,{z^{18}} + 928614\,{z^{19}} +1950390\,{z^{20}} +
O\bigl(z^{21}\bigr),
\end{eqnarray}
the second moment of the correlation functions
\begin{equation}
\chi\left<x^2\right>_G = {1\over4} \sum_x \bigl\{
(9 x_1^2 + 3 x_2^2) G_e(x) + [(3 x_1 - 1)^2 + 3 x_2^2] G_o(x)\bigr\},
\end{equation}
and the second-moment definition of the correlation length
\begin{eqnarray}
M^2_G &=& {1\over\left<x^2\right>_G}
= \case{4}{3}\,{z^{-1}} -4 +\case{8}{3}\,{z} -8\,{z^6}
+\case{40}{3}\,{{z^7}} +8\,{z^8} -16\,{z^9}
-\; 88\,{z^{10}} +96\,{z^{11}} \nonumber \\
&-&\; 144\,{z^{12}} +
\case{584}{3}\,{{z^{13}}} -40\,{z^{14}}
- 200\,{z^{15}} -5520\,{z^{16}} +5848\,{z^{17}}
-4208\,{z^{18}}
+ O\bigl(z^{19}\bigr).
\end{eqnarray}

We were able to obtain 26th-order results in terms of potentials for
the free energy per site:
\begin{eqnarray}
F &=& \case{3}{2}\,{{z^2}} +{z^6} +3\,{z^{10}} +9\,{z^{12}}
+12\,{z^{14}} +114\,{z^{16}} +\case{829}{3}\,{{z^{18}}} +
1080\,{z^{20}} +5754\,{z^{22}}
\nonumber \\ &&\quad+\;\case{34015}{2}\,{{z^{24}}} +
87396\,{z^{26}} + O\bigl(z^{28}\bigr).
\end{eqnarray}

The internal energy (per link) and the specific heat can be obtained by
\begin{equation}
E(\beta) =  1 - {1\over 3}{\partial F(\beta)\over \partial\beta} =
1 - G((1,0)),
\end{equation}
\begin{equation}
C(\beta) =  - \beta^2\,{\partial E\over\partial\beta} =
{1\over 3}\beta^2 {\partial^2 F(\beta)\over \partial\beta^2}\,.
\end{equation}
The same caveats of the square lattice case apply to the relationship
between $F$ and $G((1,0))$.

\appendix

\section{Values of selected potentials}
\label{values-potentials}

In Sect.~\ref{computing-potentials} we reported a rather general form
for potentials related to ${\rm W}(2,L)$ by bubble insertions.  Here
we present all potentials needed for a 12th-order computation, and
some generalizations.  We also list all other potentials we computed;
their expressions are too long and cumbersome to be reported here, and
they are available upon request from the authors.

Let us introduce a shorthand notation for bubble insertion.  A
sequence of $r$ $[1;2]$ bubbles will be denoted by
\begin{equation}
{\scr I}_r(a_1,b_1,...,a_r,b_r) \equiv [1;2,a_1,b_1]...[1;2,a_r,b_r];
\end{equation}
the arguments $(a_1,b_1,...,a_r,b_r)$ will often be left understood.
A sequence of $r$ $[2;2]$ bubbles and $s$ $[1;3]$ bubbles (with
$[1;2]$ splittings along the $n=3$ line) will be denoted by
\begin{eqnarray}
{\scr I}_{r,s}&&
(a_{1,1},b_{1,1},a_{2,1},b_{2,1},...,a_{1,r},b_{1,r},a_{2,r},b_{2,r};
 p_1,a^1_1,b^1_1,...,a^1_{u_1},b^1_{u_1};...;
 p_s,a^s_1,b^s_1,...,a^s_{u_s},b^s_{u_s}) \equiv \nonumber \\
&&\quad[2,a_{1,1},b_{1,1};2,a_{2,1},b_{2,1}]...
[2,a_{1,r},b_{1,r};2,a_{2,r},b_{2,r}] \nonumber \\
&&\times\quad
[1;3,p_1,{\scr I}_{u_1}(a^1_1,b^1_1,...,a^1_{u_1},b^1_{u_1})]...
[1;3,p_s,{\scr I}_{u_s}(a^s_1,b^s_1,...,a^s_{u_s},b^s_{u_s})];
\end{eqnarray}
the arguments $(a_{1,1},...,b^s_{u_s})$ will also be left understood.

The shorthand notation
\begin{equation}
{\scr Z}_{(r)}(p,q) \equiv z_{(r)}^p d_{(r)}^q.
\end{equation}
will also be used.

Potentials related to ${\rm W}(3,L)$ are easily obtained from the identity
\begin{equation}
{\rm W}(3,L,{\scr I}_r) = {\rm W}(2,a_r,b_r,[1;3,L,{\scr I}_{r-1}]).
\end{equation}
One must however compute explicitly the $r=0$ case
\begin{eqnarray}
{\rm W}(3,L) &=& {1\over N^2} \left[
{\scr Z}_+^{(3)}(L,2) + {\scr Z}_-^{(3)}(L,2) +
{\scr Z}_{2,1;0}(L,2) +
{\scr Z}_+^{(2;1)}(L,2) + {\scr Z}_-^{(2;1)}(L,2)\right]
\nonumber \\ &&\quad-\; 2 N^2 {\rm W}(2,L) - {2\over3}N^4.
\end{eqnarray}

We have also computed ${\rm W}(4,L,{\scr I}_{r,s})$, whose structure
is similar to ${\rm W}(2,L,b,{\scr I}_{r,s})$ presented in
Sect.~\ref{computing-potentials}.  More generally, we must mention
that a generating functional for ${\rm W}(n,L)$ can easily be
constructed by exploiting the general strong-coupling solution of the
chiral chain problem \cite{Brower-Rossi-Tan-1,Brower-Rossi-Tan-2,%
Green-Samuel-2}.  It is easy to get convinced that
\begin{equation}
{1\over2N^2}\ln\sum_{n}\sum_{(r)}
{\scr Z}_{(r)}(L,2) z^{nL} =
\sum_n {\rm W}(n,L) z^{nL},
\end{equation}
where the sum on the l.h.s.\ is extended to all representations with
the same value of $n$.  In principle, generating functionals for
potentials with bubble insertions can also be constructed.

Let us now consider the first nontrivial superskeleton ${\rm Y}$:
\begin{eqnarray}
{\rm Y}(2,a_1,b_1;1;1;1;1;2,a_2,b_2) &=&
z_{1;1}^{a_1+a_2} B_{1;1}^{b_1+b_2} +
\case{1}{2} N^2 z_{1;1}^{a_1} B_{1;1}^{b_1-1}
\left(z_+^{a_2} B_+^{b_2+1} +
z_-^{a_2} B_-^{b_2+1}\right) \nonumber \\
&+& \case{1}{2} N^2 z_{1;1}^{a_2} B_{1;1}^{b_2-1}
\left(z_+^{a_1} B_+^{b_1+1} +
z_-^{a_1} B_-^{b_1+1}\right) - 2 N^2 2^{b_1+b_2};
\end{eqnarray}
this quantity was first introduced in Ref.~\cite{Rossi-Vicari-chiral2}
for the special choice $b_1=b_2=0$, and it was termed
$\widetilde{\rm W}_{a_1,a_2}$.  We have also computed more general
objects of the form ${\rm Y}(2,a_1,b_1,{\scr I}_{r_1,s_1};1;1;1;1;
2,a_2,b_2,{\scr I}_{r_2,s_2})$.

\begin{eqnarray}
&&{\rm Y}(1;1;1;2,a_1,b_1;2,a_2,b_2;2,a_3,b_3) \nonumber \\
&=&
{\scr Z}_{1;1}(a_1+a_2+a_3,1)(d_{1;1}-1)B_{1;1}^{b_1+b_2+b_3} \nonumber \\
&+& \Bigl\{z_{1;1}^{a_1} B_{1;1}^{b_1}
\Bigl[\bigl({\scr Z}_+(a_2,1) B_+^{b_2} +
{\scr Z}_-(a_2,1) B_-^{b_2} \bigr)
\bigl({\scr Z}_+(a_3,1) B_+^{b_3} +
{\scr Z}_-(a_3,1) B_-^{b_3} \bigr) \nonumber \\
&&\qquad-\;
\bigl({\scr Z}_+(a_2+a_3,1) B_+^{b_2+b_3} +
{\scr Z}_-(a_2+a_3,1) B_-^{b_2+b_3}\bigr)\Bigr] \nonumber \\
&&\quad-\; 2 N^2 2^{b_2+b_3} {\rm W}(2,a_1,b_1)\Bigr\} +
\hbox{permutations of indices} - 4 N^4 2^{b_1+b_2+b_3};
\end{eqnarray}
the case $b_1=b_2=b_3=0$ was termed ${\rm W}_{a_1,a_2,a_3}$
in Ref.~\cite{Rossi-Vicari-chiral2}.
% We have also computed the generalization
% ${\rm Y}(1;1;1;2,a_1,b_1;2,a_2,b_2;2,a_3,b_3,[3,p;1])$.

\begin{eqnarray}
&&{\rm Y}(3,p;2,a_1,b_1;1;2,a_2,b_2;1;1) \nonumber \\
&=& N \Bigl[
{\scr Z}_+^{(3)}(p,1) z_+^{a_1+a_2} B_+^{b_1+b_2} +
{\scr Z}_-^{(3)}(p,1) z_-^{a_1+a_2} B_-^{b_1+b_2} \nonumber \\
&&\qquad+\; \case{1}{4} {\scr Z}_{2,1;0}(p,1) \bigl(
  z_+^{a_1+a_2} B_+^{b_1+b_2}
+ 3 z_+^{a_1} B_+^{b_1} z_-^{a_2} B_-^{b_2}
+ 3 z_+^{a_2} B_+^{b_2} z_-^{a_1} B_-^{b_1}
+ z_-^{a_1+a_2} B_-^{b_1+b_2}\bigl) \nonumber \\
&&\qquad+\; {\scr Z}_+^{(2;1)}(p,1) \bigl(
z_{1;1}^{a_1} B_{1;1}^{b_1} z_+^{a_2} B_+^{b_2} +
z_{1;1}^{a_2} B_{1;1}^{b_2} z_+^{a_1} B_+^{b_1} +
z_{1;1}^{a_1+a_2} B_{1;1}^{b_1+b_2} \bigr) \nonumber \\
&&\qquad+\; {\scr Z}_-^{(2;1)}(p,1) \bigl(
z_{1;1}^{a_1} B_{1;1}^{b_1} z_-^{a_2} B_-^{b_2} +
z_{1;1}^{a_2} B_{1;1}^{b_2} z_-^{a_1} B_-^{b_1} +
z_{1;1}^{a_1+a_2} B_{1;1}^{b_1+b_2} \bigr)\Bigr] \nonumber \\
&&\quad-\; 2 N^2 {\rm W}(2,p+a_1,b_1) 2^{b_2} -
2 N^2 {\rm W}(2,p+a_2,b_2) 2^{b_1} \nonumber \\
&&\quad-\; 2 N^2 {\rm W}(2,p,0) 2^{b_1+b_2} -
4 N^4 2^{b_1+b_2}.
\end{eqnarray}
We also computed a few generalizations:
%\[
%\begin{array}{ll}
${\rm Y}(3,p,{\scr I}_{r};2,a_1,b_1;1;2,a_2,b_2;1;1)$, %\ &
%{\rm Y}(3,p_1;2,a_1,b_1,[3,p_2;1];1;2,a_2,b_2;1;1), \\
%{\rm Y}(3,p,[4,q;1];2,a_1;1;2,a_2;1;1), \ &
${\rm Y}(3,p,{\scr I}_{r};2,a_1,b_1;2,a_2,b_2;1;1;2,a_3,b_3)$. %, \\
%{\rm Y}(3,p_1;3,p_2;2,a_1,b_1;1;2,a_2,b_2;1), \ &
%{\rm Y}(4,q;3,p;2,a_1;2,a_2;1;1).
%\end{array}
%\]

Finally, we computed, for a few special values of the indices,
${\rm Y}(3,p_1;2,a_1;1;2,a_2;1;3,p_2)$.

The second nontrivial superskeleton is ${\rm X}$.  We computed
\begin{eqnarray}
&&{\rm X}(2,a_1,b_1;1;2,a_2,b_2;1;1;1;1;1) \nonumber \\
&=& \Bigl({\scr Z}_+(a_1,1)B_+^{b_1} + {\scr Z}_-(a_1,1)B_-^{b_1} +
{\scr Z}_{1;1}(a_1,1)B_{1;1}^{b_1}\Bigr)
\nonumber \\ &&\qquad\times\;
\Bigl({\scr Z}_+(a_2,1)B_+^{b_2} + {\scr Z}_-(a_2,1)B_-^{b_2} +
{\scr Z}_{1;1}(a_2,1)B_{1;1}^{b_2}\Bigr) \nonumber \\
&&\quad+\; {N^4\over16}\,d_{1;1}
\Bigl({\scr Z}_+(a_1,-1)B_+^{b_1} + {\scr Z}_-(a_1,-1)B_-^{b_1} +
{4\over N^2}\,{\scr Z}_{1;1}(a_1,-1)B_{1;1}^{b_1}\Bigr) \nonumber \\
&&\qquad\;\times\;
\Bigl({\scr Z}_+(a_2,-1)B_+^{b_2} + {\scr Z}_-(a_2,-1)B_-^{b_2} +
{4\over N^2}\,{\scr Z}_{1;1}(a_2,-1)B_{1;1}^{b_2}\Bigr) \nonumber \\
&&\quad+\; N^2 {\scr Z}_{1;1}(a_1+a_2,-1) B_{1;1}^{b_1+b_2}(d_{1;1}+4)
\nonumber \\ &&\quad-\;
2 N^2\bigl[2^{b_1} {\rm W}(2,a_2,b_2) + 2^{b_2} {\rm W}(2,a_1,b_1)\bigr]
- 2^{b_1+b_2}(4 N^4 + 2 N^2).
\end{eqnarray}
%We also obtained expressions for
%\[
%\begin{array}{ll}
%{\rm X}(2,a_1,b_1;1;1;1;1;1;2,a_2,b_2;2,a_3,b_3),\ &
%{\rm X}(1;1;1;1;2,a_1,b_1;2,a_2,b_2;2,a_3,b_3;2,a_4,b_4), \\
%{\rm X}(2,a_1,b_1;1;2,a_2,b_2;1;3,p;1;1;1),\ &
%{\rm X}(2,a_1;2,a_2;1;1;1;2,a_3;1;1,a_4), \\
%{\rm X}(2,a_1,b_1;1;1;1;3,1;1;2,a_2,b_2;2,a_3,b_3),\ &
%{\rm X}(3,1;2,a_1;1;2,a_2;1;1;1;1).
%\end{array}
%\]
%
%We also computed several potentials corresponding to higher
%nontrivial superskeletons:
%\[
%\begin{array}{ll}
%{\rm H}(1;2,a_1,b_1;1;2,a_2,b_2;1;1;1;1;2,a_3,b_3), \ &
%{\rm H}(2,a_1,b_1;1;2,a_2,b_2;1;1;1;1;1;2,a_3,b_3), \\
%{\rm H}(2,a_1,b_1;2,a_2,b_2;1;1;1;2,a_3,b_3;1;2,a_4,b_4;1), \ &
%{\rm H}(2,a_1;2,a_2;2,a_3;1;1;2,a_4;2,a_5;1;1), \\
%{\rm H}(3,p;2,a_1;1;1;2,a_2;1;1;2,a_3;1), \ &
%{\rm H}(2,a_1;3,p;2,a_2;1;1;1;1;1;2,a_3), \\
%{\rm H}(3,p;2,a_1;1;2,a_2;1;1;1;1;2,a_3), \ &
%{\rm L}(2,a_1,b_1;1;1;1;2,a_2,b_2;1;1;1;1;1), \\
%{\rm L}(2,a_1;1;2,a_2;1;1;2,a_3;1;1;1;1), \ &
%{\rm L}(1;2,a_1;1;2,a_2;1;2,a_3;1;1;1;1), \\
%{\rm R}(1;1;2,a_1;1;1;1;2,a_2;1;1).
%\end{array}
%\]

\section{List of potentials}
\label{directory-potentials}

We list here all the potentials appearing for the first time at each
order of the strong-coupling expansion; we identify potentials
differing only for the values of $L$ and for the number $b$ of $[1;1]$
bubbles provided that $b>0$ (i.e.\ we don't identify $b=0$ with
$b\ne0$).  In the following formulae, $L$ will indicate a generic
value, and multiple occurrences of $L$ in the same expressions
indicate any combinations of $L$s (i.e.\ they can take different
values).

5th order:
\begin{eqnarray}
&&
  {\rm W}(2,L,b).
\end{eqnarray}

9th order:
\begin{eqnarray}
&&
  {\rm W}(3,L,[1;2,L]),\;{\rm W}(3,L,[1;2,L,b]),\;
  {\rm Y}(2,L; 1; 1; 1; 1; 2,L).
\end{eqnarray}

10th order:
\begin{eqnarray}
&&
  {\rm Y}(2,L; 2,L; 1; 1; 2,L; 1).
\end{eqnarray}

11th order:
\begin{eqnarray}
&&
  {\rm W}(2,L,[2,L;2,L,b]),\;{\rm Y}(3,L; 2,L; 1; 2,L; 1; 1),\;
  {\rm Y}(2,L; 1; 1; 1; 1; 2,L,b).
\end{eqnarray}

12th order:
\begin{eqnarray}
&&
  {\rm W}(2,L,b,[2,L;2,L]),\;
  {\rm W}(3,L,{{[1;2,L]}^2}),\;
  {\rm Y}(2,L; 2,L; 1; 1; 2,L,b; 1),\;
\nonumber \\ &&
  {\rm Y}(2,L,b; 1; 1; 1; 1; 2,L),\;
  {\rm X}(2,L; 1; 2,L; 1; 1; 1; 1; 1).
\end{eqnarray}

13th order:
\begin{eqnarray}
&&
  {\rm W}(3,L,[1;2,L,[1;3,L]]),\;
  {\rm W}(4,L,[1;3,L]),\;
  {\rm W}(4,L,[2,L;2,L,b]),\;
  {\rm W}(2,L,b,[2,L;2,L,b]),\;
\nonumber \\ &&
  {\rm Y}(3,L; 2,L; 1; 2,L,b; 1; 1),\;
  {\rm Y}(3,L; 2,L; 2,L; 1; 1; 2,L),\;
  {\rm Y}(2,L; 1; 1; 1; 1; [1;3,L]),\;
\nonumber \\ &&
  {\rm Y}(2,L,b; 1; 1; 1; 1; 2,L,b),\;
  {\rm Y}(2,L,b; 2,L; 1; 1; 2,L; 1),\;
  {\rm X}(2,L; 1; 1; 1; 1; 1; 2,L; 2,L),\;
\nonumber \\ &&
  {\rm H}(2,L; 1; 2,L; 1; 1; 1; 1; 1; 2,L).
\end{eqnarray}

14th order:
\begin{eqnarray}
&&
  {\rm W}(3,L,[1;2,L,b,[1;3,L]]),\;
  {\rm W}(4,L,[1;3,L,[1;2,L]]),\;
  {\rm W}(2,L,[1;{{[1;2,L]}^2}]),\;
\nonumber \\ &&
  {\rm W}(2,L,[2,L,b;2,L,b]),\;
  {\rm W}(3,L,[1;2,L]\,[1;2,L,b]),\;
  {\rm Y}(3,L; 3,L; 2,L; 1; 2,L; 1),\;
\nonumber \\ &&
  {\rm Y}(3,L; 2,L,b; 1; 2,L; 1; 1),\;
  {\rm Y}(2,L; 1; 1; 1; [1;2,L]; 2,L),\;
  {\rm Y}(2,L; 2,L; 1; 1; [1;3,L]; 1),\;
\nonumber \\ &&
  {\rm Y}(2,L; 2,L,b; 1; 1; 2,L; 1),\;
  {\rm Y}(2,L,[1;3,L]; 1; 1; 1; 1; 2,L),\;
  {\rm X}(1; 1; 1; 1; 2,L; 2,L; 2,L; 2,L),\;
\nonumber \\ &&
  {\rm X}(2,L; 1; 2,L; 1; 3,L; 1; 1; 1),\;
  {\rm X}(2,L; 1; 2,L,b; 1; 1; 1; 1; 1),\;
  {\rm X}(2,L,b; 1; 1; 1; 1; 1; 2,L; 2,L),\;
\nonumber \\ &&
  {\rm H}(1; 2,L; 1; 2,L; 1; 1; 1; 1; 2,L),\;
  {\rm H}(2,L; 2,L; 1; 1; 1; 2,L; 1; 2,L; 1),\;
\nonumber \\ &&
  {\rm R}(1; 1; 2,L; 1; 1; 1; 2,L; 1; 1).
\end{eqnarray}

15th order:
\begin{eqnarray}
&&
  {\rm W}(3,L,[1;2,L,[2,L;2,L]]),\;
  {\rm W}(3,L,[1;2,L,[2,L;2,L,b]]),\;
  {\rm W}(3,L,[3,L;2,L,b]),\;
\nonumber \\ &&
  {\rm W}(3,L,[[1;2,L];2,L]),\;
  {\rm W}(4,L,[2,L,b;2,L,b]),\;
  {\rm W}(3,L,{{[1;2,L]}^3}),\;
\nonumber \\ &&
  {\rm Y}(3,L; 2,L; 1; [1;3,L]; 1; 1),\;
  {\rm Y}(3,L; 2,L; 1; 2,L; 1; 3,L),\;
  {\rm Y}(3,L; 2,L; 2,L; 1; 1; 2,L,b),\;
\nonumber \\ &&
  {\rm Y}(3,L; 2,L,b; 1; 2,L,b; 1; 1),\;
  {\rm Y}(3,L; 2,L,b; 2,L; 1; 1; 2,L),\;
  {\rm Y}(4,L; 3,L; 2,L; 2,L; 1; 1),\;
\nonumber \\ &&
  {\rm Y}(2,L; 1; 1; 1; 1; [2,L;2,L]),\;
  {\rm Y}(2,L; 1; 1; 1; 1; 2,L,[1;3,L]),\;
  {\rm Y}(2,L; 2,L; 1; [1;2,L]; 2,L; 1),\;
\nonumber \\ &&
  {\rm Y}(2,L,b; 1; 1; 1; 1; [1;3,L]),\;
  {\rm Y}(2,L,b; 2,L; 1; 1; 2,L,b; 1),\;
  {\rm Y}(3,L,[1;2,L]; 2,L; 1; 2,L; 1; 1),\;
\nonumber \\ &&
  {\rm Y}(2,L,[1;3,L]; 2,L; 1; 1; 2,L; 1),\;
  {\rm X}(1; 1; 1; 1; 2,L; 2,L; 2,L; 2,L,b),\;
\nonumber \\ &&
  {\rm X}(3,L; 2,L; 1; 2,L; 1; 1; 1; 1),\;
  {\rm X}(2,L; 1; 1; 1; 1; 1; 2,L; 2,L,b),\;
\nonumber \\ &&
  {\rm X}(2,L; 1; 1; 1; 3,L; 1; 2,L; 2,L),\;
  {\rm X}(2,L; 2,L; 1; 1; 1; 2,L; 1; 2,L),\;
\nonumber \\ &&
  {\rm H}(3,L; 2,L; 1; 1; 2,L; 1; 1; 2,L; 1),\;
  {\rm H}(2,L; 1; 2,L; 1; 1; 1; 1; 1; 2,L,b),\;
\nonumber \\ &&
  {\rm H}(2,L; 1; 2,L; 3,L; 1; 1; 1; 1; 2,L),\;
  {\rm H}(2,L; 1; 2,L,b; 1; 1; 1; 1; 1; 2,L),\;
\nonumber \\ &&
  {\rm H}(2,L; 3,L; 2,L; 1; 1; 1; 1; 1; 2,L),\;
  {\rm H}(2,L; 2,L; 2,L; 1; 1; 2,L; 2,L; 1; 1),\;
\nonumber \\ &&
  {\rm L}(1; 2,L; 1; 2,L; 1; 2,L; 1; 1; 1; 1),\;
  {\rm L}(2,L; 1; 1; 1; 2,L; 1; 1; 1; 1; 1),\;
\nonumber \\ &&
  {\rm L}(2,L; 1; 2,L; 1; 1; 2,L; 1; 1; 1; 1).
\end{eqnarray}

\section{Square lattice results for finite $\bbox{N}$}
\label{N-square-results}

We list in the present Appendix the values of the quantities defined
in Sect.~\ref{Greens-functions} without further comments.  The
definition of these quantities and the range of validity of the
results presented here is discussed in Sect.~\ref{Greens-functions}.

% ============================================================
% machine-generated from PropagatorVal.tex
%
\begin{eqnarray}
A_{0}
 &=&\; 1-4\,z
+4\,{z^2}
-4\,{z^3}
+12\,{z^4}
-28\,{z^5}
+4\,{z^6}\,{{-17+13\,{N^2}}\over{}{{N^2}-1}}
\nonumber \\ &+&\; 12\,{z^7}\,{{-13+22\,{N^2}-11\,{N^4}}\over{}
{{{\bigl({N^2}-1\bigr)}^2}}}
+4\,{z^8}\,{{101-158\,{N^2}+81\,{N^4}}\over{}{{{\bigl({N^2}-1\bigr)}^2}}}
\nonumber \\ &+&\; 4\,{z^9}\,{{-243+358\,{N^2}-227\,{N^4}}\over{}
{{{\bigl({N^2}-1\bigr)}^2}}}
\nonumber \\ &+&\; 4\,{z^{10}}\,{{2444-6391\,{N^2}+7193\,{N^4}-
3577\,{N^6}+505\,{N^8}}\over{}{\bigl({N^2}-4\bigr)\,{{\bigl({N^2}-1\bigr)}^3}}}
\nonumber \\ &+&\; 4\,{z^{11}}\,(-23568
+89928\,{N^2}
-154753\,{N^4}
+145700\,{N^6}
\nonumber \\ &&\quad -\; 74058\,{N^8}
+17674\,{N^{10}}
-1571\,{N^{12}})
\nonumber \\ &&\quad \times\;{1\over{{{\bigl({N^2}-4\bigr)}^2}\,
{{\bigl({N^2}-1\bigr)}^4}}}
\nonumber \\ &+&\; 4\,{z^{12}}\,(59312
-227352\,{N^2}
+401499\,{N^4}
-379396\,{N^6}
\nonumber \\ &&\quad +\; 189036\,{N^8}
-43842\,{N^{10}}
+3821\,{N^{12}})
\nonumber \\ &&\quad \times\;{1\over{{{\bigl({N^2}-4\bigr)}^2}\,
{{\bigl({N^2}-1\bigr)}^4}}}
\nonumber \\ &+&\; 4\,{z^{13}}\,(-145008
+560056\,{N^2}
-1018759\,{N^4}
+986360\,{N^6}
\nonumber \\ &&\quad -\; 519366\,{N^8}
+128954\,{N^{10}}
-12235\,{N^{12}})
\nonumber \\ &&\quad \times\;{1\over{{{\bigl({N^2}-4\bigr)}^2}\,
{{\bigl({N^2}-1\bigr)}^4}}}
\nonumber \\ &+&\; 4\,{z^{14}}\,(3296016
-16590056\,{N^2}
+38621009\,{N^4}
\nonumber \\ &&\quad -\; 50952462\,{N^6}
+40255375\,{N^8}
-18791194\,{N^{10}}
+4822317\,{N^{12}}
\nonumber \\ &&\quad -\; 617438\,{N^{14}}
+29153\,{N^{16}})
\nonumber \\ &&\quad \times\;{1\over{\bigl({N^2}-9\bigr)\,
{{\bigl({N^2}-4\bigr)}^2}\,{{\bigl({N^2}-1\bigr)}^5}}}
\nonumber \\ &+&\; 4\,{z^{15}}\,(-73339344
+453776040\,{N^2}
-1300249437\,{N^4}
\nonumber \\ &&\quad +\; 2195554892\,{N^6}
-2367482622\,{N^8}
+1669383326\,{N^{10}}
-759240644\,{N^{12}}
\nonumber \\ &&\quad +\; 212701682\,{N^{14}}
-34650066\,{N^{16}}
+2929064\,{N^{18}}
-98283\,{N^{20}})
\nonumber \\ &&\quad \times\;{1\over{{{\bigl({N^2}-9\bigr)}^2}\,
{{\bigl({N^2}-4\bigr)}^2}\,{{\bigl({N^2}-1\bigr)}^6}}}
+ O\bigl(z^{16}\bigr),
\end{eqnarray}

\begin{eqnarray}
A_{1}
 &=&\; z
+{z^3}
+7\,{z^5}
+4\,{z^6}\,{{{N^2}+1}\over{}{{N^2}-1}}
+3\,{z^7}\,{{13-22\,{N^2}+11\,{N^4}}\over{}{{{\bigl({N^2}-1\bigr)}^2}}}
\nonumber \\ &+&\; 8\,{z^8}\,{{-5-2\,{N^2}+4\,{N^4}}\over{}
{{{\bigl({N^2}-1\bigr)}^2}}}
+{z^9}\,{{251-294\,{N^2}+243\,{N^4}}\over{}{{{\bigl({N^2}-1\bigr)}^2}}}
\nonumber \\ &+&\; 4\,{z^{10}}\,{{-284-13\,{N^2}+441\,{N^4}-
369\,{N^6}+81\,{N^8}}\over{}{\bigl({N^2}-4\bigr)\,{{\bigl({N^2}-1\bigr)}^3}}}
\nonumber \\ &+&\; {z^{11}}\,(24208
-76936\,{N^2}
+129833\,{N^4}
-134180\,{N^6}
+76114\,{N^8}
\nonumber \\ &&\quad -\; 19562\,{N^{10}}
+1819\,{N^{12}})
\nonumber \\ &&\quad \times\;{1\over{{{\bigl({N^2}-4\bigr)}^2}\,
{{\bigl({N^2}-1\bigr)}^4}}}
\nonumber \\ &+&\; 8\,{z^{12}}\,{{-3632+2616\,{N^2}+3565\,{N^4}-
8734\,{N^6}+7962\,{N^8}-2821\,{N^{10}}+315\,{N^{12}}}\over{}
{{{\bigl({N^2}-4\bigr)}^2}\,{{\bigl({N^2}-1\bigr)}^4}}}
\nonumber \\ &+&\; {z^{13}}\,(148848
-456760\,{N^2}
+879351\,{N^4}
-922184\,{N^6}
\nonumber \\ &&\quad +\; 540230\,{N^8}
-146906\,{N^{10}}
+14859\,{N^{12}})
\nonumber \\ &&\quad \times\;{1\over{{{\bigl({N^2}-4\bigr)}^2}\,
{{\bigl({N^2}-1\bigr)}^4}}}
\nonumber \\ &+&\; 4\,{z^{14}}\,(-422352
+796520\,{N^2}
-475949\,{N^4}
-864192\,{N^6}
\nonumber \\ &&\quad +\; 2055327\,{N^8}
-1703150\,{N^{10}}
+634345\,{N^{12}}
-102496\,{N^{14}}
+5771\,{N^{16}})
\nonumber \\ &&\quad \times\;{1\over{\bigl({N^2}-9\bigr)\,
{{\bigl({N^2}-4\bigr)}^2}\,{{\bigl({N^2}-1\bigr)}^5}}}
\nonumber \\ &+&\; {z^{15}}\,(75941712
-403857000\,{N^2}
+1114656565\,{N^4}
\nonumber \\ &&\quad -\; 1899839324\,{N^6}
+2117909054\,{N^8}
-1584540590\,{N^{10}}
+779561548\,{N^{12}}
\nonumber \\ &&\quad -\; 235498610\,{N^{14}}
+40734002\,{N^{16}}
-3591112\,{N^{18}}
+123883\,{N^{20}})
\nonumber \\ &&\quad \times\;{1\over{{{\bigl({N^2}-9\bigr)}^2}\,
{{\bigl({N^2}-4\bigr)}^2}\,{{\bigl({N^2}-1\bigr)}^6}}}
+ O\bigl(z^{16}\bigr),
\end{eqnarray}

\begin{eqnarray}
A_{2,0}
 &=&\;
-{z^6}\,{{{N^2}+1}\over{}{{N^2}-1}}
+2\,{z^8}\,{{4+2\,{N^2}-3\,{N^4}}\over{}{{{\bigl({N^2}-1\bigr)}^2}}}
-4\,{z^9}\,{{1+8\,{N^2}+2\,{N^4}}\over{}{{{\bigl({N^2}-1\bigr)}^2}}}
\nonumber \\ &+&\; {z^{10}}\,{{212-25\,{N^2}-355\,{N^4}+255\,{N^6}-
57\,{N^8}}\over{}{\bigl({N^2}-4\bigr)\,{{\bigl({N^2}-1\bigr)}^3}}}
\nonumber \\ &+&\; 4\,{z^{11}}\,{{-5-100\,{N^2}+137\,{N^4}-12\,{N^6}-
29\,{N^8}}\over{}{{{\bigl({N^2}-1\bigr)}^4}}}
\nonumber \\ &+&\;
2\,{z^{12}}\,{{2608-1944\,{N^2}+195\,{N^4}+1992\,{N^6}-
4278\,{N^8}+2001\,{N^{10}}-250\,{N^{12}}}\over{}
{{{\bigl({N^2}-4\bigr)}^2}\,{{\bigl({N^2}-1\bigr)}^4}}}
\nonumber \\ &+&\; 8\,{z^{13}}\,{{60+1485\,{N^2}-1735\,{N^4}+
520\,{N^6}+396\,{N^8}-144\,{N^{10}}}\over{}{\bigl({N^2}-4\bigr)\,
{{\bigl({N^2}-1\bigr)}^4}}}
\nonumber \\ &+&\; {z^{14}}\,(303120
-666632\,{N^2}
+1554257\,{N^4}
-1748152\,{N^6}
\nonumber \\ &&\quad +\; 205601\,{N^8}
+831342\,{N^{10}}
-468499\,{N^{12}}
+86566\,{N^{14}}
-5155\,{N^{16}})
\nonumber \\ &&\quad \times\;{1\over{\bigl({N^2}-9\bigr)\,
{{\bigl({N^2}-4\bigr)}^2}\,{{\bigl({N^2}-1\bigr)}^5}}}
\nonumber \\ &+&\; 4\,{z^{15}}\,(-3984
-68344\,{N^2}
+247735\,{N^4}
-387636\,{N^6}
\nonumber \\ &&\quad +\; 321005\,{N^8}
-97944\,{N^{10}}
-30420\,{N^{12}}
+21362\,{N^{14}}
-2908\,{N^{16}})
\nonumber \\ &&\quad \times\;{1\over{{{\bigl({N^2}-4\bigr)}^2}\,
{{\bigl({N^2}-1\bigr)}^6}}}
+ O\bigl(z^{16}\bigr),
\end{eqnarray}

\begin{eqnarray}
A_{2,2}
 &=&\;
-2\,{z^8}\,{{{N^2}+1}\over{}{{N^2}-1}}
-2\,{z^9}\,{{1+8\,{N^2}+2\,{N^4}}\over{}{{{\bigl({N^2}-1\bigr)}^2}}}
\nonumber \\ &+&\;
2\,{z^{10}}\,{{-9-7\,{N^2}+9\,{N^4}-12\,{N^6}}\over{}
{{{\bigl({N^2}-1\bigr)}^3}}}
\nonumber \\ &+&\; 2\,{z^{11}}\,{{-5-112\,{N^2}+155\,{N^4}-12\,{N^6}-
35\,{N^8}}\over{}{{{\bigl({N^2}-1\bigr)}^4}}}
\nonumber \\ &+&\; 2\,{z^{12}}\,{{64+54\,{N^2}+100\,{N^4}-52\,{N^6}-
121\,{N^8}}\over{}{{{\bigl({N^2}-1\bigr)}^4}}}
\nonumber \\ &+&\; 4\,{z^{13}}\,{{60+1917\,{N^2}-1843\,{N^4}+
640\,{N^6}+606\,{N^8}-204\,{N^{10}}}\over{}{\bigl({N^2}-4\bigr)\,
{{\bigl({N^2}-1\bigr)}^4}}}
\nonumber \\ &+&\; 2\,{z^{14}}\,(-7392
+8144\,{N^2}
-55774\,{N^4}
+92342\,{N^6}
-35882\,{N^8}
\nonumber \\ &&\quad -\; 16810\,{N^{10}}
+10745\,{N^{12}}
-1412\,{N^{14}})
\nonumber \\ &&\quad \times\;{1\over{{{\bigl({N^2}-4\bigr)}^2}\,
{{\bigl({N^2}-1\bigr)}^5}}}
\nonumber \\ &+&\; 2\,{z^{15}}\,(-4336
-96968\,{N^2}
+298881\,{N^4}
-461772\,{N^6}
\nonumber \\ &&\quad +\; 413767\,{N^8}
-131684\,{N^{10}}
-48836\,{N^{12}}
+32296\,{N^{14}}
-4264\,{N^{16}})
\nonumber \\ &&\quad \times\;{1\over{{{\bigl({N^2}-4\bigr)}^2}\,
{{\bigl({N^2}-1\bigr)}^6}}}
+ O\bigl(z^{16}\bigr),
\end{eqnarray}

\begin{eqnarray}
A_{3,1}
 &=&\; {z^9}\,{{1+8\,{N^2}+2\,{N^4}}\over{}{2\,{{\bigl({N^2}-1\bigr)}^2}}}
+{z^{11}}\,{{5+100\,{N^2}-137\,{N^4}+12\,{N^6}+29\,{N^8}}\over{}
{2\,{{\bigl({N^2}-1\bigr)}^4}}}
\nonumber \\ &+&\;
2\,{z^{12}}\,{{14\,{N^2}+93\,{N^4}+13\,{N^6}}\over{}
{{{\bigl({N^2}-1\bigr)}^3}}}
\nonumber \\ &+&\; {z^{13}}\,{{-60-1485\,{N^2}+1735\,{N^4}-520\,{N^6}-
396\,{N^8}+144\,{N^{10}}}\over{}{\bigl({N^2}-4\bigr)\,
{{\bigl({N^2}-1\bigr)}^4}}}
\nonumber \\ &+&\; 2\,{z^{14}}\,{{11-25\,{N^2}+1585\,{N^4}-
2366\,{N^6}+714\,{N^8}+241\,{N^{10}}}\over{}{{{\bigl({N^2}-1\bigr)}^5}}}
\nonumber \\ &+&\; {z^{15}}\,(4368
+68152\,{N^2}
-230047\,{N^4}
+463796\,{N^6}
-576029\,{N^8}
\nonumber \\ &&\quad +\; 309400\,{N^{10}}
-21332\,{N^{12}}
-20786\,{N^{14}}
+3612\,{N^{16}})
\nonumber \\ &&\quad \times\;{1\over{2\,{{\bigl({N^2}-4\bigr)}^2}\,
{{\bigl({N^2}-1\bigr)}^6}}}
\nonumber \\ &&\quad +\; O\bigl(z^{16}\bigr),
\end{eqnarray}

\begin{eqnarray}
A_{3,3}
 &=&\; 2\,{z^{11}}\,{{2\,{N^2}+{N^4}}\over{}{{{\bigl({N^2}-1\bigr)}^2}}}
+4\,{z^{12}}\,{{2\,{N^2}+9\,{N^4}+{N^6}}\over{}{{{\bigl({N^2}-1\bigr)}^3}}}
\nonumber \\ &+&\; 4\,{z^{13}}\,{{18\,{N^2}+5\,{N^6}+10\,{N^8}}\over{}
{{{\bigl({N^2}-1\bigr)}^4}}}
\nonumber \\ &+&\; 4\,{z^{14}}\,{{1+5\,{N^2}+223\,{N^4}-334\,{N^6}+
106\,{N^8}+35\,{N^{10}}}\over{}{{{\bigl({N^2}-1\bigr)}^5}}}
\nonumber \\ &+&\; 2\,{z^{15}}\,{{5+300\,{N^2}-255\,{N^4}+866\,{N^6}-
1612\,{N^8}+455\,{N^{10}}+274\,{N^{12}}}\over{}{{{\bigl({N^2}-1\bigr)}^6}}}
\nonumber \\ &+&\; O\bigl(z^{16}\bigr),
\end{eqnarray}

\begin{eqnarray}
A_{4,0}
&=&\;
-{z^{12}}\,{{4\,{N^2}+33\,{N^4}+5\,{N^6}}\over{}
{2\,{{\bigl({N^2}-1\bigr)}^3}}}
\nonumber \\ &+&\; {z^{14}}\,{{-2+10\,{N^2}-241\,{N^4}+362\,{N^6}-
105\,{N^8}-37\,{N^{10}}}\over{}{{{\bigl({N^2}-1\bigr)}^5}}}
\nonumber \\ &-&\; 3\,{z^{15}}\,{{1+2\,{N^2}+39\,{N^4}+296\,{N^6}+
28\,{N^8}}\over{}{{{\bigl({N^2}-1\bigr)}^4}}}
+ O\bigl(z^{16}\bigr),
\end{eqnarray}

\begin{eqnarray}
A_{4,2}
&=&\;
-{z^{12}}\,{{2\,{N^2}+9\,{N^4}+{N^6}}\over{}
{{{\bigl({N^2}-1\bigr)}^3}}}
\nonumber \\ &+&\; {z^{14}}\,{{-1-5\,{N^2}-207\,{N^4}+
306\,{N^6}-98\,{N^8}-31\,{N^{10}}}\over{}{{{\bigl({N^2}-1\bigr)}^5}}}
\nonumber \\ &-&\; 2\,{z^{15}}\,{{1+2\,{N^2}+
69\,{N^4}+360\,{N^6}+32\,{N^8}}\over{}{{{\bigl({N^2}-1\bigr)}^4}}}
+ O\bigl(z^{16}\bigr),
\end{eqnarray}

\begin{eqnarray}
A_{4,4}
 &=&\;
-2\,{z^{14}}\,{{4\,{N^4}+{N^6}}\over{}
{{{\bigl({N^2}-1\bigr)}^3}}}
-2\,{z^{15}}\,{{15\,{N^4}+32\,{N^6}+2\,{N^8}}\over{}
{{{\bigl({N^2}-1\bigr)}^4}}}
+ O\bigl(z^{16}\bigr),
\end{eqnarray}

\begin{eqnarray}
A_{5,1} &=& {z^{15}}\,{{1+2\,{N^2}+39\,{N^4}+296\,{N^6}+28\,{N^8}}\over{}
{4\,{{\bigl({N^2}-1\bigr)}^4}}}
+ O\bigl(z^{16}\bigr),
\end{eqnarray}

\begin{eqnarray}
&&A_{5,3}={z^{15}}\,{{15\,{N^4}+32\,{N^6}+2\,{N^8}}\over{}
{2\,{{\bigl({N^2}-1\bigr)}^4}}}
+ O\bigl(z^{16}\bigr),
\end{eqnarray}
%
% ============================================================

% ============================================================
% machine-generated from MassesVal.tex
%
\begin{eqnarray}
\mu_s
&=&\; -\log{}(z)-2\,z
-\case{2}{3}\,{{z^3}}
-2\,{z^4}
+2\,{z^5}\,{{1-21\,{N^2}}\over{}{5\,\bigl({N^2}-1\bigr)}}
\nonumber \\ &+&\; 2\,{z^6}\,{{-5+6\,{N^2}-4\,{N^4}}\over{}
{{{\bigl({N^2}-1\bigr)}^2}}}
+2\,{z^7}\,{{6+114\,{N^2}-155\,{N^4}}\over{}
{7\,{{\bigl({N^2}-1\bigr)}^2}}}
\nonumber \\ &+&\;
2\,{z^8}\,{{22-64\,{N^2}+58\,{N^4}-35\,{N^6}}\over{}
{{{\bigl({N^2}-1\bigr)}^3}}}
\nonumber \\ &+&\; 2\,{z^9}\,{{-104-1286\,{N^2}+9748\,{N^4}-
15331\,{N^6}+8810\,{N^8}-1594\,{N^{10}}}\over{}
{9\,\bigl({N^2}-4\bigr)\,{{\bigl({N^2}-1\bigr)}^4}}}
\nonumber \\ &+&\;
2\,{z^{10}}\,{{-1600+8768\,{N^2}-14596\,{N^4}+
13482\,{N^6}-8645\,{N^8}+2500\,{N^{10}}-260\,{N^{12}}}\over{}
{{{\bigl({N^2}-4\bigr)}^2}\,{{\bigl({N^2}-1\bigr)}^4}}}
\nonumber \\ &+&\; 2\,{z^{11}}\,(-864
-20592\,{N^2}
+266170\,{N^4}
-740050\,{N^6}
\nonumber \\ &&\quad +\; 869302\,{N^8}
-508486\,{N^{10}}
+141781\,{N^{12}}
-14389\,{N^{14}})
\nonumber \\ &&\quad \times\;{1\over
{11\,{{\bigl({N^2}-4\bigr)}^2}\,{{\bigl({N^2}-1\bigr)}^5}}}
\nonumber \\ &+&\; 2\,{z^{12}}\,(-23392
+184400\,{N^2}
-571030\,{N^4}
+920690\,{N^6}
\nonumber \\ &&\quad -\; 927373\,{N^8}
+643901\,{N^{10}}
-289681\,{N^{12}}
+68171\,{N^{14}}
-6577\,{N^{16}})
\nonumber \\ &&\quad \times\;{1\over
{3\,{{\bigl({N^2}-4\bigr)}^2}\,{{\bigl({N^2}-1\bigr)}^6}}}
+ O\bigl(z^{13}\bigr),
\end{eqnarray}

\begin{eqnarray}
\mu_d
&=&\; -\log{}(2z)
-{z^2}
+{z^4}\,{{2-3\,{N^2}}\over{}{{N^2}-1}}
+{z^6}\,{{-79+248\,{N^2}-238\,{N^4}}\over{}
{12\,{{\bigl({N^2}-1\bigr)}^2}}}
\nonumber \\ &+&\; {z^8}\,{{908-5299\,{N^2}+
12080\,{N^4}-13391\,{N^6}+6844\,{N^8}-1088\,{N^{10}}}\over{}
{8\,\bigl({N^2}-4\bigr)\,{{\bigl({N^2}-1\bigr)}^4}}}
\nonumber \\ &+&\; {z^{10}}\,(172816
-1265528\,{N^2}
+3877481\,{N^4}
-6436255\,{N^6}
\nonumber \\ &&\quad +\; 6193060\,{N^8}
-3331306\,{N^{10}}
+864838\,{N^{12}}
-83926\,{N^{14}})
\nonumber \\ &&\quad \times\;{1\over{80\,
{{\bigl({N^2}-4\bigr)}^2}\,{{\bigl({N^2}-1\bigr)}^5}}}
+ O\bigl(z^{12}\bigr),
\end{eqnarray}

\begin{eqnarray}
\chi
&=&\; 1+4\,z
+12\,{z^2}
+36\,{z^3}
+100\,{z^4}
+284\,{z^5}
\nonumber \\ &+&\; 4\,{z^6}\,{{-195+199\,{N^2}}\over{}{{N^2}-1}}
+4\,{z^7}\,{{543-1106\,{N^2}+569\,{N^4}}\over{}
{{{\bigl({N^2}-1\bigr)}^2}}}
\nonumber \\ &+&\; 12\,{z^8}\,{{493-1022\,{N^2}+
537\,{N^4}}\over{}{{{\bigl({N^2}-1\bigr)}^2}}}
\nonumber \\ &+&\; 4\,{z^9}\,{{4067-8550\,{N^2}+
4643\,{N^4}}\over{}{{{\bigl({N^2}-1\bigr)}^2}}}
\nonumber \\ &+&\; 4\,{z^{10}}\,{{44100-149869\,{N^2}+
182355\,{N^4}-90083\,{N^6}+13323\,{N^8}}\over{}
{\bigl({N^2}-4\bigr)\,{{\bigl({N^2}-1\bigr)}^3}}}
\nonumber \\ &+&\; 4\,{z^{11}}\,(481168
-2256648\,{N^2}
+4239673\,{N^4}
-4010564\,{N^6}
\nonumber \\ &&\quad +\; 1961034\,{N^8}
-452890\,{N^{10}}
+38875\,{N^{12}})
\nonumber \\ &&\quad \times\;{1\over
{{{\bigl({N^2}-4\bigr)}^2}\,{{\bigl({N^2}-1\bigr)}^4}}}
\nonumber \\ &+&\; 4\,{z^{12}}\,(1299728
-6161736\,{N^2}
+11712625\,{N^4}
\nonumber \\ &&\quad -\; 11228076\,{N^6}
+5568300\,{N^8}
-1301614\,{N^{10}}
+112879\,{N^{12}})
\nonumber \\ &&\quad \times\;{1\over
{{{\bigl({N^2}-4\bigr)}^2}\,{{\bigl({N^2}-1\bigr)}^4}}}
\nonumber \\ &+&\; 4\,{z^{13}}\,(3526000
-16894776\,{N^2}
+32535287\,{N^4}
\nonumber \\ &&\quad -\; 31650616\,{N^6}
+15957638\,{N^8}
-3784938\,{N^{10}}
+332699\,{N^{12}})
\nonumber \\ &&\quad \times\;{1\over
{{{\bigl({N^2}-4\bigr)}^2}\,{{\bigl({N^2}-1\bigr)}^4}}}
\nonumber \\ &+&\; 4\,{z^{14}}\,(85479984
-509240632\,{N^2}
+1277936387\,{N^4}
\nonumber \\ &&\quad -\; 1742114314\,{N^6}
+1385750301\,{N^8}
-641889782\,{N^{10}}
+163736431\,{N^{12}}
\nonumber \\ &&\quad -\; 20707322\,{N^{14}}
+976227\,{N^{16}})
\nonumber \\ &&\quad \times\;{1\over
{\bigl({N^2}-9\bigr)\,{{\bigl({N^2}-4\bigr)}^2}\,{{\bigl({N^2}-1\bigr)}^5}}}
\nonumber \\ &+&\; 4\,{z^{15}}\,(2078977104
-14813458920\,{N^2}
+45842748421\,{N^4}
\nonumber \\ &&\quad -\; 80400814700\,{N^6}
+87447748126\,{N^8}
-60663055822\,{N^{10}}
+26640570340\,{N^{12}}
\nonumber \\ &&\quad -\; 7157578626\,{N^{14}}
+1112718466\,{N^{16}}
-90043336\,{N^{18}}
+2904339\,{N^{20}})
\nonumber \\ &&\quad \times\;{1\over
{{{\bigl({N^2}-9\bigr)}^2}\,{{\bigl({N^2}-4\bigr)}^2}\,
{{\bigl({N^2}-1\bigr)}^6}}}
+ O\bigl(z^{16}\bigr),
\end{eqnarray}

\begin{eqnarray}
M^2_G
&=&\; {z^{-1}}-4+3\,z
+2\,{z^3}
-4\,{z^4}\,{{{N^2}+1}\over{}{{N^2}-1}}
+2\,{z^5}\,{{-5-4\,{N^2}+6\,{N^4}}\over{}
{{{\bigl({N^2}-1\bigr)}^2}}}
\nonumber \\ &+&\; 8\,{z^6}\,{{6+2\,{N^2}-5\,{N^4}}\over{}
{{{\bigl({N^2}-1\bigr)}^2}}}
+4\,{z^7}\,{{-32-42\,{N^2}+21\,{N^4}}\over{}
{{{\bigl({N^2}-1\bigr)}^2}}}
\nonumber \\ &+&\; 8\,{z^8}\,{{184+98\,{N^2}-376\,{N^4}+
203\,{N^6}-37\,{N^8}}\over{}{\bigl({N^2}-4\bigr)\,
{{\bigl({N^2}-1\bigr)}^3}}}
\nonumber \\ &+&\; 2\,{z^9}\,{{-6576-3656\,{N^2}+
30629\,{N^4}-32810\,{N^6}+14735\,{N^8}-3245\,{N^{10}}+
275\,{N^{12}}}\over{}{{{\bigl({N^2}-4\bigr)}^2}\,{{\bigl({N^2}-1\bigr)}^4}}}
\nonumber \\ &+&\; 8\,{z^{10}}\,{{4016+4136\,{N^2}-
16789\,{N^4}+15608\,{N^6}-7937\,{N^8}+2257\,{N^{10}}-
238\,{N^{12}}}\over{}{{{\bigl({N^2}-4\bigr)}^2}\,{{\bigl({N^2}-1\bigr)}^4}}}
\nonumber \\ &+&\; 2\,{z^{11}}\,(-34560
-63904\,{N^2}
+148416\,{N^4}
-142210\,{N^6}
\nonumber \\ &&\quad +\; 74870\,{N^8}
-19631\,{N^{10}}
+1658\,{N^{12}})
\nonumber \\ &&\quad \times\;{1\over
{{{\bigl({N^2}-4\bigr)}^2}\,{{\bigl({N^2}-1\bigr)}^4}}}
\nonumber \\ &+&\; 8\,{z^{12}}\,(186048
+87904\,{N^2}
-727204\,{N^4}
+1149812\,{N^6}
\nonumber \\ &&\quad -\; 1175339\,{N^8}
+716157\,{N^{10}}
-232150\,{N^{12}}
+34766\,{N^{14}}
-1906\,{N^{16}})
\nonumber \\ &&\quad \times\;{1\over
{\bigl({N^2}-9\bigr)\,{{\bigl({N^2}-4\bigr)}^2}\,{{\bigl({N^2}-1\bigr)}^5}}}
\nonumber \\ &+&\; 2\,{z^{13}}\,(-14774400
+8055072\,{N^2}
+58712088\,{N^4}
\nonumber \\ &&\quad -\; 170052698\,{N^6}
+261060513\,{N^8}
-232073848\,{N^{10}}
+117200494\,{N^{12}}
\nonumber \\ &&\quad -\; 33867953\,{N^{14}}
+5360045\,{N^{16}}
-433255\,{N^{18}}
+13878\,{N^{20}})
\nonumber \\ &&\quad \times\;{1\over
{{{\bigl({N^2}-9\bigr)}^2}\,{{\bigl({N^2}-4\bigr)}^2}\,
{{\bigl({N^2}-1\bigr)}^6}}}
+ O\bigl(z^{14}\bigr),
\end{eqnarray}

\begin{eqnarray}
F
&=&\; 2\,{z^2}
+2\,{z^4}
+4\,{z^6}
+{z^8}\,{{14-30\,{N^2}+19\,{N^4}}\over{}
{{{\bigl({N^2}-1\bigr)}^2}}}
\nonumber \\ &+&\; 8\,{z^{10}}\,{{7-16\,{N^2}+12\,{N^4}}\over{}
{{{\bigl({N^2}-1\bigr)}^2}}}
\nonumber \\ &+&\; 2\,{z^{12}}\,(5952
-30144\,{N^2}
+63364\,{N^4}
-67746\,{N^6}
+37500\,{N^8}
\nonumber \\ &&\quad -\; 9589\,{N^{10}}
+906\,{N^{12}})
\nonumber \\ &&\quad \times\;{1\over
{3\,{{\bigl({N^2}-4\bigr)}^2}\,{{\bigl({N^2}-1\bigr)}^4}}}
\nonumber \\ &+&\; 4\,{z^{14}}\,(4704
-25200\,{N^2}
+55366\,{N^4}
-62068\,{N^6}
+36768\,{N^8}
\nonumber \\ &&\quad -\; 10012\,{N^{10}}
+1009\,{N^{12}})
\nonumber \\ &&\quad \times\;{1\over
{{{\bigl({N^2}-4\bigr)}^2}\,{{\bigl({N^2}-1\bigr)}^4}}}
\nonumber \\ &+&\; {z^{16}}\,(15230592
-119651328\,{N^2}
+409792072\,{N^4}
\nonumber \\ &&\quad -\; 799817292\,{N^6}
+974422200\,{N^8}
-760569676\,{N^{10}}
+375693595\,{N^{12}}
\nonumber \\ &&\quad -\; 112460534\,{N^{14}}
+19258826\,{N^{16}}
-1690814\,{N^{18}}
+58471\,{N^{20}})
\nonumber \\ &&\quad \times\;{1\over
{2\,{{\bigl({N^2}-9\bigr)}^2}\,{{\bigl({N^2}-4\bigr)}^2}\,
{{\bigl({N^2}-1\bigr)}^6}}}
\nonumber \\ &+&\; 8\,{z^{18}}\,(237447936
-2045848320\,{N^2}
+7796836128\,{N^4}
\nonumber \\ &&\quad -\; 17299894704\,{N^6}
+24693730379\,{N^8}
-23648019056\,{N^{10}}
+15403609647\,{N^{12}}
\nonumber \\ &&\quad -\; 6787736700\,{N^{14}}
+1995934103\,{N^{16}}
-381749639\,{N^{18}}
+45159907\,{N^{20}}
\nonumber \\ &&\quad -\; 2974083\,{N^{22}}
+82898\,{N^{24}})
\nonumber \\ &&\quad \times\;{1\over
{3\,{{\bigl({N^2}-9\bigr)}^2}\,{{\bigl({N^2}-4\bigr)}^4}\,
{{\bigl({N^2}-1\bigr)}^6}}}
+ O\bigl(z^{20}\bigr).
\end{eqnarray}
%
% ============================================================

\section{The Gaussian model on the honeycomb lattice}
\label{honeycomb-gaussian}

There are a few subtleties in the analysis of models on the honeycomb
lattice, that are best illustrated by considering a simple Gaussian
model of random walks.  The essential point is related to the fact
that lattice sites are not all related by a translation group: only
points at an even distance (in the number of lattice links) are
related by such a symmetry.  As a consequence, it is convenient to
define even and odd fields $\phi_e$, $\phi_o$, according to the parity
of the corresponding lattice sites with respect to an arbitrarily
chosen origin, and even and odd correlation functions
$G_e = \left<\phi_e\phi_e\right> = \left<\phi_o\phi_o\right>$,
$G_o = \left<\phi_e\phi_o\right> = \left<\phi_o\phi_e\right>$.

Let us represent the Cartesian coordinates of the (finite, periodic)
lattice sites by
\begin{mathletters}
\begin{eqnarray}
(x,y)_e &=& \left(\case{3}{2} m, \case{1}{2} \sqrt{3} n\right) a, \\
(x,y)_o &=& \left(\case{3}{2} m + 1, \case{1}{2} \sqrt{3} n\right) a,
\end{eqnarray}
where $m$ and $n$ are integer numbers satisfying the conditions $0\le
m < L_1$, $0\le n < 2 L_2$, and $m+n$ is even.  The total number of
lattice points is $2 L_1{\times} L_2$, the number of links is
$3 L_1{\times} L_2$, and the number of plaquettes is
$L_1{\times} L_2$.
\end{mathletters}

The finite-lattice Fourier transform is consistently defined by
\begin{equation}
G_e(p) = \sum_{x\ {\rm even}} e^{ip\cdot x} G_e(x),
\end{equation}
and similarly for $G_o(p)$; the set of momenta is
\begin{equation}
p = {2\pi\over a}\left({2\over3}\,{\tilde m\over L_1},
{1\over\sqrt{3}}\,{\tilde n\over L_2}\right),
\end{equation}
with $\tilde m$ and $\tilde n$ integers, and  $0\le\tilde m<L_1$,
$0\le\tilde n<L_2$.

In a random walk model where walks of length $\nu$ are weighted by a
factor $\beta^\nu$, it is easy to establish from the recursion
relations the following relationships:
\begin{mathletters}
\begin{eqnarray}
G_e(p) &=& \beta e^{-ip_1} \left[1 +
2\cos\case{1}{2}\sqrt{3}p_2 \, e^{3ip_1/2}\right] G_o(p) + 1, \\
G_o(p) &=& \beta e^{ip_1} \left[1 +
2\cos\case{1}{2}\sqrt{3}p_2 \, e^{-3ip_1/2}\right] G_e(p),
\end{eqnarray}
\end{mathletters}
where $G_e(p)$ and $G_o(p)$ are the Fourier transform
of $G_e(x)$  and $G_o(x)$ respectively.
As a consequence, we obtain the even momentum-space Green's functions
in the form
\begin{equation}
G_e(p) = {1\over 1 - \beta^2\left[1 + 4 \cos^2 \case{1}{2}\sqrt{3}p_2
+ 4 \cos\case{1}{2}\sqrt{3}p_2\cos\case{3}{2}p_1\right]}\,.
\label{Gep}
\end{equation}

The critical value of $\beta$ is easily found to be
$\beta_c = \case{1}{3}$, and as a consequence we find the massless
lattice (even) propagator
\begin{equation}
\Delta_e(p) =  {9\over 8 - 4 \cos^2 \case{1}{2}\sqrt{3}p_2
+ 4 \cos\case{1}{2}\sqrt{3}p_2\cos\case{3}{2}p_1}
\;\mathop{\longrightarrow}_{p\to0} \;{2\over p^2}\,.
\end{equation}
The odd propagator is simply
\begin{equation}
\Delta_o(p) =  \case{1}{3}e^{ip_1}\left[1 +
2 \cos\case{1}{2}\sqrt{3}p_2 \, e^{-3ip_1/2}\right]
\Delta_e(p).
\end{equation}

The structure of the propagator in the Gaussian model offers an
important indication about the possibility of exponentiation in
wall-wall correlations.  Let us indeed recall that exponentiation
corresponds to a simple structure
\begin{equation}
{1\over A - B \cos p}
\end{equation}
in the corresponding propagator.  Let us now observe that
Eq.~(\ref{Gep}) implies
\begin{eqnarray}
G_e(p_1,0) &=&
{1\over 1 - \beta^2\left[5 + 4 \cos\case{3}{2}p_1\right]}, \\
G_e(0,p_2) + G_o(0,p_2) &=&
{1 \over 1 - \beta\left[1 + 2 \cos\case{1}{2}\sqrt{3}p_2\right]}.
\end{eqnarray}
We have therefore the possibility of defining two different
exponentiated ``wall-wall'' correlation functions, i.e.
\begin{eqnarray}
G^{{\rm(w)}}_1(x) &=& \sum_y G_e(x,y), \\
G^{{\rm(w)}}_2(x) &=& \sum_y \left[G_e(x,y) + G_o(x,y)\right].
\end{eqnarray}

Even in more general models, in the strong-coupling domain, for
sufficiently large lattice distance exponentiation will hold for the
correlation functions $G^{{\rm(w)}}_1(x)$ and $G^{{\rm(w)}}_2(x)$.  If
we take into account the discrete rotational symmetry of the honeycomb
lattice we may easily recognize that the above correlations can be
referred to directions differing by a $\pi/6$ angle, and this is the
maximal violation of the full rotational symmetry one can find on this
kind of lattice.  Therefore the ratio of the two different correlation
lengths one may define is an optimal measurement of the violation of
rotational invariance in the model under examination, in analogy with
the side/diagonal mass ratio of the square lattice.

\section{Honeycomb lattice results for finite $\bbox{N}$}
\label{N-honeycomb-results}

We list in the present Appendix the values of the quantities defined
in Sect.~\ref{honeycomb-lattice} without further comments.  The
definition of these quantities presented here is discussed in
Sect.~\ref{honeycomb-lattice}.

% ============================================================
% machine-generated from MassHexVal.tex
%

\begin{eqnarray}
\mu_1 = &\case{2}{3}&\Bigl[
-\log{}2z^2
-{z^2}
-\case{1}{2}\,{z^4}
+{z^6}\,{{-14+46\,{N^2}-41\,{N^4}}\over{}{6\,{{\bigl({N^2}-1\bigr)}^2}}}
\nonumber \\ &+&\; {z^8}\,{{9-27\,{N^2}+35\,{N^4}-37\,{N^6}}\over{}
  {4\,{{\bigl({N^2}-1\bigr)}^3}}}
\nonumber \\ &+&\;
{z^{10}}\,{{-194+1136\,{N^2}-2539\,{N^4}+2546\,{N^6}
  -1174\,{N^8}}\over{}{20\,{{\bigl({N^2}-1\bigr)}^4}}}
\nonumber \\ &+&\; {z^{12}}\,(-6208
+49952\,{N^2}
-175780\,{N^4}
+360896\,{N^6}
-469546\,{N^8}
\nonumber \\ &&\quad +\; 395168\,{N^{10}}
-205477\,{N^{12}}
+56246\,{N^{14}}
-6169\,{N^{16}})
\nonumber \\ &&\quad \times\;{1\over{24\,{{\bigl({N^2}-4\bigr)}^2}\,
  {{\bigl({N^2}-1\bigr)}^6}}}
\nonumber \\ &+&\; {z^{14}}\,(-204800
+2013696\,{N^2}
-8684352\,{N^4}
+21407856\,{N^6}
\nonumber \\ &&\quad -\; 33518156\,{N^8}
+34888595\,{N^{10}}
-24526859\,{N^{12}}
+11475108\,{N^{14}}
\nonumber \\ &&\quad -\; 3337802\,{N^{16}}
+545557\,{N^{18}}
-39943\,{N^{20}})
\nonumber \\ &&\quad \times\;{1\over{56\,{{\bigl({N^2}-4\bigr)}^3}\,
  {{\bigl({N^2}-1\bigr)}^7}}}
\nonumber \\ &+&\; {z^{16}}\,(-253184
+2885888\,{N^2}
-14879968\,{N^4}
+46028112\,{N^6}
\nonumber \\ &&\quad -\; 95149021\,{N^8}
+138282732\,{N^{10}}
-144193790\,{N^{12}}
+108442897\,{N^{14}}
\nonumber \\ &&\quad -\; 58120117\,{N^{16}}
+21234008\,{N^{18}}
-4926461\,{N^{20}}
+646061\,{N^{22}}
-36361\,{N^{24}})
\nonumber \\ &&\quad \times\;{1\over{8\,{{\bigl({N^2}-4\bigr)}^4}\,
  {{\bigl({N^2}-1\bigr)}^8}}}\Bigr]
+ O\bigl(z^{18}\bigr),
\end{eqnarray}

\begin{eqnarray}
\mu_2 = &\case{2}{3}&\sqrt{3}\Bigl[-\log{}z-z
+\case{1}{2}\,{z^2}
-\case{1}{3}\,{z^3}
-\case{3}{4}\,{{z^4}}
-\case{1}{5}\,{z^5}
-{z^6}\,{{1+11\,{N^2}}\over{}{6\,\bigl({N^2}-1\bigr)}}
\nonumber \\ &+&\; 2\,{z^7}\,{{-4+{N^2}-18\,{N^4}}
  \over{}{7\,{{\bigl({N^2}-1\bigr)}^2}}}
+{z^8}\,{{11-81\,{N^2}+{N^4}+29\,{N^6}}
  \over{}{8\,{{\bigl({N^2}-1\bigr)}^3}}}
\nonumber \\ &+&\; 5\,{z^9}\,{{2+12\,{N^2}+24\,{N^4}-11\,{N^6}}
  \over{}{9\,{{\bigl({N^2}-1\bigr)}^3}}}
\nonumber \\ &+&\; {z^{10}}\,{{19-277\,{N^2}+127\,{N^4}-319\,{N^6}}
  \over{}{10\,{{\bigl({N^2}-1\bigr)}^3}}}
\nonumber \\ &+&\; {z^{11}}\,{{-34+26\,{N^2}-622\,{N^4}+587\,{N^6}
  -452\,{N^8}}\over{}{11\,{{\bigl({N^2}-1\bigr)}^4}}}
\nonumber \\ &+&\; {z^{12}}\,{{-132+1045\,{N^2}-1813\,{N^4}
  +2158\,{N^6}-558\,{N^8}-115\,{N^{10}}-69\,{N^{12}}}
  \over{}{4\,\bigl({N^2}-4\bigr)\,{{\bigl({N^2}-1\bigr)}^5}}}
\nonumber \\ &+&\; {z^{13}}\,(-1472
+5408\,{N^2}
-24716\,{N^4}
+48156\,{N^6}
-97577\,{N^8}
\nonumber \\ &&\quad +\; 118419\,{N^{10}}
-65911\,{N^{12}}
+18318\,{N^{14}}
-2614\,{N^{16}})
\nonumber \\ &&\quad \times\;{1\over{13\,{{\bigl({N^2}-4\bigr)}^2}\,{
  {\bigl({N^2}-1\bigr)}^6}}}
\nonumber \\ &+&\; {z^{14}}\,(5568
-91344\,{N^2}
+393204\,{N^4}
-964687\,{N^6}
+1168570\,{N^8}
\nonumber \\ &&\quad -\; 845302\,{N^{10}}
+432359\,{N^{12}}
-133335\,{N^{14}}
+18275\,{N^{16}}
-885\,{N^{18}})
\nonumber \\ &&\quad \times\;{1\over{7\,{{\bigl({N^2}-4\bigr)}^3}\,
  {{\bigl({N^2}-1\bigr)}^6}}}
\nonumber \\ &+&\; {z^{15}}\,(-25024
+122896\,{N^2}
-380772\,{N^4}
+569163\,{N^6}
-1634619\,{N^8}
\nonumber \\ &&\quad +\; 3326040\,{N^{10}}
-3347466\,{N^{12}}
+1923822\,{N^{14}}
-642978\,{N^{16}}
+116764\,{N^{18}}
\nonumber \\ &&\quad -\; 7576\,{N^{20}})
\nonumber \\ &&\quad \times\;{1\over{15\,{{\bigl({N^2}-4\bigr)}^3}\,
  {{\bigl({N^2}-1\bigr)}^7}}}
\nonumber \\ &+&\; {z^{16}}\,(43712
-656912\,{N^2}
+3472772\,{N^4}
-10632331\,{N^6}
\nonumber \\ &&\quad +\; 18690504\,{N^8}
-20221284\,{N^{10}}
+15125616\,{N^{12}}
-8770834\,{N^{14}}
+4069272\,{N^{16}}
\nonumber \\ &&\quad -\; 1342540\,{N^{18}}
+260364\,{N^{20}}
-24083\,{N^{22}})
\nonumber \\ &&\quad \times\;{1\over{16\,{{\bigl({N^2}-4\bigr)}^3}\,
  {{\bigl({N^2}-1\bigr)}^8}}}\Bigr]
+ O\bigl(z^{17}\bigr),
\end{eqnarray}
%
% ============================================================

% ============================================================
% machine-generated from MiscHexVal.tex
%
\begin{eqnarray}
\chi{}
 &=&\; 1+3\,z
+6\,{z^2}
+12\,{z^3}
+24\,{z^4}
+48\,{z^5}
+90\,{z^6}
+174\,{z^7}
\nonumber \\ &+&\; 12\,{z^8}\,{{-28+29\,{N^2}}\over{}{{N^2}-1}}
+18\,{z^9}\,{{36-74\,{N^2}+39\,{N^4}}\over{}{{{\bigl({N^2}-1\bigr)}^2}}}
\nonumber \\ &+&\; 6\,{z^{10}}\,{{-203+625\,{N^2}-
649\,{N^4}+232\,{N^6}}\over{}{{{\bigl({N^2}-1\bigr)}^3}}}
\nonumber \\ &+&\; 6\,{z^{11}}\,{{388-1592\,{N^2}+
2469\,{N^4}-1725\,{N^6}+469\,{N^8}}\over{}{{{\bigl({N^2}-1\bigr)}^4}}}
\nonumber \\ &+&\; 6\,{z^{12}}\,{{736-3056\,{N^2}+
4802\,{N^4}-3407\,{N^6}+943\,{N^8}}\over{}{{{\bigl({N^2}-1\bigr)}^4}}}
\nonumber \\ &+&\; 6\,{z^{13}}\,{{1398-5850\,{N^2}+
9318\,{N^4}-6743\,{N^6}+1922\,{N^8}}\over{}{{{\bigl({N^2}-1\bigr)}^4}}}
\nonumber \\ &+&\; 6\,{z^{14}}\,{{-10520+47078\,{N^2}-
82728\,{N^4}+70812\,{N^6}-28891\,{N^8}+3951\,{N^{10}}}\over{}
{\bigl({N^2}-4\bigr)\,{{\bigl({N^2}-1\bigr)}^4}}}
\nonumber \\ &+&\; 6\,{z^{15}}\,(79712
-380208\,{N^2}
+730934\,{N^4}
-716200\,{N^6}
\nonumber \\ &&\quad +\; 372088\,{N^8}
-91009\,{N^{10}}
+8228\,{N^{12}})
\nonumber \\ &&\quad \times\;{1\over{{{\bigl({N^2}-4\bigr)}^2}\,
{{\bigl({N^2}-1\bigr)}^4}}}
\nonumber \\ &+&\; 6\,{z^{16}}\,(600192
-3641312\,{N^2}
+9388472\,{N^4}
-13344618\,{N^6}
\nonumber \\ &&\quad +\; 11337734\,{N^8}
-5813592\,{N^{10}}
+1719475\,{N^{12}}
-267280\,{N^{14}}
+16906\,{N^{16}})
\nonumber \\ &&\quad \times\;{1\over
{{{\bigl({N^2}-4\bigr)}^3}\,{{\bigl({N^2}-1\bigr)}^5}}}
\nonumber \\ &+&\; 6\,{z^{17}}\,(4531200
-33355776\,{N^2}
+107774336\,{N^4}
\nonumber \\ &&\quad -\; 200423616\,{N^6}
+236373604\,{N^8}
-183407604\,{N^{10}}
+93969088\,{N^{12}}
\nonumber \\ &&\quad -\; 31097639\,{N^{14}}
+6341524\,{N^{16}}
-722188\,{N^{18}}
+35215\,{N^{20}})
\nonumber \\ &&\quad \times\;{1\over
{{{\bigl({N^2}-4\bigr)}^4}\,{{\bigl({N^2}-1\bigr)}^6}}}
\nonumber \\ &+&\; 6\,{z^{18}}\,(-8505600
+71553024\,{N^2}
-268314976\,{N^4}
\nonumber \\ &&\quad +\; 590267696\,{N^6}
-843497305\,{N^8}
+818280951\,{N^{10}}
-546354607\,{N^{12}}
\nonumber \\ &&\quad +\; 249146769\,{N^{14}}
-75461014\,{N^{16}}
+14404142\,{N^{18}}
-1562138\,{N^{20}}
\nonumber \\ &&\quad +\; 73433\,{N^{22}})
\nonumber \\ &&\quad \times\;{1\over
{{{\bigl({N^2}-4\bigr)}^4}\,{{\bigl({N^2}-1\bigr)}^7}}}
\nonumber \\ &+&\; 6\,{z^{19}}\,(16021504
-151615488\,{N^2}
+647496704\,{N^4}
\nonumber \\ &&\quad -\; 1646428032\,{N^6}
+2769093400\,{N^8}
-3234820688\,{N^{10}}
+2679887383\,{N^{12}}
\nonumber \\ &&\quad -\; 1577573745\,{N^{14}}
+650694030\,{N^{16}}
-182224525\,{N^{18}}
+32769509\,{N^{20}}
\nonumber \\ &&\quad -\; 3399903\,{N^{22}}
+154769\,{N^{24}})
\nonumber \\ &&\quad \times\;{1\over
{{{\bigl({N^2}-4\bigr)}^4}\,{{\bigl({N^2}-1\bigr)}^8}}}
\nonumber \\ &+&\; 6\,{z^{20}}\,(-270452736
+2603295744\,{N^2}
-11339998976\,{N^4}
\nonumber \\ &&\quad +\; 29517395456\,{N^6}
-51057737736\,{N^8}
+61725236828\,{N^{10}}
-53378649928\,{N^{12}}
\nonumber \\ &&\quad +\; 33218024089\,{N^{14}}
-14765553520\,{N^{16}}
+4594385593\,{N^{18}}
-966784918\,{N^{20}}
\nonumber \\ &&\quad +\; 129788872\,{N^{22}}
-9932849\,{N^{24}}
+325065\,{N^{26}})
\nonumber \\ &&\quad \times\;{1\over
{\bigl({N^2}-9\bigr)\,{{\bigl({N^2}-4\bigr)}^4}\,{{\bigl({N^2}-1\bigr)}^8}}}
+ O\bigl(z^{21}\bigr),
\end{eqnarray}

\begin{eqnarray}
M^2_G
&=&\; \case{4}{3}\,{z^{-1}}-4+\case{8}{3}\,{z}
-8\,{z^6}\,{{{N^2}+1}\over{}{{N^2}-1}}
+8\,{z^7}\,{{-7-10\,{N^2}+5\,{N^4}}\over{}{3\,{{\bigl({N^2}-1\bigr)}^2}}}
\nonumber \\ &+&\; 8\,{z^8}\,{{-3-7\,{N^2}+4\,{N^4}+{N^6}}\over{}
{{{\bigl({N^2}-1\bigr)}^3}}}
+8\,{z^9}\,{{-12-20\,{N^2}+29\,{N^6}-6\,{N^8}}\over{}
{3\,{{\bigl({N^2}-1\bigr)}^4}}}
\nonumber \\ &+&\; 8\,{z^{10}}\,{{12-6\,{N^2}+15\,{N^4}+
8\,{N^6}-11\,{N^8}}\over{}{{{\bigl({N^2}-1\bigr)}^4}}}
\nonumber \\ &+&\; 8\,{z^{11}}\,{{-82-34\,{N^2}+165\,{N^4}-
247\,{N^6}+36\,{N^8}}\over{}{3\,{{\bigl({N^2}-1\bigr)}^4}}}
\nonumber \\ &+&\; 8\,{z^{12}}\,{{-160-432\,{N^2}+698\,{N^4}-
409\,{N^6}-2\,{N^8}-18\,{N^{10}}}\over{}{\bigl({N^2}-4\bigr)\,
{{\bigl({N^2}-1\bigr)}^4}}}
\nonumber \\ &+&\; 8\,{z^{13}}\,{{-1776-12392\,{N^2}+3665\,{N^4}+
1850\,{N^6}-1811\,{N^8}-666\,{N^{10}}+73\,{N^{12}}}\over{}
{3\,{{\bigl({N^2}-4\bigr)}^2}\,{{\bigl({N^2}-1\bigr)}^4}}}
\nonumber \\ &+&\; 8\,{z^{14}}\,(3328
+9920\,{N^2}
+23760\,{N^4}
-41452\,{N^6}
+12242\,{N^8}
\nonumber \\ &&\quad +\; 5969\,{N^{10}}
-5077\,{N^{12}}
+657\,{N^{14}}
-5\,{N^{16}})
\nonumber \\ &&\quad \times\;{1\over{{{\bigl({N^2}-4\bigr)}^3}\,
{{\bigl({N^2}-1\bigr)}^5}}}
\nonumber \\ &+&\; 8\,{z^{15}}\,(-114432
+400128\,{N^2}
-1536160\,{N^4}
+1733488\,{N^6}
\nonumber \\ &&\quad -\; 323935\,{N^8}
-602516\,{N^{10}}
+449683\,{N^{12}}
-124964\,{N^{14}}
+5895\,{N^{16}}
\nonumber \\ &&\quad +\; 1675\,{N^{18}}
-75\,{N^{20}})
\nonumber \\ &&\quad \times\;{1\over{3\,{{\bigl({N^2}-4\bigr)}^4}\,
{{\bigl({N^2}-1\bigr)}^6}}}
\nonumber \\ &+&\; 8\,{z^{16}}\,(-91136
+441856\,{N^2}
-1078144\,{N^4}
+768256\,{N^6}
\nonumber \\ &&\quad +\; 450812\,{N^8}
-1054246\,{N^{10}}
+839396\,{N^{12}}
-464465\,{N^{14}}
+194981\,{N^{16}}
\nonumber \\ &&\quad -\; 54586\,{N^{18}}
+9491\,{N^{20}}
-690\,{N^{22}})
\nonumber \\ &&\quad \times\;{1\over
{{{\bigl({N^2}-4\bigr)}^4}\,{{\bigl({N^2}-1\bigr)}^7}}}
\nonumber \\ &+&\; 8\,{z^{17}}\,(-425728
+1766144\,{N^2}
-1985440\,{N^4}
-5221520\,{N^6}
\nonumber \\ &&\quad +\; 17290561\,{N^8}
-24613836\,{N^{10}}
+24266712\,{N^{12}}
-17725908\,{N^{14}}
\nonumber \\ &&\quad +\; 9035120\,{N^{16}}
-2989557\,{N^{18}}
+594069\,{N^{20}}
-63766\,{N^{22}}
+2193\,{N^{24}})
\nonumber \\ &&\quad \times\;{1\over
{3\,{{\bigl({N^2}-4\bigr)}^4}\,{{\bigl({N^2}-1\bigr)}^8}}}
\nonumber \\ &+&\; 8\,{z^{18}}\,(-1661184
+3794944\,{N^2}
+7844000\,{N^4}
\nonumber \\ &&\quad -\; 54955728\,{N^6}
+119244695\,{N^8}
-171667137\,{N^{10}}
+175851305\,{N^{12}}
\nonumber \\ &&\quad -\; 120836371\,{N^{14}}
+53853632\,{N^{16}}
-15114634\,{N^{18}}
+2512157\,{N^{20}}
\nonumber \\ &&\quad -\; 229837\,{N^{22}}
+9716\,{N^{24}}
-526\,{N^{26}})
\nonumber \\ &&\quad \times\;{1\over{\bigl({N^2}-9\bigr)\,
{{\bigl({N^2}-4\bigr)}^4}\,{{\bigl({N^2}-1\bigr)}^8}}}
+ O\bigl(z^{19}\bigr),
\end{eqnarray}

\begin{eqnarray}
F
&=&\; \case{3}{2}\,{{z^2}}
+{z^6}
+3\,{z^{10}}
\nonumber \\ &+&\;
{z^{12}}\,{{4-18\,{N^2}+32\,{N^4}-27\,{N^6}+18\,{N^8}}\over{}
{2\,{{\bigl({N^2}-1\bigr)}^4}}} +12 \,{z^{14}}
\nonumber \\ &+&\;
6\,{z^{16}}\,{{3-16\,{N^2}+32\,{N^4}-29\,{N^6}+19\,{N^8}}\over{}
{{{\bigl({N^2}-1\bigr)}^4}}}
\nonumber \\ &+&\; {z^{18}}\,(49920
-268032\,{N^2}
+595744\,{N^4}
-732080\,{N^6}
\nonumber \\ &&\quad +\; 536515\,{N^8}
-239924\,{N^{10}}
+65162\,{N^{12}}
-10174\,{N^{14}}
+829\,{N^{16}})
\nonumber \\ &&\quad \times\;{1\over{3\,{{\bigl({N^2}-4\bigr)}^4}\,
{{\bigl({N^2}-1\bigr)}^4}}}
\nonumber \\ &+&\; 3\,{z^{20}}\,(92
-890\,{N^2}
+3756\,{N^4}
-9052\,{N^6}
+13836\,{N^8}
\nonumber \\ &&\quad -\; 14023\,{N^{10}}
+9380\,{N^{12}}
-3690\,{N^{14}}
+720\,{N^{16}})
\nonumber \\ &&\quad \times\;{1\over{2\,{{\bigl({N^2}-1\bigr)}^8}}}
\nonumber \\ &+&\; 3\,{z^{22}}\,(36864
-368640\,{N^2}
+1666560\,{N^4}
-4505344\,{N^6}
\nonumber \\ &&\quad +\; 8111248\,{N^8}
-10267152\,{N^{10}}
+9373736\,{N^{12}}
-6230112\,{N^{14}}
\nonumber \\ &&\quad +\; 3011024\,{N^{16}}
-1020866\,{N^{18}}
+227573\,{N^{20}}
-30248\,{N^{22}}
+1918\,{N^{24}})
\nonumber \\ &&\quad \times\;{1\over{{{\bigl({N^2}-4\bigr)}^4}\,
{{\bigl({N^2}-1\bigr)}^8}}}
\nonumber \\ &+&\; {z^{24}}\,(7215630336
-96566722560\,{N^2}
+592839316224\,{N^4}
\nonumber \\ &&\quad -\; 2214635962752\,{N^6}
+5639017965384\,{N^8}
-10385973650808\,{N^{10}}
\nonumber \\ &&\quad +\; 14322131449584\,{N^{12}}
-15072994434276\,{N^{14}}
+12212762091344\,{N^{16}}
\nonumber \\ &&\quad -\; 7627059006688\,{N^{18}}
+3656946091530\,{N^{20}}
-1335009872497\,{N^{22}}
\nonumber \\ &&\quad +\; 366664198242\,{N^{24}}
-74601540139\,{N^{26}}
+11013519148\,{N^{28}}
-1144634371\,{N^{30}}
\nonumber \\ &&\quad +\; 79672354\,{N^{32}}
-3373733\,{N^{34}}
+68030\,{N^{36}})
\nonumber \\ &&\quad \times\;{1\over{4\,{{\bigl({N^2}-9\bigr)}^4}\,
{{\bigl({N^2}-4\bigr)}^4}\,{{\bigl({N^2}-1\bigr)}^{10}}}}
\nonumber \\ &+&\; 3\,{z^{26}}\,(-275712
+3201536\,{N^2}
-17022176\,{N^4}
\nonumber \\ &&\quad +\; 54825776\,{N^6}
-119171333\,{N^8}
+184722213\,{N^{10}}
-210428660\,{N^{12}}
\nonumber \\ &&\quad +\; 178609366\,{N^{14}}
-112928266\,{N^{16}}
+52388456\,{N^{18}}
-17133329\,{N^{20}}
\nonumber \\ &&\quad +\; 3699171\,{N^{22}}
-474540\,{N^{24}}
+29132\,{N^{26}})
\nonumber \\ &&\quad \times\;{1\over{{{\bigl({N^2}-4\bigr)}^4}\,
{{\bigl({N^2}-1\bigr)}^9}}}
+ O\bigl(z^{28}\bigr).
\end{eqnarray}
%
% ============================================================

% ========================= REFERENCES =========================

\end{document}